\documentclass[longauth]{aa} 
\pdfoutput=1
\usepackage{graphicx}
\usepackage{txfonts}
\usepackage{pdflscape}

\usepackage{natbib,twoopt}
\usepackage{amsmath} 
\usepackage[breaklinks=true]{hyperref} 
\bibpunct{(}{)}{;}{a}{}{,} 
\makeatletter
\newcommandtwoopt{\citeads}[3][][]{\href{http://adsabs.harvard.edu/abs/#3}%
{\def\hyper@linkstart##1##2{}%
\let\hyper@linkend\@empty\citealp[#1][#2]{#3}}}
\newcommandtwoopt{\citepads}[3][][]{\href{http://adsabs.harvard.edu/abs/#3}%
{\def\hyper@linkstart##1##2{}%
\let\hyper@linkend\@empty\citep[#1][#2]{#3}}}
\newcommandtwoopt{\citetads}[3][][]{\href{http://adsabs.harvard.edu/abs/#3}%
{\def\hyper@linkstart##1##2{}%
\let\hyper@linkend\@empty\citet[#1][#2]{#3}}}
\newcommandtwoopt{\citeyearads}[3][][]%
{\href{http://adsabs.harvard.edu/abs/#3}
{\def\hyper@linkstart##1##2{}%
\let\hyper@linkend\@empty\citeyear[#1][#2]{#3}}}
\makeatother

\usepackage{color}
\definecolor{mygreen}{RGB}{0,128,0}

\hypersetup{colorlinks=true,linkcolor=blue,citecolor=blue,urlcolor=blue}

\usepackage{etoolbox}
\makeatletter
\patchcmd\@combinedblfloats{\box\@outputbox}{\unvbox\@outputbox}{}{%
   \errmessage{\noexpand\@combinedblfloats could not be patched}%
}%
 \makeatother

\newcommand{\gacs}[1]{{\footnotesize\texttt{#1}}}
\usepackage{wasysym}

\begin{document}

   \title{\textit{Gaia} Data Release 2 -- The astrometric solution}

\author{
L.~Lindegren\inst{\ref{inst:lund}}\fnmsep\thanks{Corresponding author: L. Lindegren\newline
e-mail: \href{mailto:lennart@astro.lu.se}{\tt lennart@astro.lu.se}}
\and J.~Hern{\'a}ndez\inst{\ref{inst:esac}}
\and A.~Bombrun\inst{\ref{inst:hespaceesac}}
\and S.~Klioner\inst{\ref{inst:tud}}
\and U.~Bastian\inst{\ref{inst:ari}}
\and M.~Ramos-Lerate\inst{\ref{inst:vitrocisetesac}}
\and A.~de~Torres\inst{\ref{inst:hespaceesac}}
\and H.~Steidelm\"{u}ller\inst{\ref{inst:tud}}
\and C.~Stephenson\inst{\ref{inst:vegaesac}}
\and D.~Hobbs\inst{\ref{inst:lund}}
\and U.~Lammers\inst{\ref{inst:esac}}
\and M.~Biermann\inst{\ref{inst:ari}}
\and R.~Geyer\inst{\ref{inst:tud}}
\and T.~Hilger\inst{\ref{inst:tud}}
\and D.~Michalik\inst{\ref{inst:lund}}
\and U.~Stampa\inst{\ref{inst:ari}}
\and P.J.~McMillan\inst{\ref{inst:lund}}
\and J.~Casta{\~n}eda\inst{\ref{inst:ieec}}
\and M.~Clotet\inst{\ref{inst:ieec}}  
\and G.~Comoretto\inst{\ref{inst:vegaesac}}
\and M.~Davidson\inst{\ref{inst:uoe}} 
\and C.~Fabricius\inst{\ref{inst:ieec}} 
\and G.~Gracia\inst{\ref{inst:poesac}}
\and N.C.~Hambly\inst{\ref{inst:uoe}} 
\and A.~Hutton\inst{\ref{inst:auroraesac}}
\and A.~Mora\inst{\ref{inst:auroraesac}} 
\and J.~Portell\inst{\ref{inst:ieec}} 
\and F.~van~Leeuwen\inst{\ref{inst:ioa}}
\and U.~Abbas\inst{\ref{inst:oato}} 
\and A.~Abreu\inst{\ref{inst:deimosesac}}
\and M.~Altmann\inst{\ref{inst:ari},\ref{inst:paris}}
\and A.~Andrei\inst{\ref{inst:rio}} 
\and E.~Anglada\inst{\ref{inst:sercoesac}}
\and  L.~Balaguer-N\'u\~nez\inst{\ref{inst:ieec}}
\and C.~Barache\inst{\ref{inst:paris}}
\and U.~Becciani\inst{\ref{inst:catania}}
\and S.~Bertone\inst{\ref{inst:oato},\ref{inst:paris},\ref{inst:bern}}
\and L.~Bianchi\inst{\ref{inst:eurix}}
\and S.~Bouquillon\inst{\ref{inst:paris}}
\and G.~Bourda\inst{\ref{inst:bordeaux}}
\and T.~Br{\"u}semeister\inst{\ref{inst:ari}}
\and B.~Bucciarelli\inst{\ref{inst:oato}}
\and D.~Busonero\inst{\ref{inst:oato}}
\and R.~Buzzi\inst{\ref{inst:oato}}
\and R.~Cancelliere\inst{\ref{inst:torinocs}}
\and T.~Carlucci\inst{\ref{inst:paris}}
\and P.~Charlot\inst{\ref{inst:bordeaux}}
\and N.~Cheek\inst{\ref{inst:sercoesac}}
\and M.~Crosta\inst{\ref{inst:oato}}
\and C.~Crowley\inst{\ref{inst:hespaceesac}}
\and J.~de~Bruijne\inst{\ref{inst:estec}}
\and F.~de~Felice\inst{\ref{inst:padova}}
\and R.~Drimmel\inst{\ref{inst:oato}} 
\and P.~Esquej\inst{\ref{inst:rheaesac}}
\and A.~Fienga\inst{\ref{inst:oca2}}
\and E.~Fraile\inst{\ref{inst:rheaesac}}
\and M.~Gai\inst{\ref{inst:oato}} 
\and N.~Garralda\inst{\ref{inst:ieec}}
\and J.J.~Gonz{\'a}lez-Vidal\inst{\ref{inst:ieec}}  
\and R.~Guerra\inst{\ref{inst:esac}}
\and M.~Hauser\inst{\ref{inst:ari},\ref{inst:mpia}}
\and W.~Hofmann\inst{\ref{inst:ari}}
\and B.~Holl\inst{\ref{inst:geneva}}
\and S.~Jordan\inst{\ref{inst:ari}}
\and M.G.~Lattanzi\inst{\ref{inst:oato}}
\and H.~Lenhardt\inst{\ref{inst:ari}}
\and S.~Liao\inst{\ref{inst:oato},\ref{inst:shanghai},\ref{inst:beijing}}
\and E.~Licata\inst{\ref{inst:eurix}}
\and T.~Lister\inst{\ref{inst:las-cumbres}}
\and W.~L{\"o}ffler\inst{\ref{inst:ari}}
\and J.~Marchant\inst{\ref{inst:liverpool}}
\and J.-M.~Martin-Fleitas\inst{\ref{inst:auroraesac}}
\and R.~Messineo\inst{\ref{inst:altec}}
\and F.~Mignard\inst{\ref{inst:oca1}} 
\and R.~Morbidelli\inst{\ref{inst:oato}}
\and E.~Poggio\inst{\ref{inst:torinofisica},\ref{inst:oato}}
\and A.~Riva\inst{\ref{inst:oato}}
\and N.~Rowell\inst{\ref{inst:uoe}}
\and E.~Salguero\inst{\ref{inst:atgesac}}
\and M.~Sarasso\inst{\ref{inst:oato}}
\and E.~Sciacca\inst{\ref{inst:catania}}
\and H.~Siddiqui\inst{\ref{inst:vegaesac}}
\and R.L.~Smart\inst{\ref{inst:oato}}
\and A.~Spagna\inst{\ref{inst:oato}}
\and I.~Steele\inst{\ref{inst:liverpool}}
\and F.~Taris\inst{\ref{inst:paris}}
\and J.~Torra\inst{\ref{inst:ieec}}
\and A.~van~Elteren\inst{\ref{inst:leiden}}
\and W.~van~Reeven\inst{\ref{inst:auroraesac}}
\and A.~Vecchiato\inst{\ref{inst:oato}}
}  

\institute{
Lund Observatory, Department of Astronomy and Theoretical Physics, Lund University, Box 43, SE-22100, Lund, Sweden
\label{inst:lund}
\and 
ESA, European Space Astronomy Centre, Camino Bajo del Castillo s/n, 28691 Villanueva de la Ca{\~n}ada, Spain
\label{inst:esac}
\and
HE Space Operations BV for ESA/ESAC, Camino Bajo del Castillo s/n, 28691 Villanueva de la Ca{\~n}ada, Spain
\label{inst:hespaceesac}
\and
Lohrmann-Observatorium, Technische Universit\"{a}t Dresden, Mommsenstrasse 13, 01062 Dresden, Germany
\label{inst:tud}
\and
Astronomisches Rechen-Institut, Zentrum f\"ur Astronomie der Universit\"at Heidelberg, M\"onchhofstra{\ss}e 14, 69120 Heidelberg,
Germany
\label{inst:ari}
\and
Vitrociset Belgium for ESA/ESAC, Camino Bajo del Castillo s/n, 28691 Villanueva de la Ca{\~n}ada, Spain
\label{inst:vitrocisetesac}
\and
Telespazio Vega UK Ltd for ESA/ESAC, Camino Bajo del Castillo s/n, 28691 Villanueva de la Ca{\~n}ada, Spain
\label{inst:vegaesac}
\and
Institut de Ci\`encies del Cosmos, Universitat de Barcelona (IEEC-UB), Mart\'i Franqu\`es 1, 
E-08028 Barcelona, Spain 
\label{inst:ieec}
\and
Institute for Astronomy, School of Physics and Astronomy, University of Edinburgh, 
Royal Observatory, Blackford Hill, Edinburgh, EH9~3HJ, 
United Kingdom
\label{inst:uoe}
\and
Gaia Project Office for DPAC/ESA, Camino Bajo del Castillo s/n, 28691 Villanueva de la Ca{\~n}ada, Spain
\label{inst:poesac}
\and
Aurora Technology for ESA/ESAC, Camino Bajo del Castillo s/n, 28691 Villanueva de la Ca{\~n}ada, Spain
\label{inst:auroraesac}
\and
Institute of Astronomy, University of Cambridge, Madingley Road, Cambridge CB3~0HA, UK
\label{inst:ioa}
\and
Istituto Nazionale di Astrofisica, Osservatorio Astrofisico di Torino, Via Osservatorio 20, Pino Torinese, Torino, 10025, Italy
\label{inst:oato}
\and 
Elecnor Deimos Space for ESA/ESAC, Camino Bajo del Castillo s/n, 28691 Villanueva de la Ca{\~n}ada, Spain
\label{inst:deimosesac}
\and
SYRTE, Observatoire de Paris, Universit{\'e} PSL, CNRS, Sorbonne Universit{\'e}, LNE, 
61 avenue de l’Observatoire, 75014 Paris, France
\label{inst:paris}
\and
GPA-Observatorio National/MCT, Rua Gal. Jose Cristino 77, CEP 20921-400, Rio de Janeiro, Brazil
\label{inst:rio}
\and
Serco for ESA/ESAC, Camino Bajo del Castillo s/n, 28691 Villanueva de la Ca{\~n}ada, Spain
\label{inst:sercoesac}
\and 
INAF, Osservatorio Astrofisico di Catania, Catania, Italy
\label{inst:catania}
\and
Astronomical Institute, Bern University, Sidlerstrasse 5, CH-3012 Bern, Switzerland
\label{inst:bern}
\and
EURIX S.r.l., Corso Vittorio Emanuele II, 61, 10128 Torino, Italy
\label{inst:eurix}
\and
Laboratoire d'Astrophysique de Bordeaux, Universit{\'e} de Bordeaux, CNRS, B18N, all{\'e}e Geoffroy Saint-Hilaire, 33615 Pessac, France 
\label{inst:bordeaux}
\and
University of Torino, Department of Computer Science, Torino, Italy
\label{inst:torinocs}
\and
ESA, European Space Research and Technology Centre, Keplerlaan 1, 2200 AG, Noordwijk, The Netherlands
\label{inst:estec}
\and
University of Padova, Via Marzolo 8, Padova, I-35131, Italy
\label{inst:padova}
\and
RHEA for ESA/ESAC, Camino Bajo del Castillo s/n, 28691 Villanueva de la Ca{\~n}ada, Spain
\label{inst:rheaesac}
\and
Université Côte d'Azur, Observatoire de la C{\^o}te d’Azur, CNRS, GéoAzur, 250 rue Albert Einstein, CS 10269, 06905 Sophia Antipolis Cedex, France 
\label{inst:oca2}
\and
Max Planck Institute for Astronomy, K{\"o}nigstuhl 17, 69117 Heidelberg,
Germany
\label{inst:mpia}
\and
Observatoire Astronomique de l'Universit{\'e} de Gen{\`e}ve, Sauverny, 
Chemin des Maillettes 51, CH-1290 Versoix, Switzerland
\label{inst:geneva}
\and
Shanghai Astronomical Observatory, Chinese Academy of Sciences, 80 Nandan Rd, 200030 Shanghai, China
\label{inst:shanghai}
\and
School of Astronomy and Space Science, University of Chinese Academy of Sciences, Beijing 100049, China
\label{inst:beijing}
\and
Las Cumbres Observatory, 6740 Cortona Dr. 102, Goleta, CA 93117, United States of America 
\label{inst:las-cumbres}
\and
Astrophysics Research Institute, Liverpool John Moores University, 146 Brownlow Hill,
Liverpool L3~5RF, United Kingdom
\label{inst:liverpool}
\and
ALTEC, Corso Marche 79, Torino, 10146 Italy
\label{inst:altec}
\and
Université Côte d'Azur, Observatoire de la C{\^o}te d’Azur, CNRS, Laboratoire Lagrange, Bd de l'Observatoire, CS 34229, 06304 Nice Cedex 4, France
\label{inst:oca1}
\and
Universit{\`a} di Torino, Dipartimento di Fisica, via P.~Giuria 1, 10125, Torino, Italy
\label{inst:torinofisica}
\and
ATG for ESA/ESAC, Camino Bajo del Castillo s/n, 28691 Villanueva de la Ca{\~n}ada, Spain
\label{inst:atgesac}
\and
Sterrewacht Leiden, Leiden University, P.O.\ Box 9513, 2300 RA, Leiden, The Netherlands
\label{inst:leiden}
} 

   \date{ }

 
\abstract
  {\textit{Gaia} Data Release 2 (\textit{Gaia} DR2) contains results for 1693~million sources 
  in the magnitude range 3 to 21 based on observations collected by the European Space
  Agency \textit{Gaia} satellite during the first 22~months of its operational phase.}  
  {We describe the input data, models, and processing used for the astrometric content
  of \textit{Gaia} DR2, and the validation of these results performed within the astrometry task.}  
  {Some 320~billion centroid positions from the pre-processed astrometric CCD observations 
  were used to estimate the five astrometric parameters (positions, parallaxes, and proper motions) 
  for 1332~million sources, and approximate positions at the reference epoch J2015.5 for an additional 
  361~million mostly faint sources. These data were calculated in two steps. First, the satellite attitude 
  and the astrometric calibration parameters of the CCDs were obtained in an astrometric global iterative 
  solution for 16~million selected sources, using about 1\% of the input data. This primary solution
  was tied to the extragalactic International Celestial Reference System (ICRS) by means of quasars. 
  The resulting attitude and calibration were then used to calculate the astrometric parameters of all 
  the sources. 
  Special validation solutions were 
  used to characterise the random and systematic errors in parallax and proper motion.}  
  {For the sources with five-parameter astrometric solutions, the median uncertainty 
  in parallax and position at the reference epoch J2015.5 
  is about 0.04~mas for bright ($G<14$~mag) sources,
  0.1~mas at $G=17$~mag, and 0.7~mas at $G=20$~mag. In the proper motion components the
  corresponding uncertainties are 0.05, 0.2, and 1.2~mas~yr$^{-1}$, respectively. The optical 
  reference frame defined by \textit{Gaia} DR2 is aligned with ICRS and is non-rotating with 
  respect to the quasars to within 0.15~mas~yr$^{-1}$. From the quasars and validation solutions
  we estimate that systematics in the parallaxes depending on position, magnitude, and
  colour are generally below 0.1~mas, but the parallaxes are on the whole too small by about 0.03~mas.
  Significant spatial correlations of up to 0.04~mas in parallax and 
  0.07~mas~yr$^{-1}$ in proper motion are seen on small ($<1$~deg) and intermediate
  (20~deg) angular scales. Important statistics and information for the users of the 
  \textit{Gaia} DR2 astrometry are given in the appendices.}  
  {}


   \titlerunning{\textit{Gaia} Data Release 2 -- Astrometry} 
   \authorrunning{L.~Lindegren et al.}

   \maketitle

%

\section{Introduction} 
\label{sec:intro}

\textit{Gaia} DR2 (\citeads{DPACP-36}), the second release of data 
from the European Space Agency mission \textit{Gaia} 
(\citeads{2016A&A...595A...1G}), contains provisional 
results based on observations collected during the first 22 months since 
the start of nominal operations in July 2014. The astrometric data in \textit{Gaia} 
DR2 include the five astrometric parameters (position, parallax, 
and proper motion) for 1332~million sources, and the approximate positions at
epoch J2015.5 for an additional 361~million mostly faint sources with too few
observations for a reliable five-parameter solution.
The limiting magnitude is $G\simeq 21.0$. 
The bright limit is $G\simeq 3$, although stars with $G\lesssim 6$ generally
have inferior astrometry due to calibration issues. The data are publicly 
available in the online \textit{Gaia} Archive at 
\href{https://archives.esac.esa.int/gaia}{\tt https://archives.esac.esa.int/gaia}.

This paper gives an overview of the astrometric processing for \textit{Gaia} DR2 
and describes the main characteristics of the results. Further details
are provided in the online documentation of the \textit{Gaia} Archive 
and in specialised papers. In contrast to the Tycho-\textit{Gaia} astrometric solution 
(TGAS; \citeads{2016A&A...595A...4L}) in \textit{Gaia} DR1
(\citeads{2016A&A...595A...2G}), the present solution does not incorporate any 
astrometric information from \textsc{Hipparcos} and Tycho-2, and the results are 
therefore independent of these catalogues. Similarly to \textit{Gaia} DR1, all sources are
treated as single stars and thus representable by the five astrometric parameters.
For unresolved binaries (separation $\lesssim 100$~mas), the results thus refer to 
the photocentre, while for resolved binaries the results may refer to either component
and are sometimes spurious due to confusion of the components.
For a very small number of nearby sources, perspective effects due to their radial 
motions were taken into account. 

The input data for the astrometric solutions are summarised in Sect.~\ref{sec:data}.
A central part of the processing carried out by the \textit{Gaia} Data Processing
and Analysis Consortium (DPAC; \citeads{2016A&A...595A...1G}) is the astrometric 
global iterative solution (AGIS) described in 
\citetads[][hereafter the AGIS paper]{2012A&A...538A..78L}, and the present
results were largely computed using the models and algorithms described in that paper.
However, a few major additions have been made since 2012, and they are outlined in 
Sect.~\ref{sec:models}. Section~\ref{sec:solutions} describes the main steps of the 
solutions. The validation of the results carried out by the astrometry team of DPAC
primarily aimed at estimating the level of systematic errors; this is described in
Sect.~\ref{sec:valid}, with the main conclusions in Sect.~\ref{sec:concl}.
Three appendices give statistics and other information of potential interest to users
of the \textit{Gaia} DR2 astrometry.

\section{Data used}
\label{sec:data}

The main input to the astrometric solutions are one- or two-dimensional
measurements of the locations of point-source images on \textit{Gaia}'s CCD
detectors, derived by the image parameter determination
(Sect.~\ref{sec:imagepar}) in the pre-processing of the raw \textit{Gaia} data
\citepads{2016A&A...595A...3F}.  The CCD measurements must be assigned
to specific sources, so that all the measurements of a given source can be
considered together in the astrometric solution.  This is achieved by a
dedicated cross-matching procedure following the same overall three-step scheme
as for \textit{Gaia} DR1.  First all sources close to a detection -- the
candidate matches -- are found. This is done for the full set of observations,
using updated calibrations and an extended attitude covering also time
intervals that may later be excluded.
Next, the detections are divided into isolated groups consisting of the
smallest possible sets of detections with candidate matches to the same
sources, such that a given candidate source only appears in one group. Finally,
each group is resolved into clusters of detections and each cluster assigned to
one source.  What is done differently from \textit{Gaia} DR1 is the way the
clusters are formed. For \textit{Gaia} DR1, this involved a simple nearest-neighbour
algorithm, applied to one detection at a time, without a global view of the
group.  For \textit{Gaia} DR2, a more elaborate clustering algorithm was
used, giving better results in dense areas and performing much better for sources 
with high proper motions as it includes the detection of linear motion. The overall 
cross-match scheme is
described in \citet{DPACP-45}. For \textit{Gaia} DR2, about 52~billion detections
were processed, but 11~billion were considered spurious and therefore did not
take part in the cross matching. The remaining 41~billion transits were
matched to 2583~million sources, of which a significant number could
still be spurious. Even among the clearly non-spurious sources, many had too 
few or too poor observations to make it to the release, which therefore has 
a total of 1693~million sources.

A second important input to the astrometric solution for \textit{Gaia} DR2 is
the colour information, available for most of the sources thanks to the early
photometric processing of data from the blue and red photometers (BP and RP;
\citeads{2017A&A...599A..32V}; \citeads{DPACP-44}; \citeads{DPACP-40}). 
This processing used astrometric data
(source and attitude parameters) taken from a provisional
astrometric solution (Sect.~\ref{sec:agis21}).

Additional input data are obtained from the basic angle monitor (BAM;
Sect.~\ref{sec:bam}) and the orbit reconstruction and time synchronisation
 data provided by the Mission Operations Centre (Sect.~5.3 in
\citeads{2016A&A...595A...1G}).

\subsection{Time coverage}
\label{sec:coverage}

\textit{Gaia} DR2 is based on data collected from the start of the nominal observations on
2014 July 25 (10:30 UTC) until 2016 May 23 (11:35 UTC), or 668~days.
However, the astrometric solution for this release did not use the observations during 
the first month after commissioning, when a special scanning mode (the ecliptic pole 
scanning law, EPSL) was employed. The data for the astrometry therefore start 
on 2014 Aug 22 (21:00 UTC) and cover 640~days or 1.75~yr, with some 
interruptions mentioned below.

Hereafter we use the onboard mission timeline (OBMT) to label onboard events; it is
expressed as the number of nominal revolutions of exactly 21\,600~s (6~h) onboard time from 
an arbitrary origin. The approximate relation between OBMT (in revolutions) and barycentric
coordinate time (TCB, in Julian years) at \textit{Gaia} is
\begin{equation}\label{eq:obmt}
\text{TCB} \simeq \text{J}2015.0 + (\text{OBMT} - 1717.6256~\text{rev})/(1461~\text{rev~yr}^{-1}) \, .
\end{equation} 
The nominal observations start at OBMT 1078.38~rev. The astrometric solution used
data in the interval OBMT 1192.13--3750.56~rev, with major gaps at 
OBMT 1316.49--1389.11~rev and 2324.90--2401.56~rev due to mirror
decontamination events and the subsequent recovery of thermal equilibrium. 
Planned maintenance operations (station-keeping manoeuvres, telescope 
refocusing, etc.), micrometeoroid hits, and
other events caused additional gaps that rarely exceeded a few hours.

The reference epoch used for the astrometry in \textit{Gaia} DR2 is J2015.5 (see Sect.~3.1),
approximately half-way through the observation period used in the solution. This reference 
epoch, chosen to minimise correlations between the positions and proper motions, is 
0.5~Julian year later than the reference epoch for \textit{Gaia} DR1; this difference must be 
taken into account when comparing positional data from the two releases.

\subsection{Image parameters}
\label{sec:imagepar}

Image parameters are obtained by fitting a model profile to the photon
counts in the observation window centred on the source in the CCD pixel stream.
The model profile is a point spread function (PSF) for a two-dimensional window
and a line spread function (LSF) in the more common case of a one-dimensional
window (for details on the CCD operations, see Sect.~3.3.2 in
\citeads{2016A&A...595A...1G}).  The main image parameters are the estimated
one- or two-dimensional location of the image centroid (defined by the origin
of the fitted PSF or LSF) and the integrated flux of the image.  The image parameter
determination for \textit{Gaia} DR2 is essentially the same as for \textit{Gaia} DR1 (see
Sect.~5 in \citeads{2016A&A...595A...3F}). In particular, the fitted PSF and
LSF were assumed to be independent of time and of the colour and magnitude 
of the source, 
which means that centroid shifts depending on time, colour, and magnitude 
need to be modelled in the astrometric solution (Sect.~\ref{sec:cal}).
For \textit{Gaia} DR2, all image parameters have been re-determined in a uniform way and
recovering observations that for various reasons did not enter \textit{Gaia}
DR1. The sky background has been recalibrated, and we now have a far more
detailed calibration of the electronic bias of the CCDs \citep{DPACP-29}.
Important for sources brighter than $G\simeq 12$ is a more reliable
identification of saturated samples, which are not used in the PSF fitting. 

All observations provide an along-scan (AL) measurement, consisting of the
precise time at which the image centroid passes a fiducial line on the CCD. The
two-dimensional windows, mainly used for bright sources ($G\lesssim 13$),
provide in addition a less precise across-scan (AC) measurement from the pixel
column of the image centroid. A singe transit over the field of 
view thus generates ten AL measurements and one or ten AC measurements,
although some of them may be discarded in the subsequent processing. The 
first observation in a transit is always made with the sky mapper (SM); it is
two-dimensional, but less precise in both AL and AC than the subsequent
observations in the astrometric field (AF) because of the special readout mode of 
the SM detectors. Only AF observations are used in the astrometric solutions. 
All measurements come with a formal uncertainty estimated by the image parameter
determination. Based on the photon-noise statistics, the median formal AL 
uncertainty is about 0.06~mas per CCD observation in the AF for $G<12$~mag, 
0.20~mas at $G=15$~mag, and 3.8~mas at $G=20$~mag 
(cf.\ Fig.~\ref{fig:sigmaAlVsG}).

\begin{figure}
\centering
  \resizebox{0.9\hsize}{!}{\includegraphics{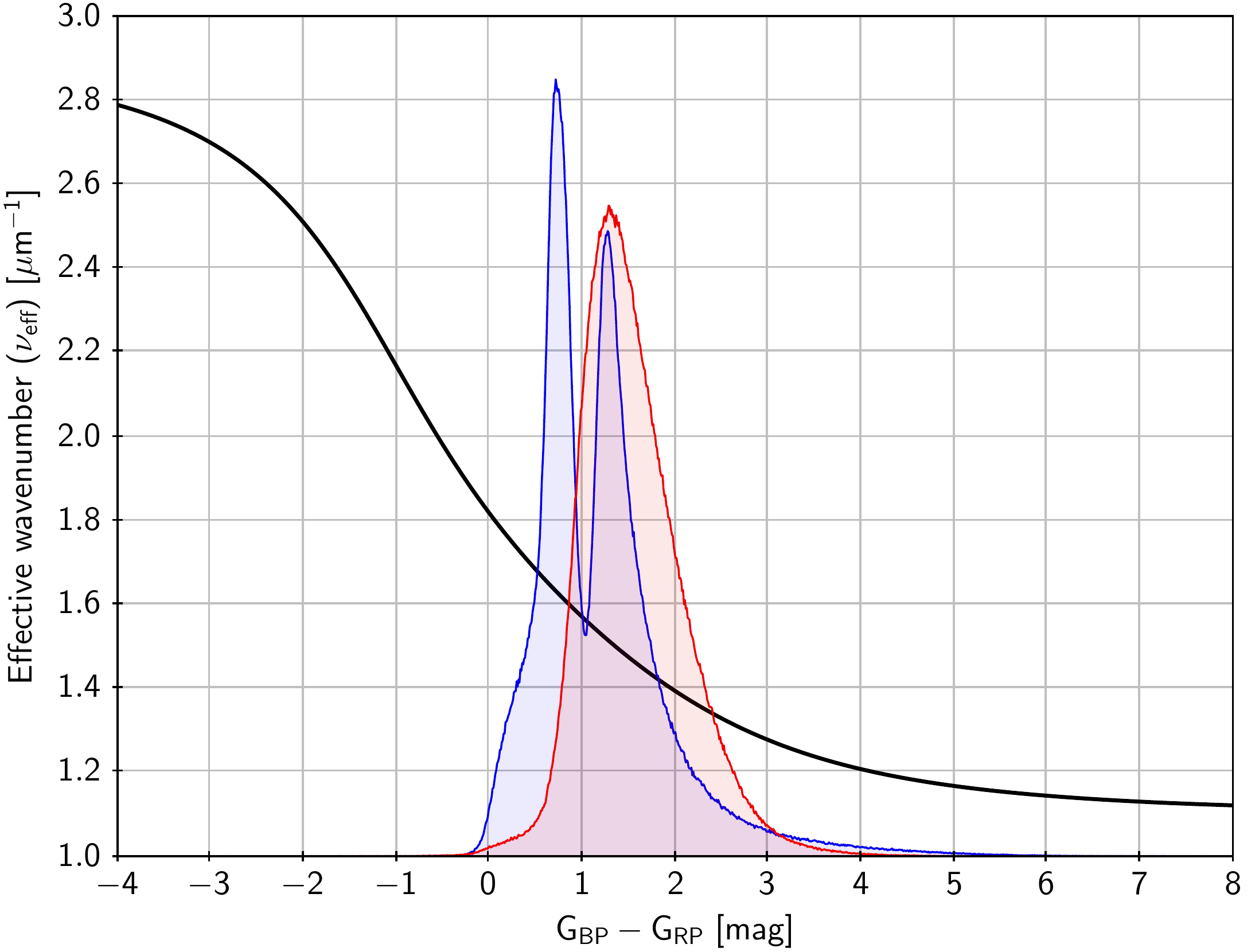}}
    \caption{Effective wavenumber as a function of colour index. The curve is the
    analytical relation in Eq.~(\ref{eq:nuEffVsC}). We also show the distribution of 
    $G_\text{BP}-G_\text{RP}$ for a random selection of bright ($G<12$~mag, bluish 
    histogram with two peaks) and faint ($G>18$~mag, reddish histogram) sources.}
    \label{fig:nuEffVsC}
\end{figure}

\subsection{Colour information}
\label{sec:cu5data}

The chromaticity calibration (Sect.~\ref{sec:cal}) requires that the effective wavenumber
$\nu_\text{eff}=\langle\lambda^{-1}\rangle$ is known for all 
primary sources. For \textit{Gaia} DR2, this quantity was computed from the mean 
integrated $G_\text{BP}$ and 
$G_\text{RP}$ magnitudes provided by the photometry pipeline \citepads{DPACP-44}, 
using the formula 
\begin{multline}\label{eq:nuEffVsC}
\nu_\text{eff}~[\mu\text{m}^{-1}] =\\
2.0 - \frac{1.8}{\pi}\arctan\bigl(0.331+0.572C-0.014C^2+0.045C^3\bigr)\, ,
\end{multline}
where $C=G_\text{BP}-G_\text{RP}$ (Fig.~\ref{fig:nuEffVsC}). The arctan 
transformation constrains $\nu_\text{eff}$ to the interval $[1.1,\,2.9]~\mu\text{m}^{-1}$
(roughly corresponding to the passband of $G$, or $\simeq\,$340--910~nm) as a safeguard
against spurious extreme values of $C$. The polynomial coefficients are based on pre-launch 
calibrations of the photometric bands and standard stellar flux libraries. In future releases,
more accurate values of $\nu_\text{eff}$ may be computed directly from the 
calibrated BP and RP spectra.

\begin{table*}[t]
\caption{Ten \textsc{Hipparcos} sources in \text{Gaia} DR2 with the largest predicted perspective acceleration.\label{tab:RV}}
\small
\begin{tabular}{lrrrl}
\hline\hline
\noalign{\smallskip}
Designation & \multicolumn{1}{c}{HIP} & \multicolumn{1}{c}{$v_r$} 
& \multicolumn{1}{c}{$\Delta$} & Name \\
& & [km~s$^{-1}$] & [mas] & \\
\noalign{\smallskip}
\hline
\noalign{\smallskip}
\textit{Gaia} DR2 4472832130942575872 & 87937 & $-$110.51 & 1.975 & Barnard's star\\
\textit{Gaia} DR2 4810594479417465600 & 24186 & 245.19 & 1.694 & Kapteyn's star\\
\textit{Gaia} DR2 2552928187080872832 & 3829 & 263.00 & 0.573 & Van Maanen 2\\
\textit{Gaia} DR2 1872046574983507456 & 104214 & $-$65.74 & 0.313 & 61 Cyg A\\
\textit{Gaia} DR2 1872046574983497216 & 104217 & $-$64.07 & 0.297 & 61 Cyg B\\
\noalign{\smallskip}
\textit{Gaia} DR2 4034171629042489088 & 57939 & $-$98.35 & 0.239 & Groombridge 1830\\
\textit{Gaia} DR2 5853498713160606720 & 70890 & $-$22.40 & 0.208 & $\alpha$ Cen C (Proxima)\\
\textit{Gaia} DR2 6412595290592307840 & 108870 & $-$40.00 & 0.163 & $\epsilon$ Ind\\
\textit{Gaia} DR2 3340477717172813568 & 26857 & 105.83 & 0.144 & Ross 47 \\
\textit{Gaia} DR2 4847957293277762560 & 15510 & 87.40 & 0.141 &  e Eri \\
\noalign{\smallskip}
\hline
\end{tabular}
\tablefoot{The table gives the assumed radial velocity, $v_r$ (taken from SIMBAD, \citeads{2000A&AS..143....9W}), for 10 of the 53 \textsc{Hipparcos} sources where
the perspective acceleration was taken into account in the astrometric solutions. 
$\Delta$ is the predicted size of the effect calculated as described in the text.
The complete table of the 53 sources is given in the \textit{Gaia} DR2 online documentation.}
\end{table*}

\subsection{BAM data}
\label{sec:bam}

The basic angle monitor (BAM) is an interferometric device measuring short-term
($\lesssim 1$~day) variations of the basic angle at $\mu$as precision 
\citepads{2016SPIE.9904E..2DM}.
Similarly to what was done for \textit{Gaia} DR1 
(Appendix~A.2 in \citeads{2016A&A...595A...4L}), 
the BAM data are here used to correct the astrometric measurements for the rapid variations 
(in particular the $\sim$1~mas amplitude 6~h oscillations) not covered by the astrometric calibration 
model. However, the corrections are considerably more detailed for \textit{Gaia} DR2, taking advantage of
several improvements in the processing and analysis of the BAM data: cosmic-ray filtering at pixel 
level of the raw BAM data; use of cross-correlation to determine very precise relative fringe phases; 
improved modelling of discontinuities and other variations that cannot be represented by the
simple harmonic model used for \textit{Gaia} DR1 \citep[cf.\ Figs.~A.2 and A.3 in][]{2016A&A...595A...4L}.
Some 370 basic-angle jumps with a median amplitude of 45~$\mu$as are corrected in this way. 
The jumps appear seemingly at random times, but at a much increased rate in the weeks following 
a decontamination event. 
The jumps, plus the smoothed BAM data between jumps, provided the
basic-angle corrector for \textit{Gaia} DR2 in the form of a spline function of time.

The spin-related distortion model (Sect.~\ref{sec:vbacfoc}) provides certain global corrections to
the BAM data, derived from the astrometric observations, but cannot replace the BAM data, which
contain a host of more detailed information such as the jumps.

\section{Models}
\label{sec:models}

\subsection{Source model}
\label{sec:source}

The \textit{Gaia} data processing is based on a consistent theory of relativistic
astronomical reference systems (\citeads{2003AJ....126.2687S}). Relevant
components of the model are gathered in the \textit{Gaia} relativity model
(GREM; \citeads{2003AJ....125.1580K}, \citeyearads{2004PhRvD..69l4001K}).
The primary coordinate system is the Barycentric Celestial Reference System 
(BCRF) with origin at the solar system barycentre and axes aligned with the 
International Celestial Reference System (ICRS). The time-like coordinate of 
the BCRS is the barycentric coordinate time (TCB).
 
The astrometric solutions described in this paper always assume that the observed centre
of the source moves with uniform space motion relative to the solar system barycentre.
(Non-linear motions caused by binarity and other perturbations require special solutions
that will be included in future \textit{Gaia} releases.) The relevant source model is described 
in Sect.~3.2 of the AGIS paper and is not repeated here. It depends on six kinematic 
parameters per source, that is,\ the standard five astrometric parameters 
$(\alpha,\,\delta,\,\varpi,\,\mu_{\alpha*}, \text{and }\,\mu_\delta),$ and the radial velocity $v_r$.
The astrometric parameters in \textit{Gaia} DR2 refer to the reference 
epoch $\text{J2015.5}=\text{JD~2457\,206.375 (TCB)}=\text{2015 July 2, 21:00:00}$ (TCB).
The positions and proper motions refer to the ICRS thanks to the special frame 
alignment procedure (Sect.~\ref{sec:frame}).

The source model allows taking into account perspective acceleration through terms 
depending on the radial velocity $v_r$. The accumulated effect over a time interval $T$ is 
$\Delta=|v_r|\,\mu\varpi T^2/A_\text{u}$, where $\mu=(\mu_{\alpha*}^2+\mu_\delta^2)^{1/2}$ 
is the total proper motion and $A_\text{u}$ is the astronomical unit. This is negligible 
except for some very nearby high-velocity stars, and for nearly all sources we ignore 
the effect by setting $v_r=0$ in the astrometric processing. Only for 53 nearby 
\textsc{Hipparcos} sources was it taken into account by assuming non-zero values
of $v_r$ taken from the literature (SIMBAD; \citeads{2000A&AS..143....9W}).
These sources were selected as having a predicted $\Delta>0.023$~mas for $T=1.75$~yr, 
calculated from \textsc{Hipparcos} astrometry \citepads{2007ASSL..350.....V}. 
(The somewhat arbitrary limit 0.023~mas corresponds to an RMS modelling
error below 0.002~mas, which is truly insignificant for this release.) The
top ten cases are listed in Table~\ref{tab:RV}. In future releases, perspective acceleration 
will be taken into account whenever possible, using radial-velocity data from 
\textit{Gaia}'s onboard spectrometer (RVS; \citeads{DPACP-47}).
We note that 34 of the 53 sources have radial velocities from
the RVS in this release,
with a median absolute deviation of 0.6~km~s$^{-1}$ from the values used here.
The absolute difference exceeds 5~km~s$^{-1}$ in only four cases, the most 
extreme being HIP~47425 $=$ \textit{Gaia}~DR2~5425628298649940608 with 
$v_r=+142\pm 21$~km~s$^{-1}$ from SIMBAD, based on
\citetads{1974PASP...86..742R}, and $v_r=+17.8\pm 0.2$~km~s$^{-1}$ in 
\textit{Gaia} DR2. In none of the cases will the error in $v_r$ cause an astrometric 
effect exceeding 0.02~mas in the present reduction. 

The final secondary solution (Sect.~\ref{sec:agis22}) requires knowledge of
$\nu_\text{eff}$ for all sources in order to take the chromaticity into account.
For most but not all sources, this is known from the photometric processing as
described in Sect.~\ref{sec:cu5data}. Given the calibrated chromaticity, it is also possible, however, to obtain an astrometric estimate of $\nu_\text{eff}$ 
for every source by formally introducing it as an additional (sixth) astrometric source
parameter.
The resulting estimate of $\nu_\text{eff}$, called pseudo-colour, is much less precise than
the $\nu_\text{eff}$ calculated from $G_\text{BP}-G_\text{RP}$ using Eq.~(\ref{eq:nuEffVsC}), 
but has the advantage that it can be obtained for every source allowing a five-parameter 
solution. Moreover, it is not affected by the BP/RP flux excess issue \citepads{DPACP-40}, 
which tends to make faint sources in crowded areas too blue as measured by the 
$G_\text{BP}-G_\text{RP}$.

To ensure the most uniform astrometric treatment of sources, the pseudo-colour was 
consistently used as a proxy for $\nu_\text{eff}$ in all cases where \textit{Gaia}~DR2
provides a five-parameter solution, that is,\ even when photometric colours are available. 
Because it is so important for the astrometry, the pseudo-colour is given in the 
\textit{Gaia} Archive as {\gacs{astrometric\_pseudo\_colour}}. Normally, it does
not provide an astrophysically useful estimate of the colour because its precision 
is much lower than the photometric data.

Our treatment of the pseudo-colour as a sixth source parameter should not 
be confused with the use of the radial proper motion $\mu_r= v_r\varpi/A_\text{u}$
in the kinematic source model (e.g.\ Eq.~2 of \citeads{2016A&A...595A...4L}). This
quantity, sometimes referred to as the ``sixth astrometric parameter'', is  
used internally in AGIS to take into account the perspective acceleration, but is never 
explicitly estimated as an astrometric parameter.

\subsection{Attitude model}
\label{sec:attitude}

The attitude specifies the orientation of the optical instrument in ICRS as a 
function of time. Mathematically, it is given by the unit quaternion $\textbf{q}(t)$.
The attitude model described in Sect.~3.3 of the AGIS paper represents the 
time-dependent components of $\textbf{q}(t)$ as cubic splines. For \textit{Gaia} DR1,
a knot interval of about 30~s was used in the splines, but it was noted that a 
much shorter knot interval (i.e.\ more flexible splines) would actually be needed 
to cope with the considerable attitude irregularities on shorter timescales, 
including a large number of ``micro-events'' such as the very frequent micro-clanks 
(see Appendices~C.4 and E.4 in \citeads{2016A&A...595A...4L}) and less frequent
micrometeoroid hits. Decreasing the knot interval of the splines
is not a good way forward, however, as it would weaken the solution by the
increased number of attitude parameters. Moreover, this cannot adequately
represent the CCD-integrated effects of the micro-events, which depend
also on the gate ($g$) used for an observation. For \textit{Gaia} DR2 the
attitude model includes a new layer, known as the corrective attitude 
$\textbf{q}_\text{\,c}(t,g)$, such that the (gate-dependent) effective 
attitude becomes
\begin{equation}\label{eq:att1}
\textbf{q}_\text{\,e}(t, \, g) = \textbf{q}_\text{\,p}(t) \, \textbf{q}_\text{\,c}(t, \, g) \, .
\end{equation}
Here $\textbf{q}_\text{\,p}(t)$ is the primary attitude: this uses the same spline 
representation as the old attitude model, and its parameters are estimated in 
the primary solution in a similar way as before, the main difference being that
the field angle residuals (Eqs. 25--26 in the AGIS paper) are now computed
using the effective attitude $\textbf{q}_\text{\,e}(t, \, g)$ for the relevant gate.
The effective attitude represents the mean pointing of the instrument during
the CCD integration interval, which is different depending on $g$.

In Eq.~(\ref{eq:att1}) the corrective attitude $\textbf{q}_\text{\,c}$ represents a 
small time- and gate-dependent rotation that takes care of attitude irregularities 
that are too fast for the spline model. It is calculated in the AGIS
pre-processor and remains fixed during subsequent astrometric solutions.
For details about its calculation, we refer to the \textit{Gaia} DR2 online 
documentation. Briefly, the procedure includes the following steps:
\begin{enumerate}
\item 
Given two successive CCD observations in the astrometric field (AF) of the 
same source, with observation times $t_k$ and $t_{k+1}$, an estimate of 
the inertial angular rate along the nominal spin axis $z$ (in the scanning 
reference system, SRS) is obtained as
\begin{equation}\label{eq:att2}
\overline{\omega}_z = -\frac{\eta_{k+1}-\eta_k}{t_{k+1}-t_k}
+ \left(\omega_x\cos\varphi+\omega_y\sin\varphi\right)\tan\zeta \, ,
\end{equation}
where $\eta_k$ and $\eta_{k+1}$ are the AL field angles calculated from a
preliminary geometrical model of the instrument. The minus sign on the first term is
due to the apparent motions of images in the direction of negative $\eta$
(see Fig.~3 in the AGIS paper). The second term takes into account the (slow)
rotation of the field that is due to the across-scan (AC) angular rates $\omega_x$
and $\omega_y$. $\varphi$ and $\zeta$ are the AL and AC instrument angles of the 
source (Fig.~2 in the AGIS paper) at a time mid-way between the two observations.
(Only approximate AC rates are needed here, as $|\tan\zeta|<0.01$.)
The bar in $\overline{\omega}_z$ signifies that it is a mean value of the 
instantaneous rate, averaged over both the CCD integration time ($\simeq\,$4.42~s 
for ungated observations) and the time between successive CCD observations 
($\simeq\,$4.86~s). 
\item
Applying Eq.~(\ref{eq:att2}) to ungated AF observations for all sources in 
the magnitude range 12 to 16 yields on average several hundred measurements 
per second of the AL angular rate. The rate measurements are binned by time, 
using a bin size of 0.2~s, and the median value calculated in each bin. This
provides an accurate time-series representation of $\overline{\omega}_z(t)$ with 
sufficient time resolution for the next step.
\item
Micro-clanks are small quasi-instantaneous changes in the physical orientation
of the instrument axes, which create trapezoidal profiles in $\overline{\omega}_z(t)$ with
a constant and known profile; for an example, see the bottom panels of Fig.~D.4
in \citetads{2016A&A...595A...4L}. In this step, micro-clanks are detected, and their 
times and amplitudes 
estimated, by locally fitting a smooth background signal plus a scaled profile to
the time-series representation of $\overline{\omega}_z(t)$. The fitted profile is 
subtracted and the procedure repeated until no more significant clank is detected.
The end result is a list of detected clanks, with their times and amplitudes, together 
with an estimate $\overline{\omega}^\text{\,nc}_z(t)$ of the rate without clanks. 
\item
Integrating $\overline{\omega}^\text{\,nc}_z(t)$ as a function of time and fitting a
cubic spline with uniform 5~s knot separation provides an estimate of the 
attitude irregularities at frequencies below $\simeq\,$0.1~Hz, including the effects
of minor micrometeoroid hits. Finally, the corrective attitude is obtained by 
adding, depending on $g$, the analytically integrated effect of the detected clanks.
\end{enumerate}
Thanks to the use of a pre-computed corrective attitude, it is possible to use 
a rather long (30~s) knot interval in the primary astrometric solution without 
causing a degradation in the accuracy. For \textit{Gaia} DR2, this procedure was 
only applied to the AL attitude component ($z$ axis). In the future,
the AC components will be similarly corrected for micro-clanks and other 
medium-frequency irregularities.

Micrometeoroid hits cause rate irregularities that are distinctly different from the 
clank profiles: they are less abrupt, of much longer duration, and have somewhat
variable profiles depending on the response of the onboard attitude control system.
Nevertheless, they could in principle be detected and handled in a similar way as 
the clanks. Currently, however, only major hits are automatically detected and 
treated simply by inserting data gaps around them. Such hits, detected from attitude 
rate disturbances exceeding a few mas~s$^{-1}$, occurred at a fairly constant rate of
about five hits per month. 
Minor hits remain undetected, but are effectively corrected by the integrated 
rate that is part of the corrective attitude.

\subsection{Calibration model}
\label{sec:cal}

The astrometric calibration model specifies the location of the fiducial ``observation line''
for a particular combination of field of view ($f$), CCD ($n$), and gate ($g$)
indices, 
as a function of the AC pixel coordinate $\mu$, time $t$, and other relevant 
quantities (Sect.~3.4 in the AGIS paper). Formally, it defines the functions
$\eta_{fng}(\mu, t, \dots)$, $\zeta_{fng}(\mu, t, \dots)$ in terms of a 
discrete set of calibration parameters, where $(\eta,\zeta)$ are the field
angles along the observation line. In the generic calibration model, these 
functions are written as sums of a number of ``effects'', which in turn are
linear combinations of basis functions with the calibration parameters as
coefficients. Table~\ref{tab:cal} gives an overview of the effects and 
number of calibration parameters used in the final primary solution for
\textit{Gaia} DR2. All calibration effects are independently modelled for the 
$2\times 62=124$ combinations of the field and CCD indices. The calibration
model for the sky mappers (SM) is similar, but not described here as the SM 
observations are not used in the astrometric solutions. 

Although \textit{Gaia} is designed to be extremely stable on short time-scales,
inevitable changes in the optics and mechanical support structure require a 
time-dependent calibration. Occasional spontaneous, minute changes in the 
instrument geometry, and major operational events such as mirror decontaminations, 
telescope refocusing, unplanned data gaps and resets, make it necessary 
to have breakpoints (discontinuities) at specific times. To accommodate both 
gradual and sudden changes, the generic calibration model allows the use of several
time axes, with different granularities, such that an independent subset of
calibration parameters is estimated for each granule. The current model
uses three time axes with 243, 14, and 10 granules spanning the length of 
the data. The first one, having the shortest granules of typically 3~days, is 
used for the most rapidly changing effects. The other two are used for effects 
that are either intrinsically less variable (e.g.\ representing the internal structure 
of the CCDs) or less critical for the solution (e.g.\ the AC calibration). The third 
axis has granules of exactly 63~days duration, tuned to the scanning law in order 
to minimise cross-talk between spin-related calibration effects and the celestial 
reference frame.
  
The current calibration model differs in many details from the one used for 
\textit{Gaia} DR1 (Appendix~A.1 in \citeads{2016A&A...595A...4L}); in 
particular, it includes colour- and magnitude-dependent terms needed to 
account for centroid shifts that are not yet calibrated in the pre-processing of the raw data.

\begin{table*}[t]
\caption{Summary of the astrometric calibration model and number of calibration
parameters in the astrometric solution for \textit{Gaia} DR2.\label{tab:cal}}
\small
\begin{tabular}{rllcccccccccr}
\hline\hline
\noalign{\smallskip}
&& \multicolumn{1}{l}{Basis functions} 
& \multicolumn{9}{c}{---------- Multiplicity of dependencies ----------} & 
\multicolumn{1}{c}{Number of} \\ 
\multicolumn{2}{l}{Effect and brief description} & 
\multicolumn{1}{l}{$K_{lm}(\tilde{\mu},\tilde{t})$} & \multicolumn{1}{c}{$K_{lm}$} 
& \multicolumn{1}{c}{$j$} & $f$ & \multicolumn{1}{c}{$n$} & 
$g$ & $b$ & $w$ & \multicolumn{1}{c}{$\nu_\text{eff}$} & $G$ &
\multicolumn{1}{c}{parameters} \\
\noalign{\smallskip}
\hline
\noalign{\smallskip}
1& AL large scale  & $lm=00,10,20,01$ & 4 & 243 & 2 & 62 & -- & -- & -- & -- & -- &  120\,528\\ 
2& AL medium scale, gate & $lm=00,10$ & 2 & 10 & 2 & 62 & 8 & 9 & -- & -- & -- & 178\,560 \\ 
3& AL large scale, window class & $lm=00,10$ & 2 & 14 & 2 & 62 & -- & -- & 3 & -- & -- & 10\,416 \\ 
4& AL large scale, window class, colour & $lm=00,10,01$ & 3 & 14 & 2 & 62 & -- & -- & 3 & 1 & -- & 15\,624 \\ 
5& AL large scale, window class, magnitude & $lm=00,10$ & 2 & 14 & 2 & 62 & -- & -- & 3 & -- & 1 &  10\,416\\ 
\noalign{\smallskip}
\hline
\noalign{\smallskip}
1& AC large scale  & $lm=00,10,20,01$ & 4 & 14 & 2 & 62 & -- & -- & -- & -- & -- &  6\,944\\ 
2& AC large scale, gate & $lm=00$ & 1 & 14 & 2 & 62 & 8 & -- & -- & -- & -- & 13\,888 \\ 
3& AC large scale, window class & $lm=00,10$ & 2 & 14 & 2 & 62 & -- & -- & 3 & -- & -- & 10\,416 \\ 
4& AC large scale, window class, colour & $lm=00,10,01$ & 3 & 14 & 2 & 62 & -- & -- & 3 & 1 & -- & 15\,624 \\ 
5& AC large scale, window class, magnitude & $lm=00,10$ & 2 & 14 & 2 & 62 & -- & -- & 3 & -- & 1 &  10\,416\\ 
\noalign{\smallskip}
\hline
\end{tabular}
\tablefoot{The column Basis functions lists the combinations of indices $l$ and $m$
used to model variations with AC coordinate on a CCD ($\tilde{\mu}$) and with time within 
a time granule ($\tilde{t}$). Multiplicity of dependencies gives the number of distinct 
functions or values for each dependency, or a dash if there is no dependency: 
basis functions ($K_{lm}$), granule index ($j$), 
field index ($f$), CCD index ($n$), gate ($g$), stitch block ($b$), window class ($w$), 
effective wavenumber ($\nu_\text{eff}$), and magnitude ($G$). The last column 
is the product of multiplicities, equal to the number of calibration parameters of the effect.}
\end{table*}

The AL calibration model is the sum of the five different effects listed in
the upper part of Table~\ref{tab:cal}, giving a total of 335\,544 AL parameters.
As explained in Appendix~A.1 of \citetads{2016A&A...595A...4L}, the variation with 
across-scan coordinate $\mu$ within a CCD, and with time $t$ within a granule, 
is modelled as a linear combination of basis functions
\begin{equation}\label{eq:Klm}
K_{lm}(\tilde{\mu},\tilde{t})=\tilde{P}_l(\tilde{\mu})\tilde{P}_m(\tilde{t}\,) \, ,
\end{equation}
where $\tilde{P}_l(x)$, $\tilde{P}_m(x)$ are the shifted Legendre polynomials%
\footnote{The shifted Legendre polynomials $\tilde{P}_n(x)$ are related to the 
(ordinary) Legendre polynomials $P_n(x)$ by $\tilde{P}_n(x)=P_n(2x-1)$.
Specifically, $\tilde{P}_0(x)=1$, $\tilde{P}_1(x)=2x-1$, and 
$\tilde{P}_2(x)=6x^2-6x+1$. In the AGIS paper and in
\citetads{2016A&A...595A...4L}, the shifted Legendre polynomials were
denoted $L^\ast_n(x)$.\label{fn2}}
of degree $l$ and $m$, orthogonal on $0\le x \le 1$ for $l\ne m$,
$\tilde{\mu}=(\mu-\mu_\text{min})/(\mu_\text{max}-\mu_\text{min})$ 
is the normalised AC pixel coordinate (with $\mu_\text{min}=13.5$ and
$\mu_\text{max}=1979.5$), and $\tilde{t}=(t-t_j)/(t_{j+1}-t_j)$ the 
normalised time within granule $j$, $t\in[t_j,t_{j+1})$.
The third and fourth columns in Table~\ref{tab:cal} list
the combination of indices $l$ and $m$ used for a particular effect, and the 
number of basis functions $K_{lm}$ used for each combination of $jfn$, and
their orders $lm$. For example, effect~1 is a linear combination of
$K_{00}$, $K_{10}$, $K_{20}$, and $K_{01}$ for each combination $jfn$. 
Similarly, effect~2 is a linear
combination of $K_{00}$ and $K_{10}$ for each combination $jfngb$.

This calibration model does not include any effects that vary on
a very short spatial scale, for instance,\ from one pixel column to the next. Such
small-scale effects do exist (see Fig.~\ref{fig:alResVsMu}), and will be included 
in future calibrations. In the present astrometric solutions, they are treated as 
random noise on the individual CCD observations.

In principle, the image parameter determination (Sect.~\ref{sec:imagepar})
should result in centroid positions that are independent of window class,%
\footnote{Window classes (WC) 0, 1, and 2 are different sampling schemes
of pixels around a detected source, decided by an onboard algorithm mainly 
based on the brightness of the source: WC0 (for $G\lesssim 13$) is a 
two-dimensional sampling, from which both the AL and AC centroid locations
can be determined on ground, while WC1 ($13\lesssim G\lesssim 16$) and 
WC2 ($G\gtrsim 16$) give one-dimensional arrays of 18 and 12 samples, 
respectively, allowing only the AL location to be determined.
\label{footnote1}}
colour, and magnitude. For the current solution, this was not the case, and 
these effects were instead included in the astrometric calibration model described here.
Effect~3 describes the displacement of each window class ($w$) for a
source of reference colour ($\nu_\text{eff}=1.6$) and reference magnitude ($G=13$),
while effects~4 and 5 describe the dependence on colour and magnitude
by means of additional terms proportional to $\nu_\text{eff}-1.6$ and
$G - 13$, respectively.

Combining all five effects, the complete AL calibration model is
\begin{equation}\label{eq:calAL}
\begin{split}
\eta_{fngw}&(\mu,t,\nu_\text{eff},G) = \eta_{ng}^{(0)}(\mu)\\
&+ \Delta\eta^{(1)}_{lmjfn}K_{lm} + \Delta\eta^{(2)}_{lmjfngb}K_{lm} 
+ \Delta\eta^{(3)}_{lmjfnw}K_{lm} \\
&+ \Delta\eta^{(4)}_{lmjfnw}(\nu_\text{eff}-1.6)K_{lm} 
+ \Delta\eta^{(5)}_{lmjfnw}(G-13)K_{lm} \, ,
\end{split}
\end{equation}
where $\eta_{gn}^{(0)}(\mu)$ is the nominal observation line for CCD $n$ and 
gate $g$, and $\Delta\eta$ are the calibration parameters. For brevity, the arguments 
of $K_{lm}$ (different in each term) are suppressed and Einstein's summation convention is used for the 
repeated indices $lm$. Indices $j$ and $b$ are implicit functions of $t$ and $\mu$, 
respectively, with $j$ depending on the granularity of the time axis and $b$ depending
on the ``stitch block'' structure imprinted on the pixel geometry by the CCD manufacturing
process (cf.\ Fig.~\ref{fig:alResVsMu}).

The AC calibration model is similarly a sum of the five effects given in 
the lower part of Table~\ref{tab:cal}, giving a total of 57\,288 AC parameters.
The expression for $\zeta_{fngw}(\mu,t,\nu_\text{eff},G)$ is analogous to 
(\ref{eq:calAL}), with $\zeta$ replacing $\eta$ everywhere, except that there 
is no dependence on the stitch block index $b$. The coarse time granularity 
is used for all AC effects. 

Certain constraints among the calibration parameters are needed to
avoid degeneracies in the astrometric solution. For \textit{Gaia} DR2, only the 
basic constraints defining the origin of $\eta$ and $\zeta$ (Eqs.~16--18 in 
the AGIS paper) were used. It is known that the calibration model has additional
degeneracies, corresponding to missing constraints; these are handled internally 
by the solution algorithm (cf.\ Appendix~C.3 in the AGIS paper) and should
not affect the astrometric parameters.

\subsection{Spin-related distortion model}
\label{sec:vbacfoc}

As shown by the BAM data (Sect.~\ref{sec:bam}) and confirmed in early astrometric
solutions, the basic angle between \textit{Gaia}'s two fields of view undergoes 
very significant ($\sim$1~mas amplitude) periodic variations. The variations depend 
mainly on the phase of the 6~h spin with respect to the Sun, as given by the 
heliotropic spin phase $\Omega(t)$ (e.g.\ Fig.~1 in \citeads{2017A&A...603A..45B}).
To first order, they can be represented by
\begin{multline}\label{eq:bam}
\Delta\Gamma(t) = d(t)^{-2} 
\sum_{k=1}^8 \Bigl( \Bigl[C_{k,0}+C_{k,1}(t-t_\text{ref})\Bigr]\cos k\Omega(t) \\
+ \Bigl[S_{k,0}+S_{k,1}(t-t_\text{ref})\Bigr]\sin k\Omega(t) \Bigr) \, 
\end{multline}
(cf.\  Eqs.~A.10--A.11 in \citeads{2016A&A...595A...4L}),
where $d(t)$ is the Sun--\textit{Gaia} distance in au. Values of the Fourier
coefficients obtained by fitting Eq.~(\ref{eq:bam}) to the periodic part of the
basic-angle corrector (Sect.~\ref{sec:bam}), using $t_\text{ref}=\text{J}2015.5$, 
are given in Table~\ref{tab:vbac}.

Although the exact mechanism is not known, the large 6~h variations are believed to 
be caused by thermoelastic perturbations in the Sun-illuminated service module
of \textit{Gaia} propagating to the optomechanical structure of the payload
(\citeads{2016SPIE.9904E..2DM}). It is then almost unavoidable that the optical 
distortions in the astrometric fields also undergo periodic variations, although most
likely of much smaller amplitude. The spin-related distortion model aims at estimating, 
and hence correcting, such variations in the astrometric solution, based on the 
assumption that they are stable on long time-scales. Specifically, for
\textit{Gaia} DR2, it is assumed that the variations scale with the inverse square of
the distance to the Sun, but otherwise are strictly periodic in $\Omega(t)$. Since such 
a model in fact describes the basic-angle variations measured by the BAM rather well, 
it is not unreasonable to assume that it could also work for the optical distortion.

The spin-related distortion may be regarded as just another effect in the 
astrometric calibration model (Sect.~\ref{sec:cal}). However, the character
of the variations, requiring a single block of parameters for all observations, 
made it more convenient to implement it as a set of global parameters
(Sect.~5.4 in the AGIS paper).
 
Depending on the field index $f$ ($=+1$ for the preceding and $-1$ 
for the following field of view), 
the spin-related distortion model adds a time-dependent AL displacement to the calibration model in Sect.~\ref{sec:cal}:
\begin{equation}\label{eq:calVF3a}
\Delta\eta_f(t,\eta,\zeta)=\sum_{l\ge 0,\,m\ge 0}^{l+m\le 3}
F_{flm}(t) \tilde{P}_l(\tilde{\eta}) \tilde{P}_m(\tilde{\zeta}) \, .
\end{equation}
Here $\tilde{P}_l(x)$ and $\tilde{P}_m(x)$ are the shifted Legendre 
polynomials of degree $l$ and $m$ (see footnote~\ref{fn2}), and 
$\tilde{\eta}=(\eta-\eta_\text{min})/(\eta_\text{max}-\eta_\text{min})$,
$\tilde{\zeta}=(\zeta-\zeta_{f,\text{min}})/(\zeta_{f,\text{max}}-\zeta_{f,\text{min}})$
are normalised field angles. (The limits in $\zeta$ depend 
on $f$ because of the different AC locations of the optical centre in the preceding 
and following fields; see Fig.~3 and Eq.~14 in the AGIS paper.)
For the present third-order model ($l+m\le 3$), there are ten two-dimensional basis 
functions $\tilde{P}_l(\tilde{\eta})\tilde{P}_m(\tilde{\zeta})$ per field of view. 
The functions $F_{\!flm}(t)$ of degree $l+m>0$ are modelled as a truncated Fourier 
series in $\Omega(t)$, scaled by the inverse square of
the distance to the Sun:
\begin{multline}\label{eq:calVF3b}
F_{\!flm}(t) = d(t)^{-2} 
\sum_{k=1}^8 \Bigl(c_{fklm}\cos k\Omega(t) + s_{fklm}\sin k\Omega(t)\Bigr) \, ,\\
0 < l+m \le 3 \, .
\end{multline}
This gives 288~parameters $c_{fklm}$ and $s_{fklm}$. 
The functions $F_{-1,0,0}(t)$ and $F_{+1,0,0}(t)$, that is,\ for $f=\pm 1$ and $l=m=0$, 
require a separate treatment to avoid degeneracy. They represent time-dependent 
offsets in the two fields that are independent of the field angles $\eta$ and $\zeta$. 
The mean function $[F_{-1,0,0}(t)+F_{+1,0,0}(t)]/2$ is equivalent to a time-dependent 
AL shift of the attitude and can therefore be constrained to zero for all $t$. The 
difference $\delta\Gamma(t)=F_{-1,0,0}(t)-F_{+1,0,0}(t)$     
represents a time-dependent correction to the basic angle in
addition to the basic-angle 
corrector derived from BAM data (Sect.~\ref{sec:bam}) and the slower variations of 
the calibration model (Sect.~\ref{sec:cal}). This correction is modelled as a scaled 
Fourier series, in which the Fourier coefficients have a linear dependence on time
similar to Eq.~(\ref{eq:bam}):
\begin{multline}\label{eq:calVF3c}
\delta\Gamma(t) = d(t)^{-2} 
\sum_{k=1}^8 \Bigl( \Bigl[\delta C_{k,0}+\delta C_{k,1}(t-t_\text{ref})\Bigr]\cos k\Omega(t) \\
+ \Bigl[\delta S_{k,0}+\delta S_{k,1}(t-t_\text{ref})\Bigr]\sin k\Omega(t) \Bigr) \,  
\end{multline}
with $t_\text{ref}=\text{J}2015.5$. However, as discussed in 
Sect.~\ref{sec:zeropoint}, the parameter $\delta C_{1,0}$ is nearly 
degenerate with a global shift of the parallaxes and in the present solution
it was not estimated, meaning that it was assumed to be zero. This gave 31 parameters 
for $\delta\Gamma(t)$, and a total of 319 parameters for the complete spin-related 
distortion model. 

Results from the final solution for the parameters in Eq.~(\ref{eq:calVF3c})  
are shown in Table~\ref{tab:vbac} in the columns marked Corr. These values 
can be interpreted as corrections to the mean harmonic coefficients from 
Eq.~(\ref{eq:bam}) shown in the columns marked BAM. The statistical uncertainty 
of all values is below $1~\mu$as or $1~\mu$as~yr$^{-1}$. The main conclusion 
from this table is that the BAM data, while substantially correct, nevertheless require 
significant corrections at least for $k\le 4$. One possible interpretation 
is that the BAM accurately measures the basic-angle variations at the location
of the BAM CCD, outside the astrometric field, but that these variations are not
completely representative for the whole astrometric field. The special case of 
$\delta C_{1,0}$ and further aspects of $c_{fklm}$ and $s_{fklm}$ are discussed 
in Sect.~\ref{sec:zeropoint}.

\begin{table}
\caption{Fourier coefficients for the basic-angle variations.  
\label{tab:vbac}}
\small
\begin{tabular}{rrrrrrr}
\hline\hline
\noalign{\smallskip}
& \quad & \multicolumn{2}{c}{Coefficient [$\mu$as]}
& \quad & \multicolumn{2}{c}{Derivative [$\mu$as~yr$^{-1}$]}\\
&& BAM & Corr. && BAM & Corr.\\
\noalign{\smallskip}
\hline
\noalign{\smallskip}
$C_{1}$ && $+$909.80 &    $-$ && $+$73.34 &  $+$1.37 \\
$C_{2}$ && $-$110.50 & $-$23.38 &&  $+$1.86 &  $-$1.59 \\
$C_{3}$ && $-$68.39 &  $-$4.65 &&  $+$1.37 &  $-$0.33 \\
$C_{4}$ && $+$17.61 &  $-$2.53 &&  $-$0.79 &  $-$1.53 \\
$C_{5}$ &&  $+$2.79 &  $-$1.15 &&  $-$0.25 &  $-$2.86 \\
$C_{6}$ &&  $+$3.67 &  $+$1.47 &&  $+$0.40 &  $-$0.18 \\
$C_{7}$ &&  $+$0.12 &  $+$0.38 &&  $-$0.34 &  $+$0.42 \\
$C_{8}$ &&  $-$0.51 &  $-$0.44 &&  $-$0.01 &  $+$0.61 \\
\noalign{\smallskip}
$S_{1}$ && $+$668.41 & $-$25.42 && $+$19.78 &  $+$1.49 \\
$S_{2}$ && $-$90.95 & $+$34.46 && $-$10.68 &  $+$5.23 \\
$S_{3}$ && $-$63.47 &  $+$4.63 &&  $+$3.02 &  $+$0.97 \\
$S_{4}$ && $+$18.11 &  $+$3.32 &&  $+$1.20 &  $-$1.33 \\
$S_{5}$ &&  $-$0.11 &  $-$0.55 &&  $+$0.79 &  $-$1.61 \\
$S_{6}$ &&  $+$0.02 &  $-$1.11 &&  $-$0.69 &  $+$0.38 \\
$S_{7}$ &&  $+$0.18 &  $-$0.05 &&  $-$0.27 &  $-$0.14 \\
$S_{8}$ &&  $-$0.49 &  $+$0.25 &&  $+$0.09 &  $-$0.32 \\
\noalign{\smallskip}
\hline
\end{tabular}
\tablefoot{The columns headed BAM contain the coefficients $C_{k,0}$, $S_{k,0}$
and derivatives $C_{k,1}$, $S_{k,1}$ for a harmonic fit to the BAM data 
according to Eq.~(\ref{eq:bam}). The columns headed Corr.\ contain
the corresponding corrections $\delta C_{k,0}$, $\delta S_{k,0}$, $\delta C_{k,1}$, 
and $\delta S_{k,1}$ to the BAM data obtained in the primary astrometric solution 
using the model in Eq.~(\ref{eq:calVF3c}). The reference epoch for the
coefficients is J2015.5.}
\end{table}

\section{Astrometric solutions}
\label{sec:solutions}

The astrometric results in \textit{Gaia} DR2 were not produced in a single large
least-squares process, but were the end result of a long series of solutions
using different versions of the input data and testing different calibration
models and solution strategies. The description below ignores much of this
and only mentions the main path and milestones. As described in the AGIS 
paper, a complete astrometric solution consists of two parts, known
as the primary solution and the secondary solutions. In the primary solution,
which involves only a small fraction of the sources known as primary sources, 
the attitude and calibration parameters (and optionally the global parameters) 
are adjusted simultaneously with the astrometric parameters of the primary
sources using an iterative algorithm. The reference frame is also adjusted
using a subset of the primary sources identified as quasars. In the secondary solutions, the five 
astrometric parameters of every source are adjusted using 
fixed attitude, calibration, and global parameters from the preceding primary 
solution. The restriction on the number of primary sources comes mainly 
from practical considerations, as the primary solution is computationally 
and numerically demanding because of the large systems of equations that need 
to be solved. By contrast, the secondary solutions can be made one source 
at a time essentially by solving a system with only five unknowns (or six, if
pseudo-colour is also estimated). For consistency, the astrometric parameters 
of the primary sources are re-computed in the secondary solutions. 

For \textit{Gaia} DR2, two complete astrometric solutions were 
calculated, internally referred to as AGIS02.1 and AGIS02.2.
The published data exclusively come from AGIS02.2.

\subsection{Provisional solution (AGIS02.1)}
\label{sec:agis21}

The first complete astrometric solution based on the \textit{Gaia} DR2 input data was 
made in December 2016. This solution, known as AGIS02.1, provided a provisional 
attitude and astrometric calibration, and provisional astrometric parameters 
for about 1620~million sources. These data were used as a starting point
for the final solution (AGIS02.2) and allowed us to identify and resolve
a number of issues at an early stage. Typical differences between the
provisional and final solutions are below 0.2~mas or 0.2~mas~yr$^{-1}$.

The provisional solution was also used in some of the down-stream processing, 
notably for the wavelength calibrations of the photometric instruments 
\citep{DPACP-44} and radial-velocity spectrometer \citep{DPACP-47}.
The availability of a provisional solution more than a year before the release
was crucial for the inclusion of high-quality photometric and spectroscopic
results in \textit{Gaia} DR2.

\subsection{Final \textit{Gaia} DR2 solution (AGIS02.2)}
\label{sec:agis22}

Compared with the provisional solution, the main improvements in the 
final solution were
\begin{itemize}
\item
use of pseudo-colours in the source model (Sect.~\ref{sec:source}) to
take chromaticity into account;
\item
a more accurate corrective attitude (Sect.~\ref{sec:attitude}), based on the 
AGIS02.1 calibration;
\item
an improved basic angle corrector, including many detected jumps (Sect.~\ref{sec:bam}); 
\item
a calibration model (Sect.~\ref{sec:cal}) better tuned to the data, derived 
after detailed analysis of several test runs;
\item
inclusion of global parameters for the spin-related distortion model 
(Sect.~\ref{sec:vbacfoc}).
\end{itemize}
The main steps for producing the final solution were as follows.
\begin{enumerate}
\item
\textit{AGIS pre-processing.} This collected and converted input data for each source:
astrometric parameters from a previous solution, photometric information, radial
velocity when relevant (Sect.~\ref{sec:source}), and the image parameters from all 
the astrometric observations of the source. The corrective attitude was also computed 
at this point.
\item
\textit{Preliminary secondary solutions.} A preliminary adjustment of the parameters
for all the sources was performed, using the attitude and calibration from AGIS02.1.
The main purpose of this was to collect source statistics in order to tune the selection
of primary sources for the next step. Two secondary solutions were made for each source:
the first computed the pseudo-colour of the source, and the second re-computed
the astrometric solution using the derived pseudo-colour. This gave preliminary 
astrometric parameters and solution statistics for nearly 2500~million sources.
\item
\textit{Selection of primary sources.} About 16~million primary sources were selected 
based on the results of the previous step. The criteria for the selection were that 
(i) sources must have $G$, $G_\text{BP}$, and $G_\text{RP}$ magnitudes from the 
photometric processing; 
(ii) there should be a roughly equal number of sources with observations in each of
the three window classes; 
(iii) for each window class, there should be a roughly homogeneous coverage of
the whole sky and a good distribution in magnitude and colour; 
and (iv) within the constraints set by the previous criteria, sources with high astrometric 
weight (bright, with small excess noise and a good number of observations) were
preferentially selected. To this were added some 490\,000 probable quasars for the 
reference frame alignment (Sect.~\ref{sec:frame}). 
\item
\textit{Primary solution.} The astrometric parameters of the primary sources were
adjusted, along with the attitude, calibration, and global parameters, using a hybrid 
scheme of simple and conjugate gradient iterations (see Sect.~4.7 in the AGIS paper).
The frame rotator was used to keep the astrometric parameters and attitude on
ICRS using the subset of primary sources identified as quasars (Sect.~\ref{sec:frame}).  
\item
\textit{Final secondary solutions.} This essentially repeated step~2 with the final 
attitude, calibration, and global parameters from step~4, including a re-computation
of the pseudo-colours for all sources using the final chromaticity calibration. 
Sources failing to meet the acceptance criteria for a five-parameter solution 
(Sect.~\ref{sec:fallback}) obtained a fall-back solution at this stage.  
\item
\textit{Regeneration of attitude and calibration.} The primary solution did not use data
from the first month of nominal operations (in EPSL mode; Sect.~\ref{sec:coverage}),
and several shorter intervals of problematic observations were also skipped. In this step 
the attitude and calibration were re-computed for these intervals by updating the
corresponding parameters while keeping the source parameters fixed. This allowed
other processes, such as the photometric processing, to make use of observations in these time intervals as well.
\item
\textit{AGIS post-processing.} This converted the results into the required formats and
stored them in the main database for their subsequent use by all other processes,
including the generation of the \textit{Gaia} Archive.  
\end{enumerate}
Although not part of the astrometric processing proper, a further important step was carried 
out at the point when the astrometric data were converted from the main database into the
\textit{Gaia} Archive: the formal uncertainties of the five-parameter solutions were 
corrected for the ``DOF bug''. The background
and details of this are described in Appendix~\ref{sec:correction}. Here it is sufficient to note
that the formal astrometric uncertainties given in the \textit{Gaia} Archive, denoted 
$\sigma_{\alpha*}$, $\sigma_{\delta}$, $\sigma_{\varpi}$, $\sigma_{\mu\alpha*}$, and 
$\sigma_{\mu\delta}$, generally differ from the (uncorrected) uncertainties obtained in step~5.
When occasionally we need to refer to the latter values, we use the notation 
$\varsigma_{\alpha*}$, $\varsigma_{\delta}$, $\varsigma_{\varpi}$, $\varsigma_{\mu\alpha*}$, and 
$\varsigma_{\mu\delta}$ for the uncorrected uncertainties.

\begin{figure}
\centering
  \resizebox{0.9\hsize}{!}{\includegraphics{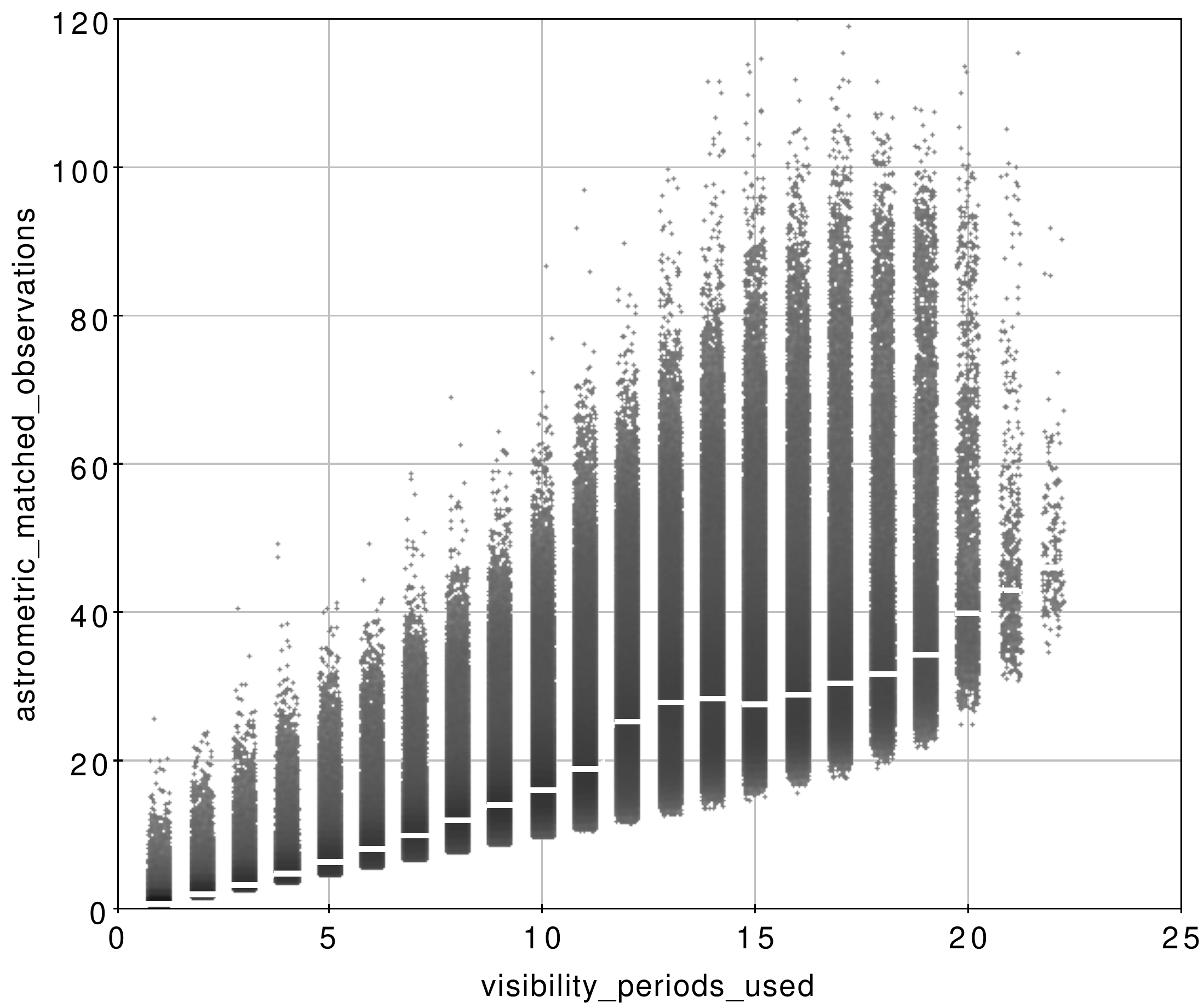}}
    \caption{Relation between the number of visibility periods and field-of-view
    transits (matched observations) per source used in the secondary astrometric
    solutions. A small random number was added to the integer number of visibility 
    periods to widen the vertical bars. The white horizontal line through each bar shows 
    the location of the median. The diagram was constructed for a random subset
    of about 2.5~million sources.}
    \label{fig:vpu}
\end{figure}

\subsection{Acceptance criteria and fall-back (two-parameter) solution}
\label{sec:fallback}

In the final secondary solution (step~5 of Sect.~\ref{sec:agis22}), a five-parameter solution 
without priors was first attempted for every source. If this solution was not of acceptable
quality, a fall-back solution for the two position parameters was tried instead. The fall-back 
solution is actually still a five-parameter solution, but with prior information added on
the parallax and proper motion components. Details of the procedure are given in
\citetads{2015A&A...583A..68M}. In the notation of that paper, the precise priors used
in the fall-back solutions of \textit{Gaia} DR2 were 
$\sigma_{\alpha*,\text{p}}=\sigma_{\delta,\text{p}}=1000$~mas for
the position, $\sigma_{\varpi,\text{p}}=10\,\sigma_{\!\varpi,\text{F90}}$ for the parallax, 
and $\sigma_{\mu,\text{p}}=10{\cal R}\,\sigma_{\!\varpi,\text{F90}}$ for the proper
motion components, with ${\cal R}=10$~yr$^{-1}$.
Compared with a 
genuine two-parameter solution, where the parallax and proper motion are constrained 
to be exactly zero, the use of priors in most cases gives a more realistic estimate of the 
positional uncertainties. The resulting parallax and proper motion values are biased by 
the priors, and therefore not published.

The criterion for accepting a five-parameter solution uses two quality indicators 
specifically constructed for this purpose:
\begin{itemize}
\item
\gacs{visibility\_periods\_used} counts the number of distinct observation epochs,
or ``visibility periods'', used in the secondary solution for a particular source. A visibility period 
is a group of observations separated from other groups by a gap of at least four days.
This statistic is a better indicator of an astrometrically well-observed source than 
for example \gacs{astrometric\_matched\_observations} (the number of field-of-view 
transits used in the solution): while a five-parameter solution is in principle possible with 
fewer than ten field-of-view transits, such a solution will be very unreliable unless the
transits are well spread out in time. As illustrated in Fig.~\ref{fig:vpu}, there are many 
sources with $>$10 transits concentrated in just a few visibility periods.
\item
\gacs{astrometric\_sigma5d\_max} is a five-dimensional equivalent to the semi-major axis
of the position error ellipse and is useful for filtering out cases where one of
the five parameters, or some linear combination of several parameters, is particularly bad. 
It is measured in mas and computed as the square root of
the largest singular value of the scaled $5\times 5$ covariance matrix of the astrometric
parameters. The matrix is scaled so as to put the five parameters on a comparable scale, 
taking into account the maximum along-scan parallax factor for the parallax and the 
time coverage of the observations for the proper motion components. If $\vec{C}$ is
the unscaled covariance matrix, the scaled matrix is $\vec{S}\vec{C}\vec{S}$, where
$\vec{S}=\text{diag}(1,1,\sin\xi,T/2,T/2)$, $\xi=45^\circ$ is the solar aspect angle
in the nominal scanning law, and $T=1.75115$~yr the time coverage of the data used
in the solution. \gacs{astrometric\_sigma5d\_max} was not corrected for the DOF bug,
as that would obscure the source selection made at an earlier stage based on the
uncorrected quantity.
\end{itemize}
The five-parameter solution was accepted if the following conditions were all met for the source:
\begin{equation}\label{eq:critP5}
\left.\begin{aligned}
\text{(i)}\quad&\text{mean magnitude}~G\le 21.0 \\ 
\text{(ii)}\quad&\gacs{visibility\_periods\_used}\ge 6 \\ 
\text{(iii)}\quad&\gacs{astrometric\_sigma5d\_max} \le \text{(1.2~mas)}\times\gamma(G)
\end{aligned} \quad \right\}
,\end{equation}
where $\gamma(G)=\max [1,\,10^{0.2(G-18)}]$. The upper limit in (iii) gradually 
increases from 1.2~mas for $G\le 18$ to 4.78~mas at $G=21$. This test was applied using
preliminary $G$ magnitudes, with the result that some sources in \textit{Gaia} DR2 have
five-parameter solutions even though they do not satisfy (iii).

If the five-parameter solution was rejected by Eq.~(\ref{eq:critP5}), a fall-back solution was attempted
as previously described. The resulting position, referring  to the epoch J2015.5, was 
accepted provided that the following conditions are all met:
\begin{equation}\label{eq:critP2}
\left.\begin{aligned}
\text{(i)}\quad&\gacs{astrometric\_matched\_observations} \ge 5 \\ 
\text{(ii)}\quad&\gacs{astrometric\_excess\_noise}<20~\text{mas} \\ 
\text{(iii)}\quad&\sigma_\text{pos, max}<100~\text{mas} 
\end{aligned} \quad \right\}
.\end{equation}
\gacs{astrometric\_excess\_noise} is the excess source noise $\epsilon_i$ introduced
in Sect.~3.6 of the AGIS paper, and $\sigma_\text{pos, max}$ is the semi-major axis 
of the error ellipse in position given by Eq.~(\ref{eq:sigmaPos}).
Sources rejected also by Eq. (\ref{eq:critP2}) are mostly spurious and no results are
published for them. 

These criteria resulted in 1335~million sources with a five-parameter solution
and 400~million with a fall-back solution, that is,\ without parallax and proper motion.
About 18~million sources were subsequently removed as duplicates, that is,\ where 
the observations of the same physical source had been split between two or more 
different source identifiers. Duplicates were identified by positional coincidence, 
using a maximum separation of 0.4~arcsec. To decide which source to keep, the 
following order of preference was used: unconditionally keep any source (quasar) 
used for the reference frame alignment; otherwise prefer a five-parameter solution 
before a fall-back solution, and keep the source with the smallest 
\gacs{astrometric\_sigma5d\_max} to break a tie. 

\textit{Gaia} DR2 finally gives five-parameter solutions for 1332~million sources,
with formal uncertainties ranging from about 0.02~mas to 2~mas in parallax and
twice that in annual proper motion. For the 361~million sources with fall-back solutions, the 
positional uncertainty at J2015.5 is about 1 to 4~mas. Further statistics are given 
in Appendix~\ref{sec:properties}.

\section{Internal validation}
\label{sec:valid}

This section summarises the results of a number of investigations carried out by
the DPAC astrometry team in order to validate the astrometric solutions. This aimed in
particular at characterising the systematic errors in parallax and proper motion,
and the realism of the formal uncertainties. Some additional quality indicators are
discussed in Appendix~\ref{sec:clean}.

\begin{figure}
\centering
  \resizebox{0.85\hsize}{!}{\includegraphics{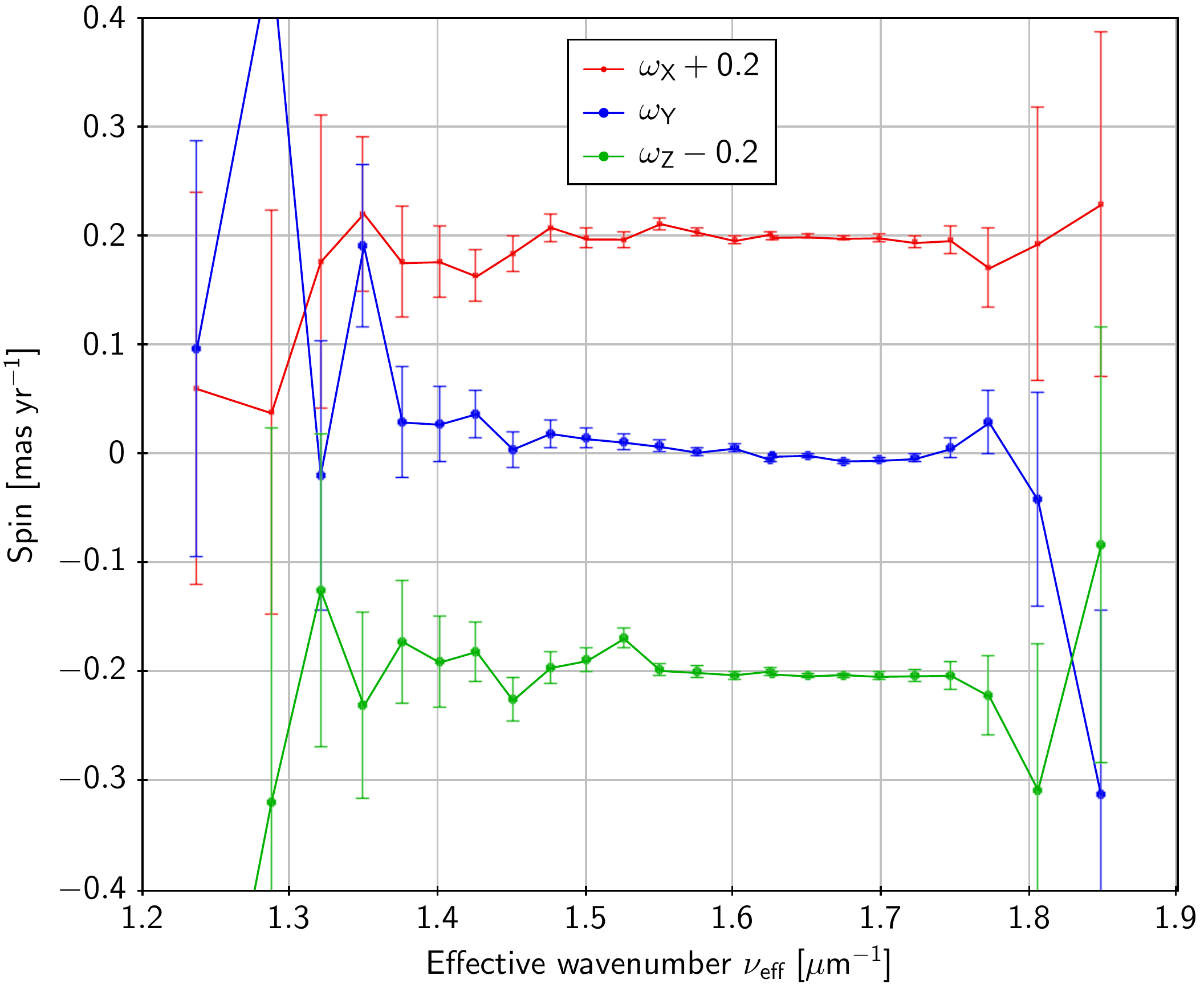}}
    \caption{Dependence of the faint reference frame on colour. The diagram
    shows the components of spin $\omega_X$, $\omega_Y$, and
$\omega_Z$ 
    around the ICRS axes, as estimated for faint ($G\simeq 15$--21) quasars 
    subdivided by effective wavenumber. The components in $X$ and $Z$
    were shifted by $\pm 0.2$~mas~yr$^{-1}$ for better visibility. Error bars
    are at 68\% confidence intervals for the estimated spin. 
    }
    \label{fig:spin1}
\end{figure}

\begin{figure}
\centering
  \resizebox{0.85\hsize}{!}{\includegraphics{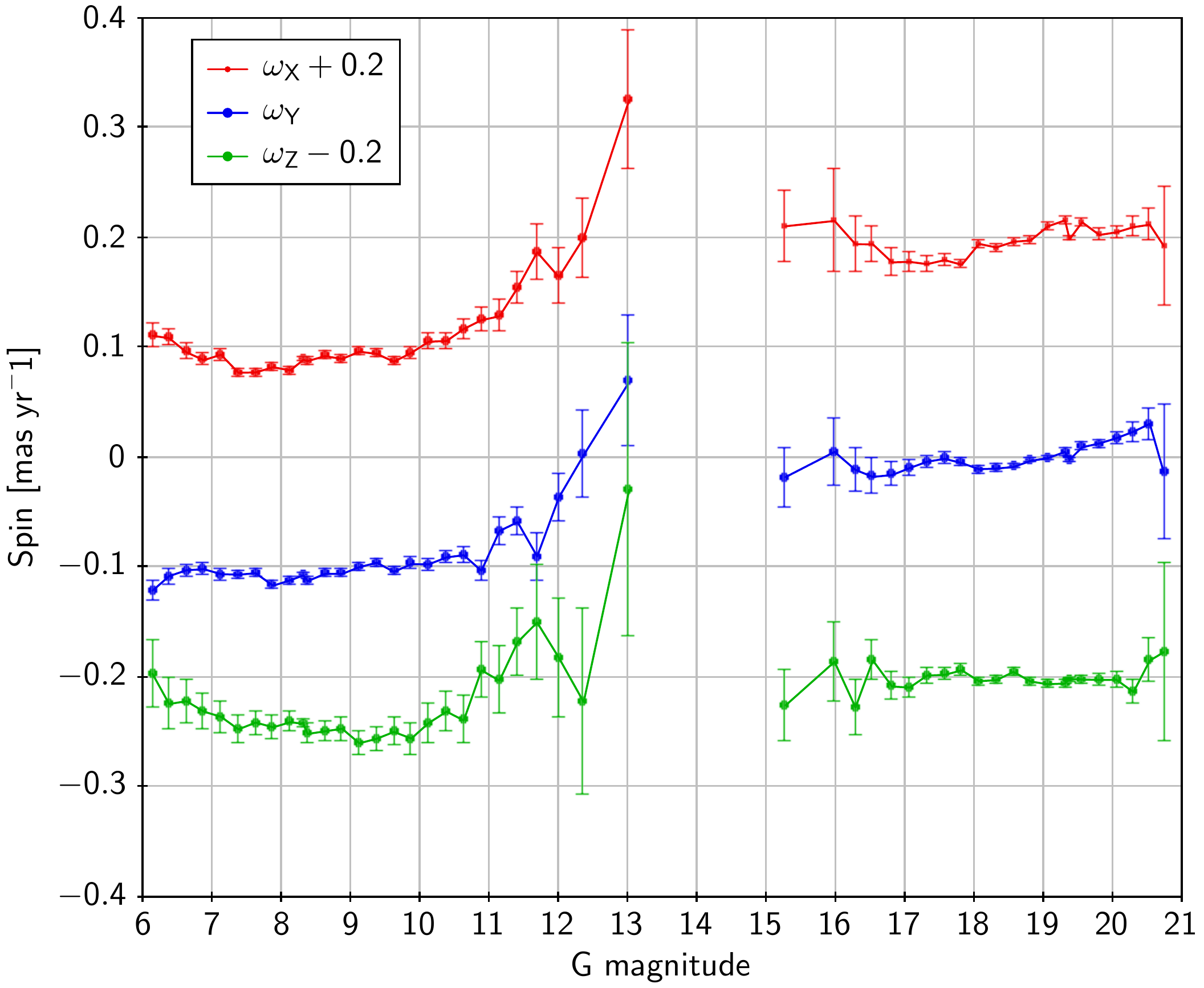}}
    \caption{Dependence of the reference frame on magnitude. The diagram
    shows the spin components as in Fig.~\ref{fig:spin1}, but subdivided by
    magnitude. The points at the faint end ($G\gtrsim 15$) are estimated
    from the proper motions of quasars. At the bright end ($G\lesssim 13$),
    the spin is estimated from the differences in stellar proper motions between
    \textit{Gaia} DR2 and the \textsc{Hipparcos} subset of TGAS in \textit{Gaia} DR1.}
    \label{fig:spin2}
\end{figure}

\subsection{Reference frame}
\label{sec:frame}

The celestial reference frame of \textit{Gaia} DR2, known as \textit{Gaia}-CRF2
\citep{DPACP-30}, is nominally aligned with ICRS and
non-rotating with respect to the distant universe. This was achieved by means 
of a subset of 492\,006 primary sources assumed to be quasars. These 
included 2843 sources provisionally identified as the optical counterparts of VLBI 
sources in a prototype version of ICRF3,
and 489\,163 sources found by cross-matching AGIS02.1 with the AllWISE AGN 
catalogue \citep{2015ApJS..221...12S,2016yCat..22210012S}. 
The unpublished prototype ICRF3 catalogue (30/06/2017, solution from GSFC) 
contains accurate VLBI positions for 4262 radio sources and  
was kindly made available to us by the IAU Working Group \emph{Third Realisation of 
International Celestial Reference Frame}.

The radius for the
positional matching was 0.1~arcsec for the VLBI sources and 1~arcsec for the
AllWISE sample. Apart from the positional coincidence, the joint application of 
the following conditions reduced the risk of contamination by Galactic stars:
\begin{equation}\label{eq:qsoSel}
\left.\begin{aligned}
\text{(i)}\quad&\gacs{astrometric\_matched\_observations} \ge 8 \\ 
\text{(ii)}\quad&\varsigma_\varpi<1~\text{mas} \\ 
\text{(iii)}\quad&|\,\varpi/\varsigma_\varpi\,| <5 \\ 
\text{(iv)}\quad&(\mu_{\alpha*}/\varsigma_{\mu\alpha*})^2+(\mu_{\delta}/\varsigma_{\mu\delta})^2 <25 \\ 
\text{(v)}\quad&|\sin b\,| > 0.1
\end{aligned} \quad \right\}
,\end{equation}
where $b$ is Galactic latitude. We used the formula
$\sin b = (-0.867666\cos\alpha-0.198076\sin\alpha)\cos\delta+0.455984\sin\delta$,
which is accurate to about 0.1~arcsec.
These conditions were applied to both samples, except that (v) was not used for
the VLBI sample where the risk of contamination is much lower thanks to 
the smaller positional match radius. 

The selection of sources for the frame rotator described above was made before the 
final solution had been computed and therefore used preliminary values for the various
quantities in Eq.~(\ref{eq:qsoSel}), including standard uncertainties ($\varsigma$) not yet
corrected for the DOF bug. The resulting subsets of sources are indicated in
the \textit{Gaia} Archive by the field \gacs{frame\_rotator\_object\_type}, which is
2 for the 2843 sources matched to the ICRF3 prototype, 3 for the 489\,163 sources
matched to the AllWISE AGN catalogue, and 0 for sources not used by the frame rotator.
The magnitude distributions of these subsets are shown in Fig.~\ref{fig:histNsrc}.
It can be noted that the AllWISE sample (labelled ``QSO'' in the diagram) contains
three bright sources ($G<12$) that are probably distant Galactic stars of unusual
colours (the brightest being the Herbig AeBe star HD~37357). These objects are not
included in the larger but cleaner quasar sample analysed in Sect.~\ref{sec:zeropoint},
obtained by applying the stricter Eq.~(\ref{eq:qsoSel1}) to the final data. 

The adjustment of the reference frame was done in the primary solution 
(step~4 of Sect.~\ref{sec:agis22}) using the frame rotator described in Sect.~6.1 
of the AGIS paper. At the end of an iteration, the frame rotator estimated the 
frame orientation parameters $[\epsilon_X,\epsilon_Y,\epsilon_Z]$ at J2015.5, 
using the VLBI sources, and the spin parameters $[\omega_X,\omega_Y,\omega_Z]$ 
using the AllWISE and VLBI sources. The attitude and the positions and proper
motions of the primary sources were then corrected accordingly. The acceleration 
parameters $[a_X,a_Y,a_Z]$ were not estimated as part of this process, as they
are expected to be insignificant compared with the current level of systematics 
(see below).

At the end of the primary solution, the attitude was thus aligned with the 
VLBI frame, and the subsequent secondary solutions (step~5 of Sect.~\ref{sec:agis22}) 
should then result in source parameters in the desired reference system. 
This was checked by a separate off-line analysis, using independent software 
and more sophisticated algorithms. This confirmed the global alignment of the 
positions with the VLBI to within $\pm 0.02$~mas per axis. This applies to the 
faint reference frame represented by the VLBI sample with a median
magnitude of $G\simeq  18.8$. The bright reference frame was checked
by means of some 20 bright radio stars with accurate VLBI positions and
proper motions collected from the literature. Unfortunately, their small number
and the sometimes large epoch difference between the VLBI observations
and \textit{Gaia}, combined with the manifestly non-linear motions of many 
of the radio stars, did not allow a good determination of the orientation error of
the bright reference frame of \textit{Gaia} DR2 at epoch J2015.5. No significant 
offset was found at an upper ($2\sigma$) limit of about $\pm 0.3$~mas per axis.     

Concerning the spin of the reference frame relative to the quasars, estimates
of $[\omega_X, \omega_Y, \omega_Z]$ using various weighting schemes 
and including also the acceleration parameters confirmed that the faint reference 
frame of \textit{Gaia} DR2 is globally non-rotating to within $\pm 0.02$~mas~yr$^{-1}$ 
in all three axes. Particular attention was given 
to a possible dependence of the spin parameters on colour (using the effective 
wavenumber $\nu_\text{eff}$) and magnitude ($G$). Figure~\ref{fig:spin1}
suggests a small systematic dependence on colour, for example,\ by 
$\pm 0.02$~mas~yr$^{-1}$ over the range $1.4 \lesssim \nu_\text{eff}\lesssim 1.8~\mu\text{m}^{-1}$ 
corresponding to roughly $G_\text{BP}-G_\text{RP}=0$ to 2~mag.
As this result was derived for quasars that are typically fainter than 15th magnitude,
it does not necessarily represent the quality of the \textit{Gaia} DR2 reference frame for
much brighter objects.

Figure~\ref{fig:spin2} indeed suggests that the bright ($G\lesssim 12$) reference 
frame of \textit{Gaia} DR2 has a significant ($\sim$0.15~mas~yr$^{-1}$) spin 
relative to the fainter quasars. The points in the left part of the diagram were 
calculated from stellar proper motion differences between the current solution and 
\textit{Gaia} DR1 (TGAS). Only 88\,091 sources in the \textsc{Hipparcos} 
subset of TGAS were used for this comparison owing to their superior precision
in TGAS. Although based on a much shorter
stretch of \textit{Gaia} observations than the present solution, TGAS provides a valuable
comparison for the proper motions thanks to its $\sim$24~yr time difference from
the \textsc{Hipparcos} epoch. If the spin difference of 0.15~mas~yr$^{-1}$
between the two catalogues were to be explained as systematics in TGAS,
it would require an alignment error of $\sim$3.6~mas in the positions either in 
TGAS at epoch J2015.0 or in \textsc{Hipparcos} at epoch J1991.25. Given the way
these catalogues were constructed, both hypotheses are very unlikely. The 
most reasonable explanation for the offsets in Fig.~\ref{fig:spin2} is therefore 
systematics in the \textit{Gaia} DR2 proper motions of the bright sources. The gradual change 
between magnitudes 12 and 10 suggests an origin in the gated observations, 
which dominate for $G\lesssim 12$, or possibly in observations of window class 0, 
which dominate for $G\lesssim 13$. 

Formally, \textit{Gaia}-CRF2 is materialised by the positions in \textit{Gaia} DR2 of the 
556\,869 sources identified as quasars in Sect.~\ref{sec:zeropoint}. A separate list of
these sources is provided in the \textit{Gaia} Archive.    
A more comprehensive analysis of \textit{Gaia}-CRF2 is given by \citet{DPACP-30}.

\begin{figure}
\centering
  \resizebox{0.9\hsize}{!}{\includegraphics{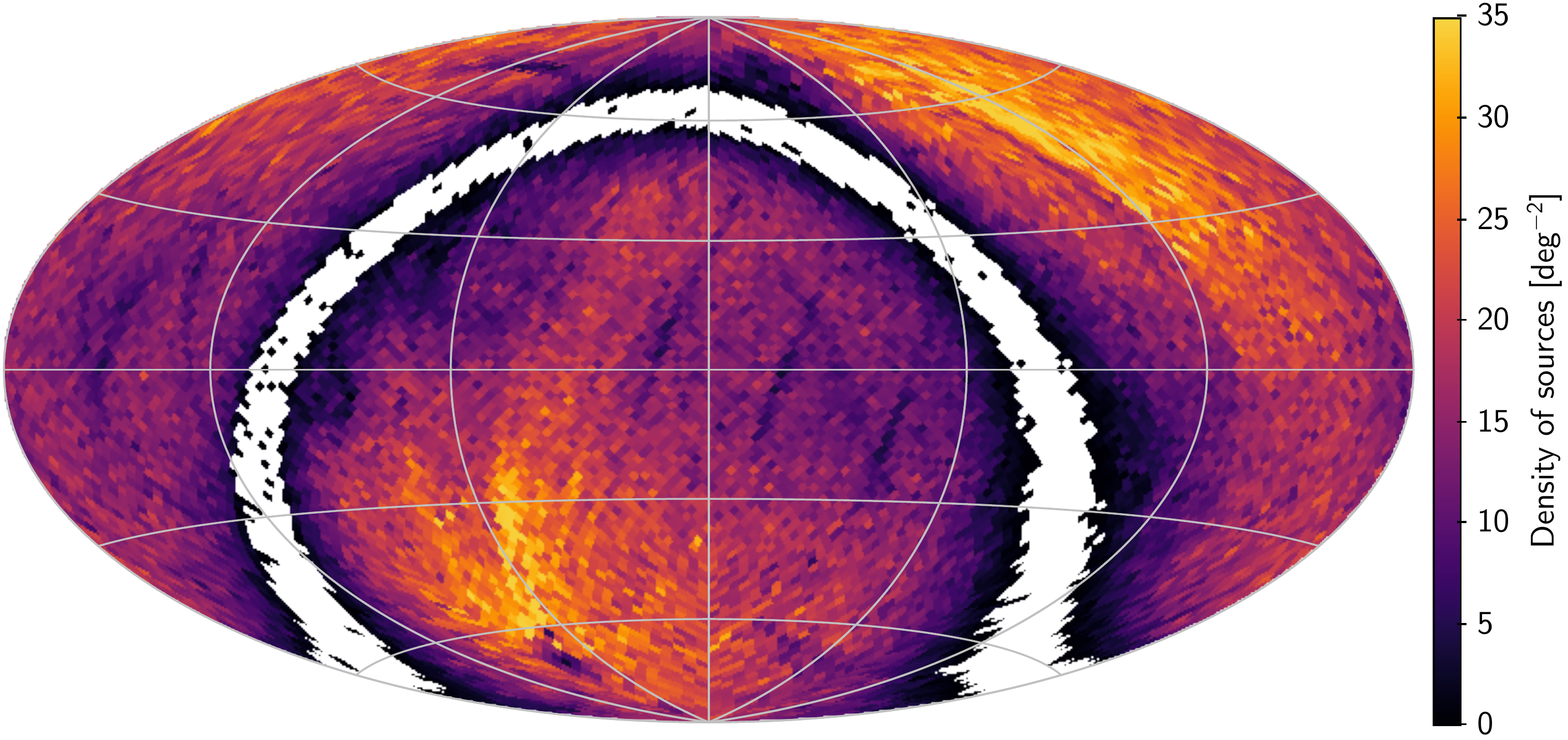}}
    \caption{Density map of the full quasar sample (union of AllWISE AGNs and
    VLBI sources) at a resolution of $1.8\times 1.8$~deg$^2$. The scatter of 
    points in the Galactic band are VLBI sources.
    This and following full-sky maps use a Hammer--Aitoff projection in equatorial (ICRS) 
    coordinates with $\alpha=\delta=0$ at the centre, north up, and 
    $\alpha$ increasing from right to left.}
    \label{fig:qsoDensityMap}
\end{figure}

\begin{figure}
\centering
  \resizebox{0.8\hsize}{!}{\includegraphics{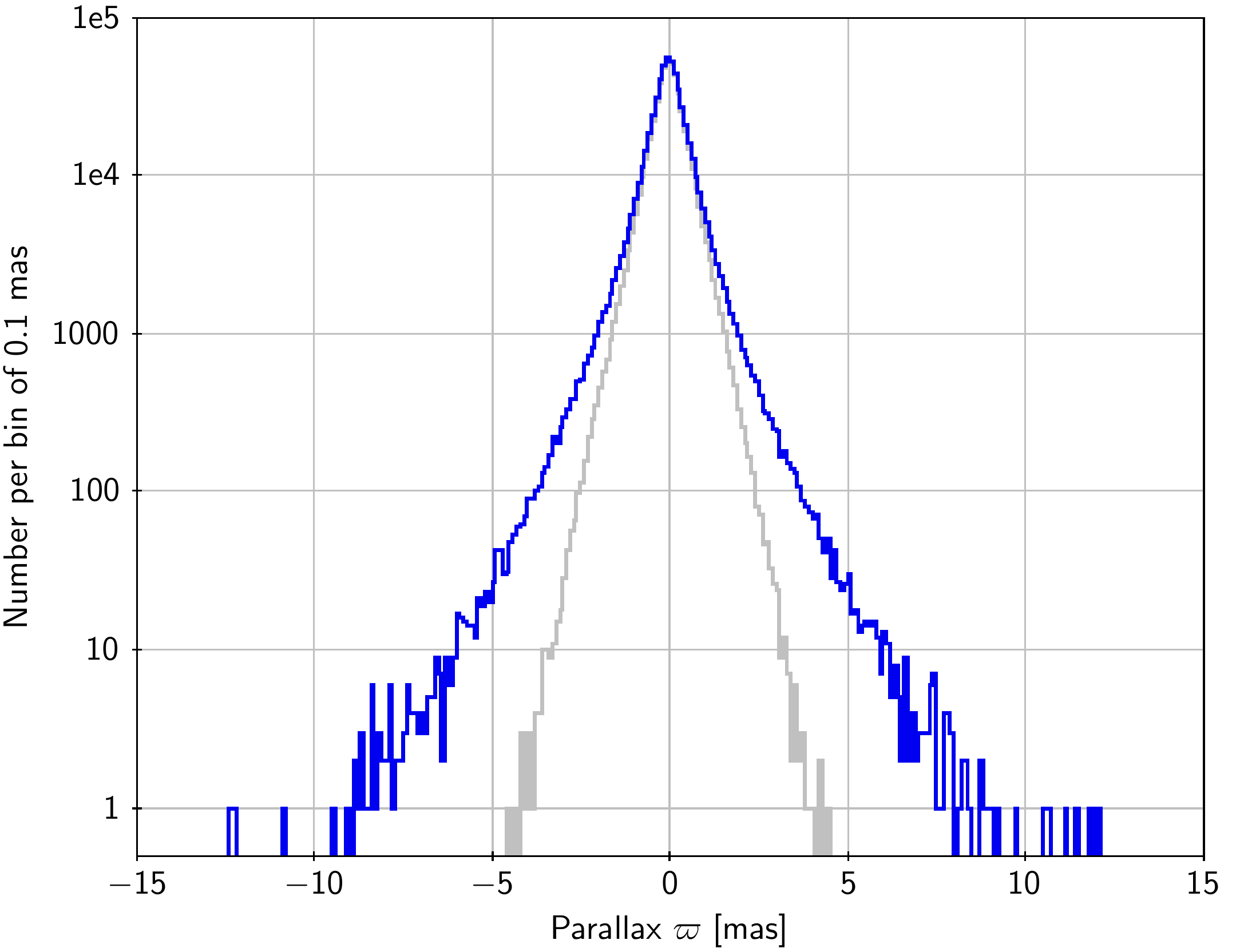}}
    \caption{Parallax distribution for 556\,869 sources identified as quasars.
    Outer (blue) curve: the whole sample; inner (grey) curve: the
    subsample of 492\,928 sources with $\sigma_\varpi<1$~mas.}
    \label{fig:qsoPlxHist}
\end{figure}

\subsection{Parallax zero point}
\label{sec:zeropoint}

Global astrometric satellites like \textsc{Hipparcos} and \textit{Gaia} are able to
measure absolute parallaxes, that is,\ without zero-point error, but this capability is
susceptible to various instrumental effects, in particular, to a certain kind of 
basic-angle variations. As discussed by 
\citetads{2017A&A...603A..45B}, periodic variations of the basic angle ($\Gamma$)
of the form $\delta\Gamma(t) = A_1 d(t)\cos\Omega(t)$, where $d(t)$ is the distance 
of \textit{Gaia} from the solar system barycentre in au and $\Omega(t)$ is the spin
phase relative to the barycentre, are observationally almost indistinguishable from 
a global parallax shift of $\delta\varpi=A_1/[2\sin\xi\sin(\Gamma/2)]\simeq 0.883A_1$.
This is clearly reminiscent of the first term in Eq.~(\ref{eq:calVF3c}).
Although $d$, $\xi,$ and $\Omega$ in that equation are heliotropic quantities, while
the present formula uses barytropic quantities, and $d$ appears with different powers 
in the two expressions, the differences are small enough to cause a near-degeneracy 
between $A_1$ and $\delta C_{1,0}$. This is the reason why the latter parameter 
was not estimated in the spin-related distortion 
model.

\begin{figure*}
\centering
  \resizebox{0.33\hsize}{!}{\includegraphics{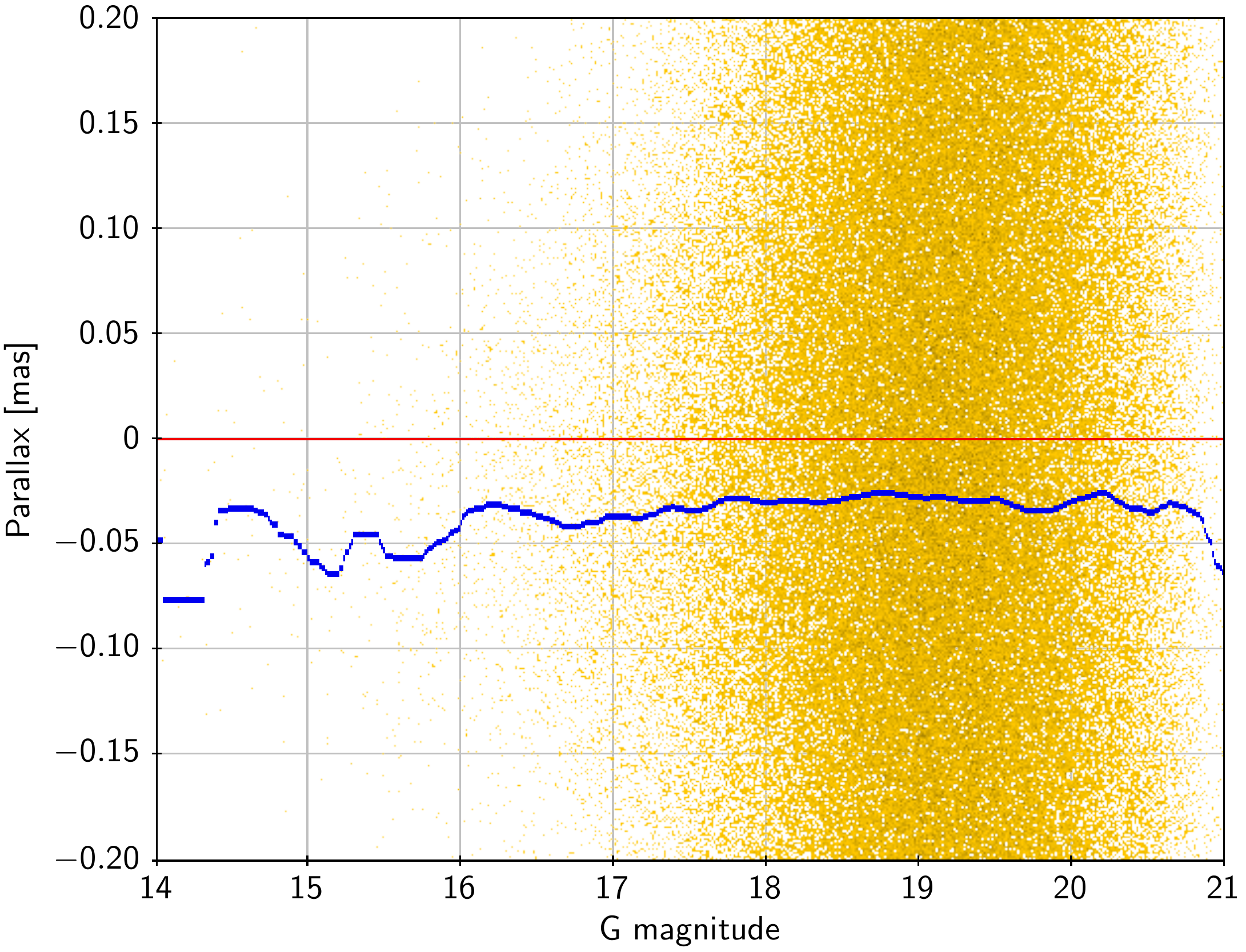}}
  \resizebox{0.33\hsize}{!}{\includegraphics{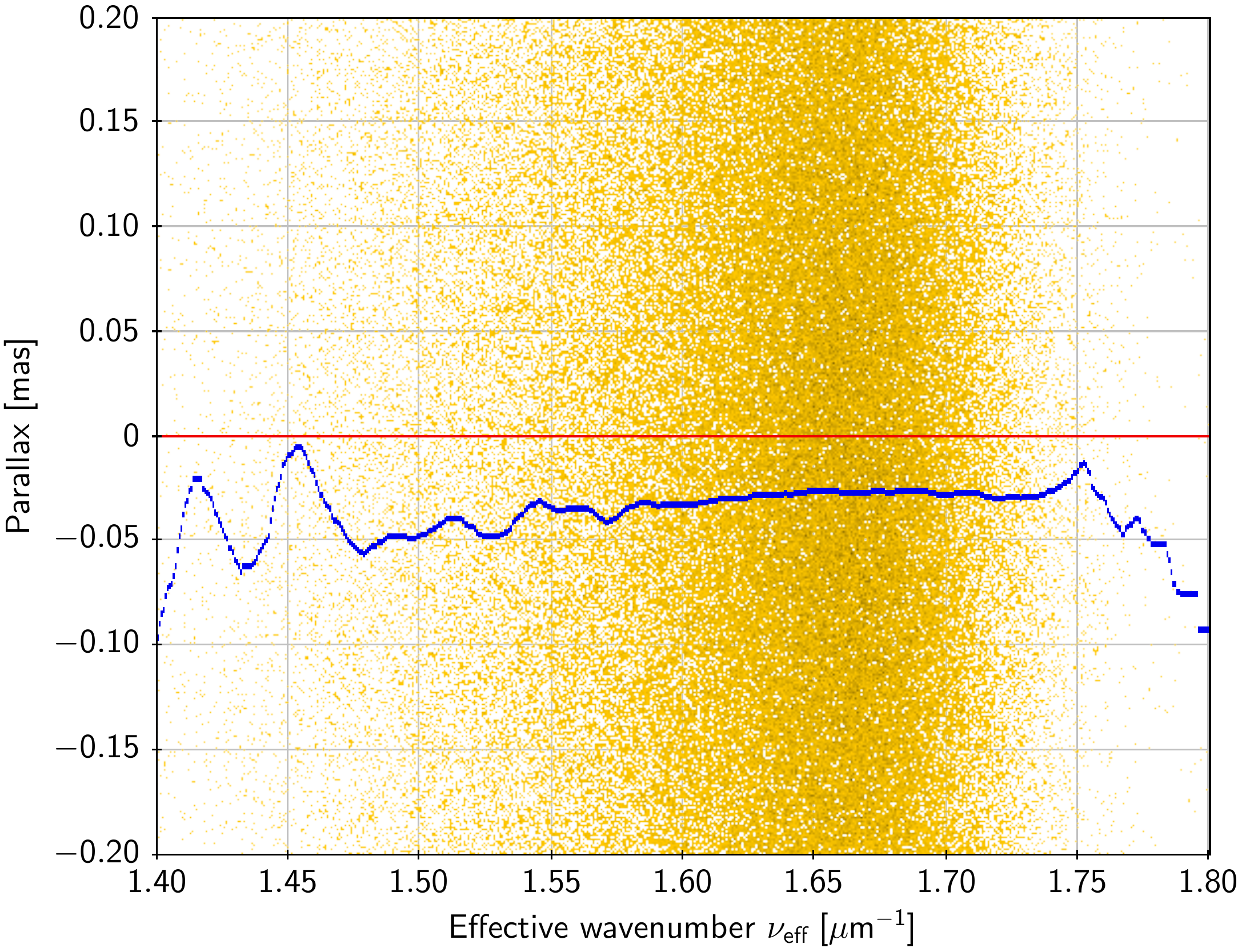}}
  \resizebox{0.33\hsize}{!}{\includegraphics{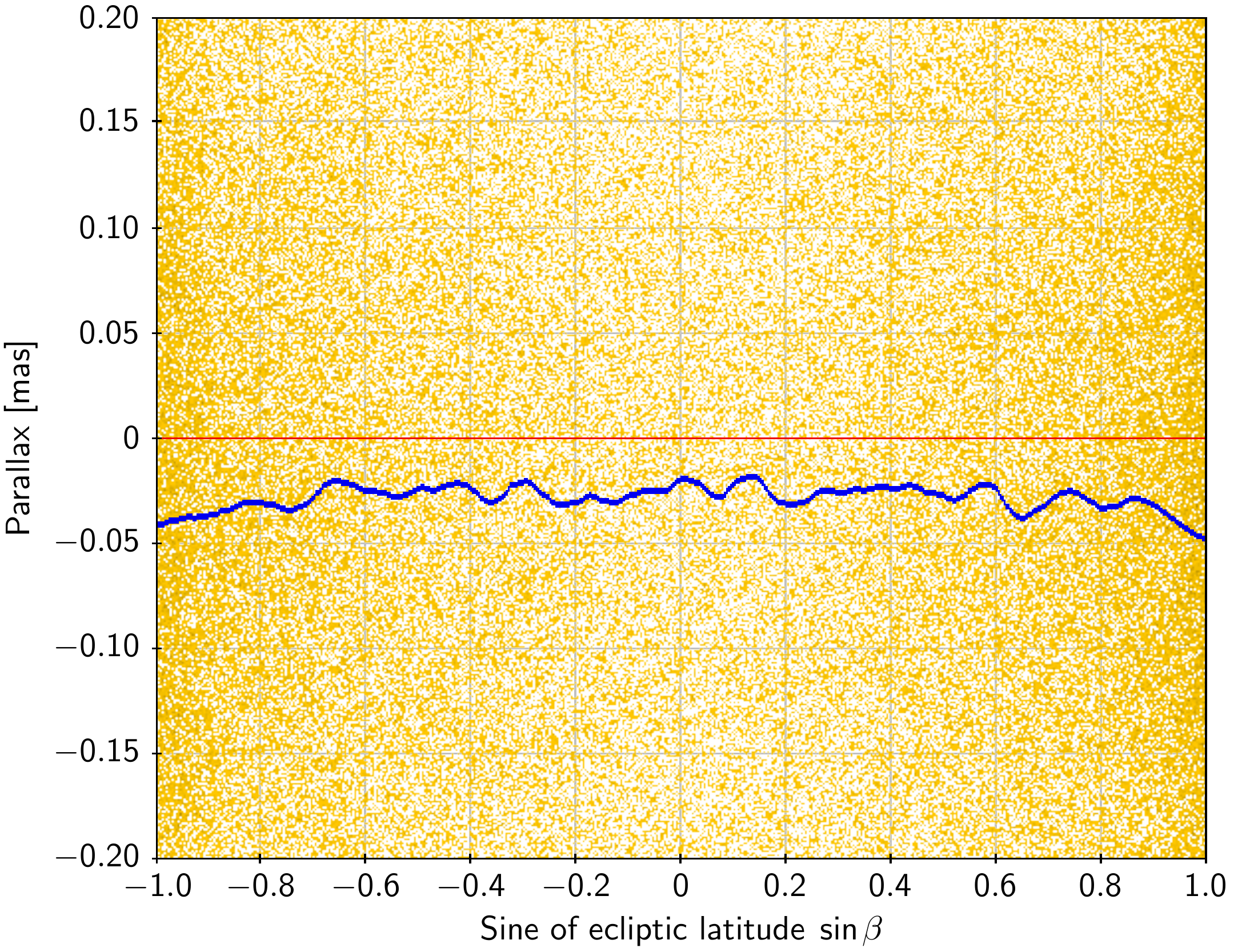}}
    \caption{Parallaxes for the full quasar sample plotted against magnitude (left), 
    colour (middle), and ecliptic latitude (right). Because of the chosen scale, only about
    one-third of the data points are shown as yellow dots; the blue curves are the
    running medians.}
    \label{fig:qsoPlxVsGetc}
\end{figure*}

\begin{figure}
\centering
  \resizebox{0.9\hsize}{!}{\includegraphics{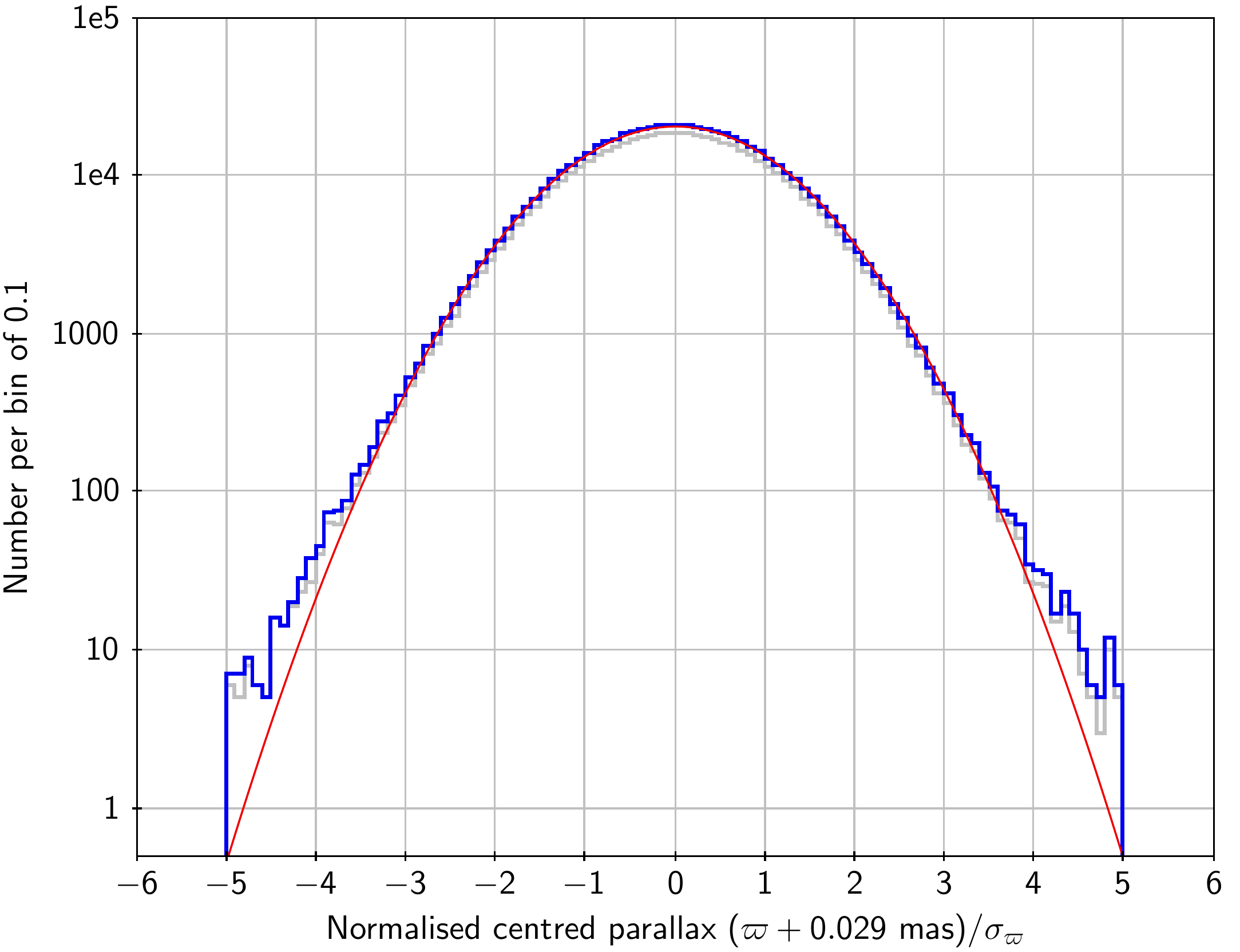}}
    \caption{Distributions of the normalised centred parallaxes for the same
    samples as in Fig.~\ref{fig:qsoPlxHist}.
    The red curve is a Gaussian distribution with the same standard deviation
    (1.081) as the normalised centred parallaxes for the full sample.}
    \label{fig:qsoNcplxHist}
\end{figure}

It is believed that the basic-angle corrector derived from BAM data 
(Sect.~\ref{sec:bam}) eliminates basic-angle variations very
efficiently, 
but a remaining small variation corresponding to the undetermined $\delta C_{1,0}$
cannot be excluded. This would then show up as a small offset in the parallaxes.
For this reason, it is extremely important to investigate the parallax zero point by
external means, that is,\ using astrophysical sources with known parallaxes. 
It is also important to check possible dependences of the zero point on other 
factors such as position, magnitude, and colour, which could be created by
errors in the calibration model.

The quasars are almost ideal for checking the parallax zero point thanks to their 
extremely small parallaxes ($<0.0025~\mu$as for redshift $z>0.1$), large number, 
availability over most of the celestial sphere, and, in most cases, nearly point-like 
appearance. The main drawbacks are their faintness and peculiar colours.

In order to create the largest possible quasar sample for validation purposes,
a new cross-match of the final \textit{Gaia} DR2 data with the AllWISE AGN catalogue 
\citepads{2015ApJS..221...12S} was made, 
choosing in each case the nearest positional match. The further selection used the criteria 
\begin{equation}\label{eq:qsoSel1}
\left.\begin{aligned}
\text{(i)}\quad&\gacs{astrometric\_matched\_observations} \ge 8 \\ 
\text{(ii)}\quad&\gacs{astrometric\_params\_solved} = 31 \\ 
\text{(iii)}\quad&\left|\,(\varpi+0.029~\text{mas})/\sigma_\varpi\,\right| <5 \\ 
\text{(iv)}\quad&(\mu_{\alpha*}/\sigma_{\mu\alpha*})^2+(\mu_{\delta}/\sigma_{\mu\delta})^2 <25 \\ 
\text{(v)}\quad&|\sin b\,| > 0.1 \\
\text{(vi)}\quad&\rho < (2~\text{arcsec})\times |\sin b\,|
\end{aligned} \quad \right\}
,\end{equation}
which is somewhat similar to Eq.~(\ref{eq:qsoSel}), but stricter and applied to the final data.
Step (ii) selects five-parameter solutions ($31=11111_2$), and
step (iii) takes into account the 
median offset of the parallaxes (see below). The combination of steps 
(v) and (vi) makes the probability of a chance match with a Galactic star generally 
lower than ${\sim}10^{-4}$ at all Galactic latitudes. A reality check of the resulting selection 
against SIMBAD revealed that the two brightest sources (at $G=8.85$ and 11.72~mag)
are stars; removing them leaves 555\,934 sources in the sample. The fraction of stars 
among the AllWISE AGN sources is estimated at $\le 0.041\%$ \citepads{2015ApJS..221...12S}, 
or $\lesssim 230$ in this sample, but only a fraction of them may pass the criteria in 
Eq.~(\ref{eq:qsoSel1}).

Applying conditions (i)--(iv) to the sources matched to the ICRF3 prototype
(Sect.~\ref{sec:frame}) gave 2820 sources, 1885 of which were already in the 
AllWISE sample. The union set thus contains a total of 556\,869 sources,
which also define the celestial reference frame of \textit{Gaia} DR2
\citep{DPACP-30}.
A density map of this quasar sample (Fig.~\ref{fig:qsoDensityMap}) shows
imprints of the \textit{Gaia} and AllWISE scanning laws as well as the effects
of Galactic extinction and confusion.
In the following, the high-precision subset of 492\,928 sources with 
$\sigma_\varpi < 1$~mas is sometimes used instead of the full quasar sample.

Figure~\ref{fig:qsoPlxHist} shows the distribution of parallaxes for the full quasar
sample and the high-precision subset. For the full sample, the mean and
median parallax is $-0.0308$~mas and $-0.0287$~mas, respectively; for the
high-precision subset, the corresponding values are $-0.0288$~mas and 
$-0.0283$~mas. For the subsequent analysis
we adopt $-0.029$~mas as the global zero point of the parallaxes.
Scatter plots of the parallaxes versus magnitude and colour (left and middle
panels of Fig.~\ref{fig:qsoPlxVsGetc}) show systematic trends with a change of 
$\sim$0.02~mas over the ranges covered by the data. A plot against ecliptic
latitude (right panel) shows a roughly quadratic variation with $\sim$0.010~mas
smaller parallaxes towards the ecliptic poles. Thus, while the global mean offset
of $-0.029$~mas is statistically well-determined, the actual offset applicable 
for a given combination of magnitude, colour, and position may be different
by several tens of $\mu$as. Spatial variations of the parallax zero point are 
further analysed in Sect.~\ref{sec:corr}.

Figure~\ref{fig:qsoNcplxHist} shows the distribution
of $(\varpi+0.029~\text{mas})/\sigma_\varpi$, that is,\ the parallaxes corrected
for the global offset and normalised by the formal uncertainties. Ideally, this
should follow a normal distribution with zero mean and unit variance.
The actual sample standard deviation of this quantity is 1.081. Similarly, 
the sample standard deviations of the normalised proper motions, 
$\mu_{\alpha*}/\sigma_{\mu\alpha*}$ and 
$\mu_{\delta}/\sigma_{\mu\delta}$, are 1.093 and 1.115, respectively.
The distributions are very close to normal, as suggested by the red curve in
Fig.~\ref{fig:qsoNcplxHist}, although it should be noted that the selection in 
Eq.~(\ref{eq:qsoSel1}) removed any point beyond $\pm 5$ units in the normalised
quantities. The conclusion is that the accidental errors are close to normal,
but with a standard deviation some 8--12\% larger than the formal uncertainties.
This applies to the faint sources ($G\gtrsim 15$) beyond the Galactic plane
($|\sin b\,|>0.1$) represented by the quasar subset.

The observations contributing to the parallax determinations are distributed
roughly uniformly over the 62 CCDs in the central $0.7^\circ\times 0.7^\circ$ 
astrometric field of the \textit{Gaia} instrument. The basic-angle variation 
relevant for the parallax zero point is therefore effectively given by the 
average variation in this field. On the other hand, the CCD generating the 
BAM data is situated about $0.7^\circ$ from the centre of the astrometric field, 
that is,\ well outside the field near one of its corners. The corrections given in
Table~\ref{tab:vbac} show that the variations measured by the BAM are 
not fully representative of the variations present in the astrometric field. 
It is noted that a parallax zero point of $-29~\mu$as corresponds to a value
$\simeq -33~\mu$as for the undetermined correction $\delta C_{1,0}$ in
Table~\ref{tab:vbac}.

Differential variations within the astrometric field depending on $\Omega$ 
are described by the global parameters $c_{fklm}$, $s_{fklm}$ in
Eq.~(\ref{eq:calVF3b}), which are estimated in the primary solution.
In principle, this allows the differential variations to be extrapolated to the 
location of the BAM. Although such a procedure is clearly problematic, it
could provide an independent estimate of the crucial parameter 
$\delta C_{1,0}$ and important consistency checks for other parameters.
A detailed investigation along these lines will only be meaningful at a later
time when other calibration errors have been substantially eliminated.
With the current solution, we note that the largest amplitudes $|c_{fklm}|$, 
$|s_{fklm}|$ are associated with the lowest temporal ($k$) and spatial 
($l+m$) orders, as would be expected for a physical instrument. Moreover, 
their sizes (0.01 to 0.05~mas) are in the approximate range needed to 
account for the corrections to the BAM data reported in Table~\ref{tab:vbac} 
as well as the global parallax offset of $-0.029$~mas. However, there could
be many other explanations for this offset; in particular, it appears that 
unmodelled AL centroid shifts related to the transverse smearing of the 
images during a CCD integration (depending on the AC rate 
$\text{d}\zeta/\text{d}t$) could be an important contributor (Sect.~\ref{sec:residual}).

\subsection{Residual analysis}
\label{sec:residual}  
  
\begin{figure}
\centering
  \resizebox{0.85\hsize}{!}{\includegraphics{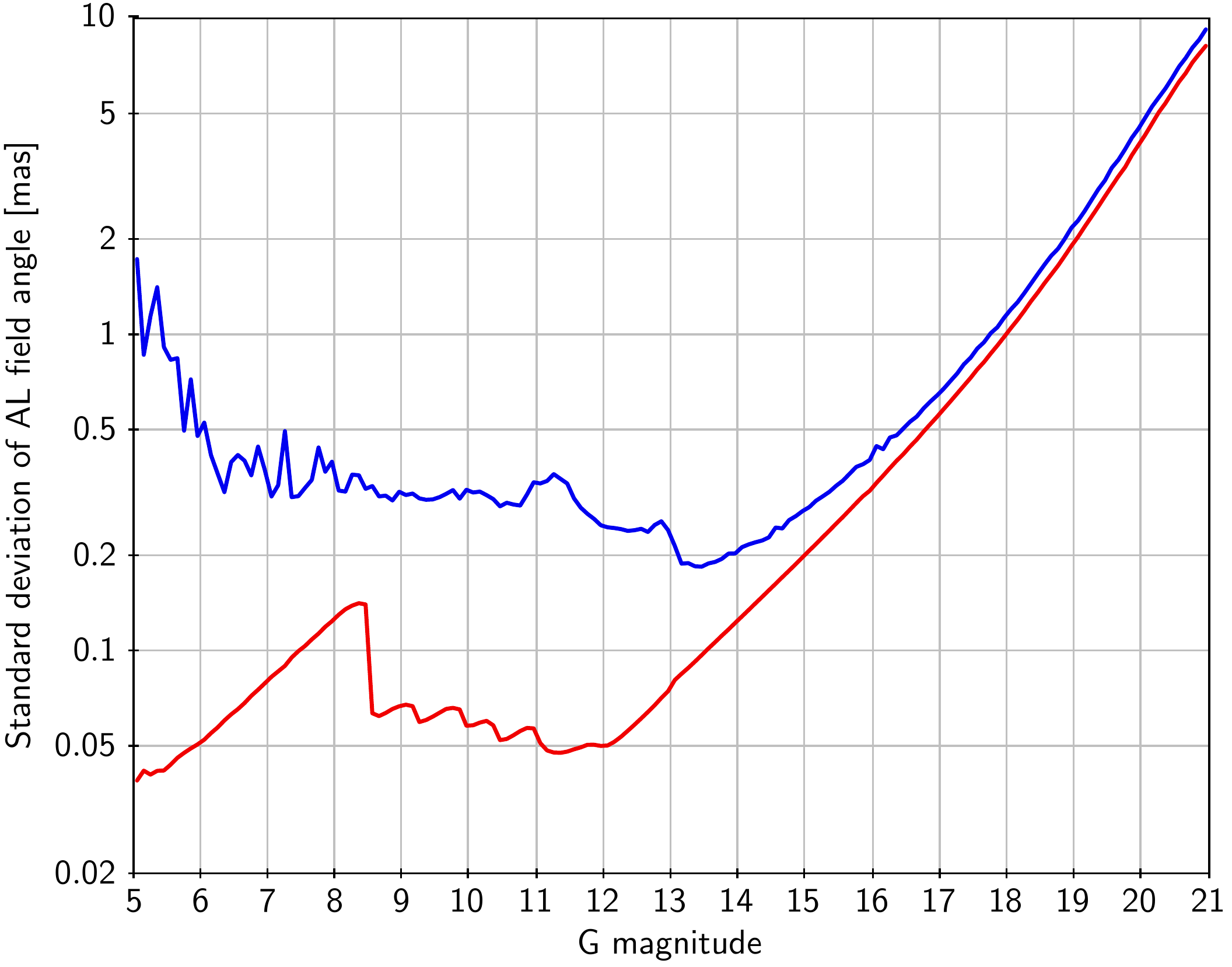}}
    \caption{Precision of along-scan astrometric measurements as a function
    of magnitude. The red (lower) curve is a running median of the formal precision from
    the image parameter determination; the blue (upper) curve is a robust estimate of the 
    actual standard deviation of the post-fit residuals. The difference between 
    the two curves represents the combination of all unmodelled errors.}
    \label{fig:sigmaAlVsG}
\end{figure}

Analysis of the astrometric residuals can reveal inadequacies in the calibration model, for 
example where a new effect needs to be added or where the time granularity of some  
effect already included in the model has insufficient resolution. It is particularly interesting
to look for model deficiencies that might explain the systematics seen in the astrometric
results, for instance,\ the parallax zero point error. In this section we first estimate the total size
of the unmodelled errors, and then give two examples of effects that contribute to the
errors in the present solutions, but could be eliminated in future releases.
 
Figure~\ref{fig:sigmaAlVsG} compares the photon-statistical uncertainties of the AL 
angular measurements with the scatter of post-fit residuals in the astrometric solution.
The red curve is the formal precision from the image parameter determination, derived from
the assumed Poissonian character of the individual CCD sample values. This curve has
three domains, depending on the number of photons ($N$) in the stellar image: for 
moderately bright sources ($G\simeq 12$--17), the centroiding precision is limited by
the photon noise in the stellar image, or $\sigma\propto N^{-1/2}$, leading to a slope of
about 0.2~dex~mag$^{-1}$; for fainter sources ($G\gtrsim 17$), the background gradually
becomes more important, leading to a higher slope in the red curve; finally, for the 
bright stars ($G\lesssim 12$), the use of the gates limits $N$ and hence the centroiding
precision to a value roughly independent of $G$.
 
The blue curve in Fig.~\ref{fig:sigmaAlVsG} is the robust scatter estimate (RSE)%
 \footnote{The RSE is a robust measure of the dispersion of a distribution, defined as 
$\left(2\!\sqrt{2}\,{\rm erf}^{-1}\!\left(4/5\right)\right)^{-1}\simeq 0.390152$ 
times the difference between the 90th and 10th percentiles. For a normal distribution, 
the RSE equals the standard deviation.\label{footnote:RSE}}
of the post-fit residuals, computed in bins of 0.1~mag. For faint sources, it agrees
reasonably well with the formal uncertainties (for $G>17$ the RSE is on average 15\%
higher than the formal uncertainties), but for brighter sources, there is a strong
discrepancy. The difference between the blue and red curves represents the 
combination of all unmodelled source, attitude, and calibration errors. The quadratic 
difference amounts to about 0.3~mas for $G\simeq 6$--12, 0.25~mas for 
$G\simeq 12$--13, and 0.15~mas for $G\gtrsim 12$. Part of this may be attributable 
to the sources (e.g.\ binarity), part to residual attitude irregularities, but a major part is 
clearly due to inadequacies of the calibration models, including the LSF and PSF models 
used for the image parameter determination. A main task in preparation for future \textit{Gaia} data
releases will be to improve these models and hence reduce the gap between the two curves.

\begin{figure}
\centering
  \resizebox{0.96\hsize}{!}{\includegraphics{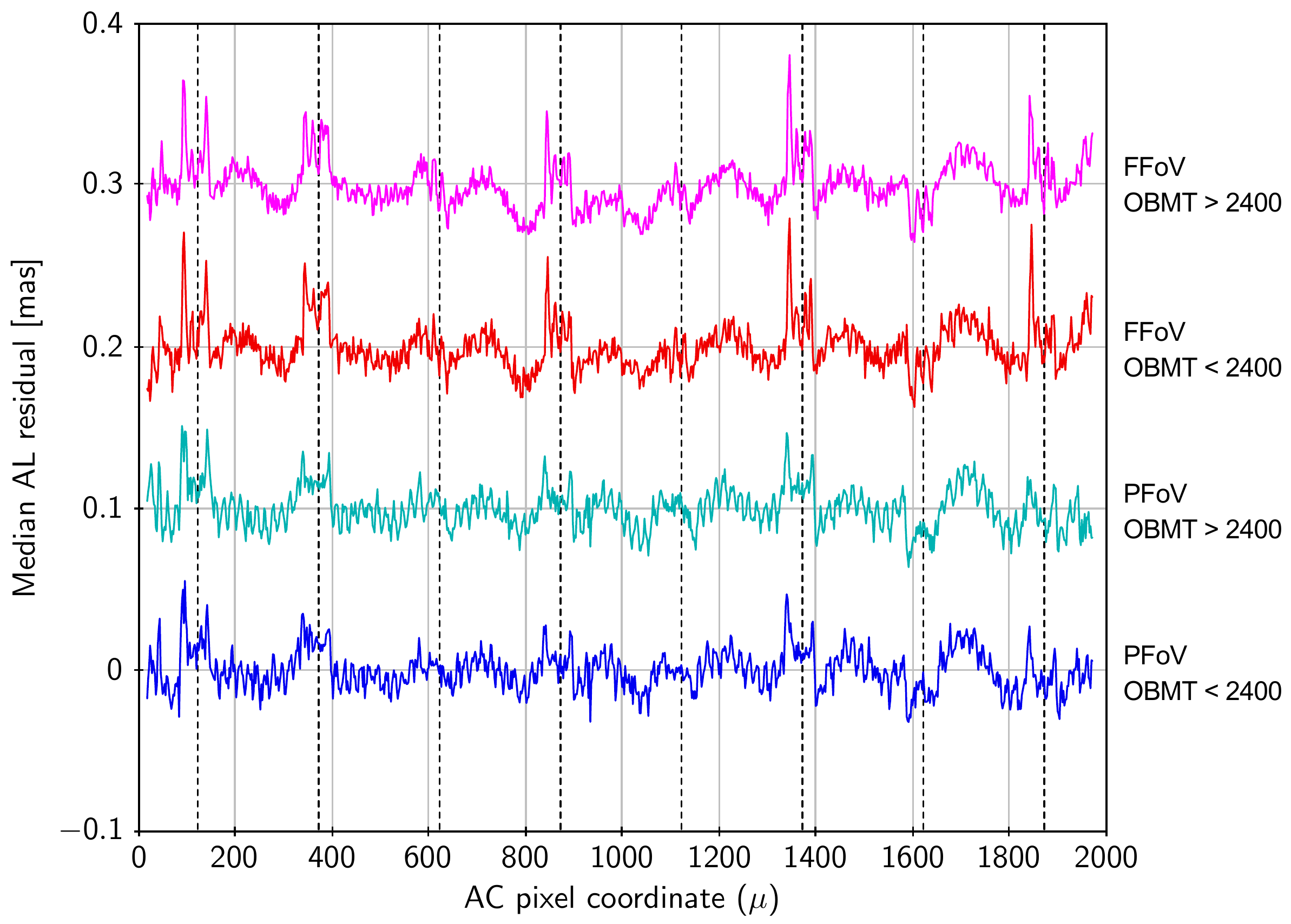}}
    \caption{Small-scale distortion for ungated observations on one of the astrometric CCDs
    (strip 7, row 4). The curves show the median AL residual for sources in the magnitude
    range $G=13$--16 plotted against the AC pixel coordinate $\mu$, and subdivided according
    to field of view (preceding PFoV, or following FFoV) and time (before or after the decontamination
    at $\text{OBMT}\simeq 2400$). For better visibility, the successive curves were vertically 
    displaced by 0.1~mas. The vertical dashed lines show the stitch block boundaries, which divide
    the 1966 pixels in blocks of 250~pixels, except for the two outermost blocks that are 
    108~pixels.}\label{fig:alResVsMu}
\end{figure}

\begin{figure}
\centering
  \resizebox{0.9\hsize}{!}{\includegraphics{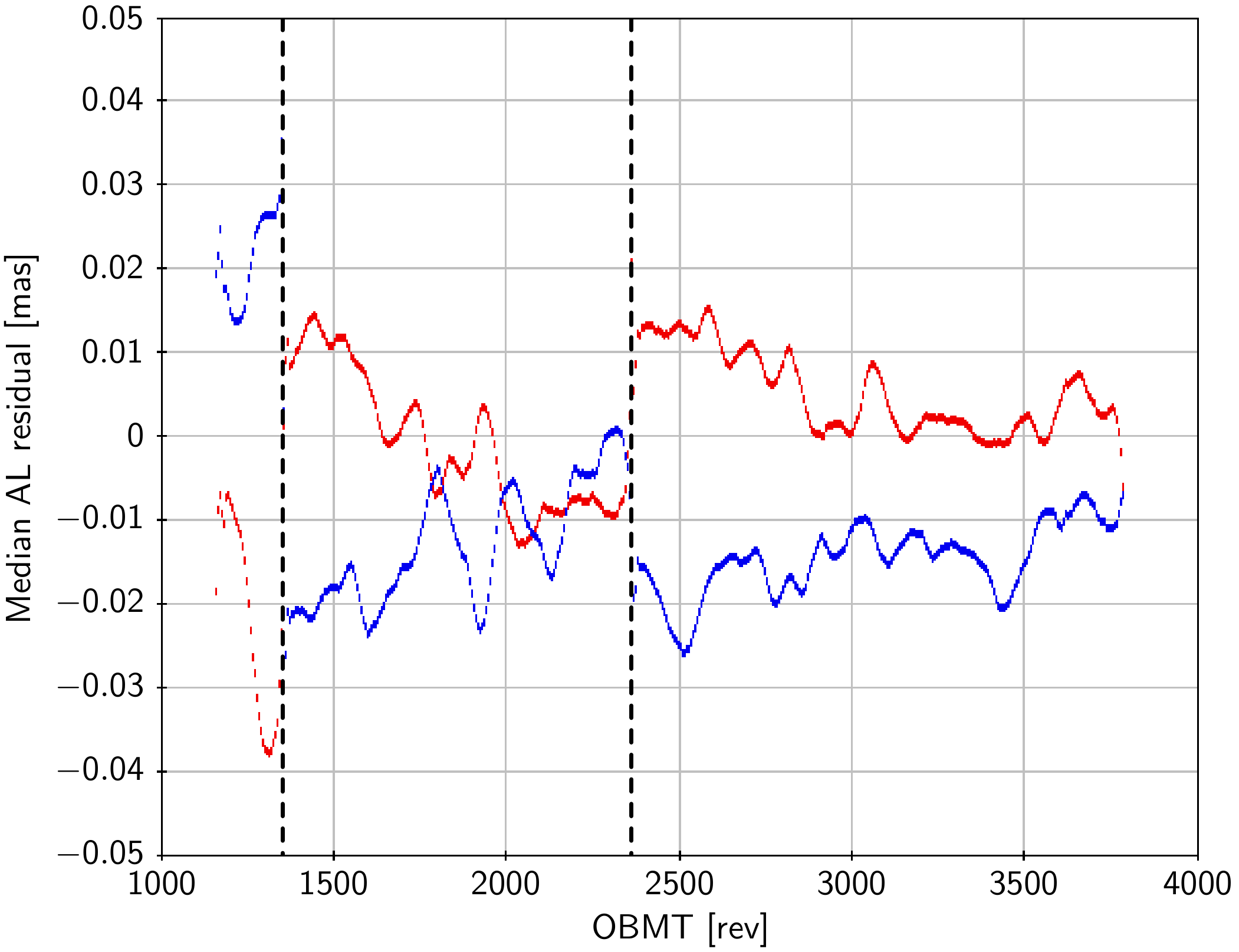}}
    \caption{Residual systematics depending on the AC scan rate. The curves show the median
    residual as a function of OBMT for observations of window class 1 ($G\simeq 13$--16) in
    the preceding field of view. The red curve is for observations with positive AC rate, and the 
    blue curve for negative AC rate. The vertical dashed lines show the approximate times
    of the two decontamination events.}
    \label{fig:alResVsAcRate}
\end{figure}

The astrometric calibration model (Sect.~\ref{sec:cal}) currently does not include small-scale
irregularities of the CCDs. To assess the importance of such errors, we plot in Fig.~\ref{fig:alResVsMu}
the median AL residual, subdivided by field of view and time, as a 
function of the AC pixel coordinate $\mu$. Comparing the four curves, it is seen that the
pattern is extremely stable in time, but slightly different in the two fields of view.
The rms amplitude is only 0.013~mas in the preceding and 0.015~mas in the following field
of view, far too small to explain the discrepancy seen in Fig.~\ref{fig:sigmaAlVsG}.
While the small-scale irregularities are therefore unimportant in the current solution, they
will be included in future calibration models.

One of the most interesting trends revealed by the residual analysis concerns a hitherto 
unmodelled dependence on the across-scan rate $\text{d}\zeta/\text{d}t$, where $\zeta$ 
is the AC field angle. In the nominal scanning law, the AC rate varies sinusoidally over the
6~hr spin period with an amplitude of about $\pm 0.18$~arcsec~s$^{-1}$, or $\pm 0.3$\% of
the constant AL rate (60~arcsec~s$^{-1}$). It is in general different in the two fields of view.  
The AC motion of stellar images by up to 0.8~arcsec during its motion across a CCD smears
the PSF in the AC direction. While this obviously has a strong effect on the AC location of the 
image, it should, to a first approximation, not affect the AL location of the centroid. However,
secondary effects involving a non-symmetric PSF or non-linear response to the photon flux 
could easily generate a small dependence
of the precise AL location on the AC rate. Figure~\ref{fig:alResVsAcRate} shows that this is
indeed the case. Test solutions including astrometric calibration terms depending on the 
AC rate show reduced levels of systematics, for example\ in terms of the $\sim\,$1~deg scale
correlations discussed in Sect.~\ref{sec:corr}. AL centroiding errors depending on the AC rate
are particularly insidious, as the AC rate exhibits a strong correlation with the AL parallax 
factor in the current nominal scanning law.

\begin{figure}
\centering
  \resizebox{0.9\hsize}{!}{\includegraphics{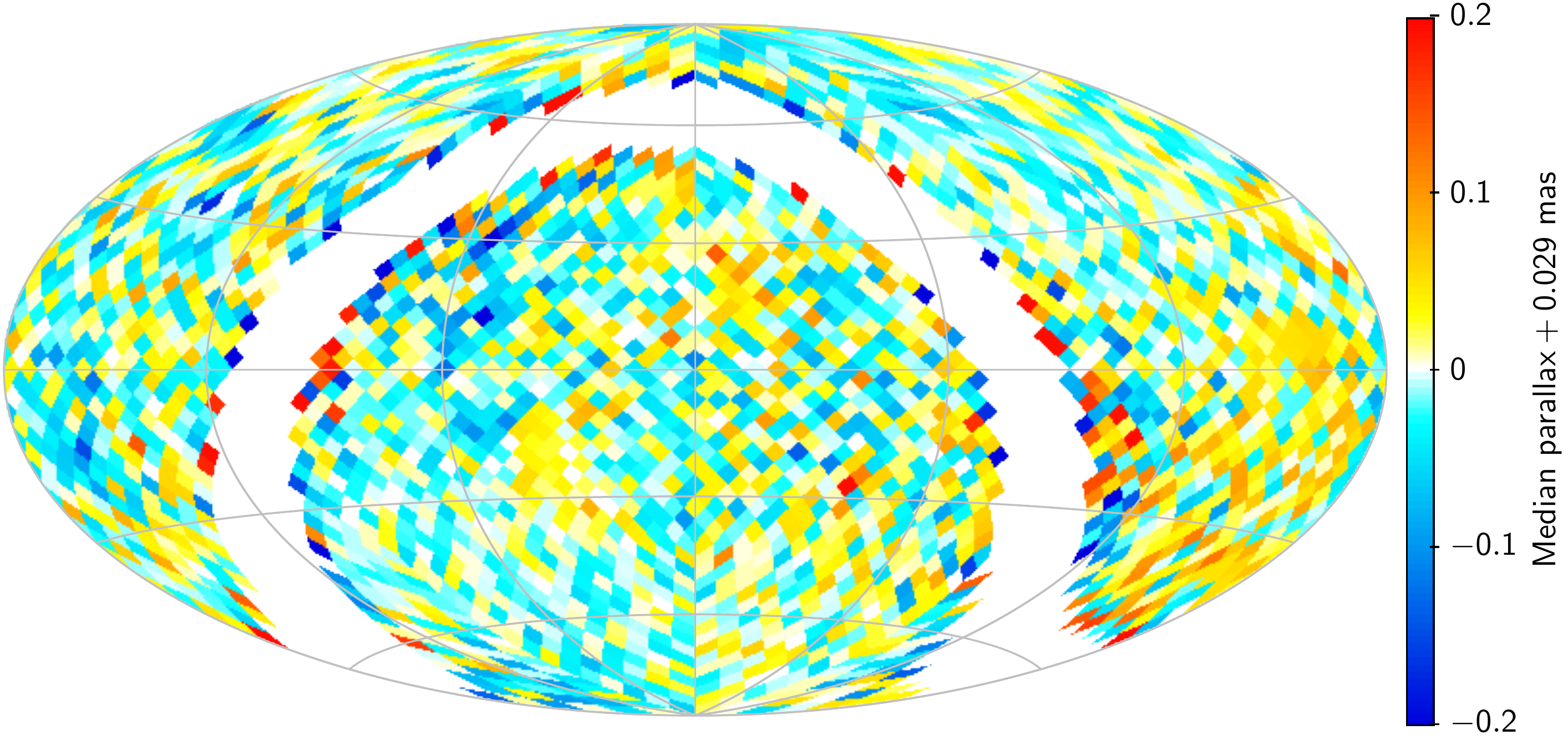}}
    \caption{Map of the median parallaxes for the full quasar sample,
    showing large-scale variations of the parallax zero point.
    See Fig.~\ref{fig:qsoDensityMap} for the coordinate system and
    density of sources.
    Median values are calculated in cells of about $3.7\times 3.7$~deg$^2$.
    Only cells with $|\sin b\,|>0.2$ are plotted.}
    \label{fig:qsoPlxMap}
\end{figure}

\begin{figure}
\centering
  \resizebox{0.9\hsize}{!}{\includegraphics{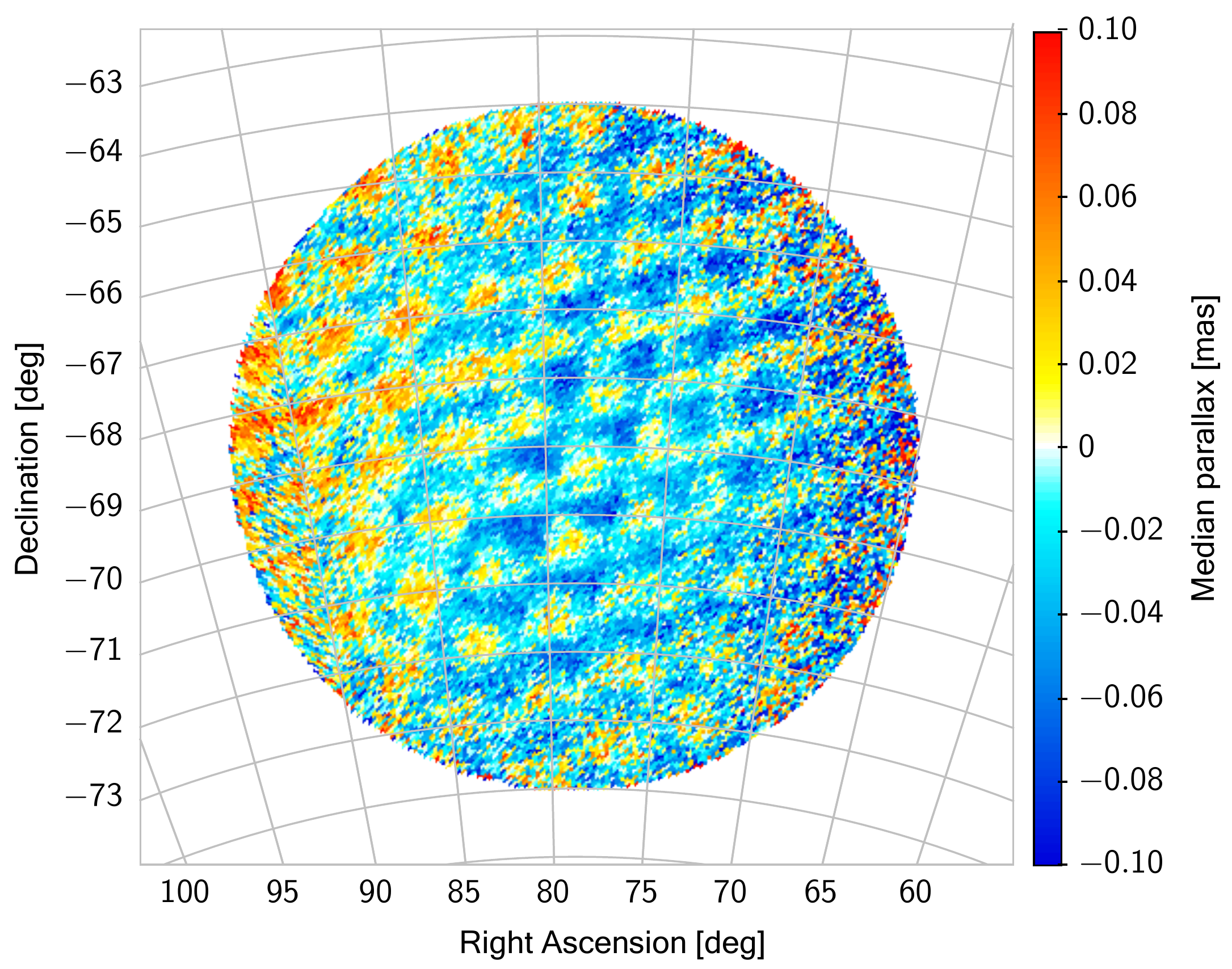}}
    \caption{Map of the median parallaxes for a sample of sources in 
    the LMC area, showing small-scale variations of the parallax zero point.
    Median values are calculated in cells of about $0.057\times 0.057$~deg$^2$.}
    \label{fig:lmcPlxMap}
\end{figure}

\subsection{Spatial correlations}
\label{sec:corr}

Figure~\ref{fig:qsoPlxMap} is a map of the median quasar parallax, adjusted 
for the median offset $-0.029$~mas, at a resolution of a few degrees. Away from
the Galactic plane, where there is a sufficient density of quasars 
(cf.\ Fig.~\ref{fig:qsoDensityMap}) for estimating a local zero point, 
there are several areas of a few tens of 
degrees where the parallaxes are systematically offset by about 
$\pm 0.05$~mas from the global mean. This demonstrates the presence of 
correlated errors on spatial scales of 10--20~deg and RMS values of a few 
tens of $\mu$as. Irregularities on smaller scales cannot be probed in this way
using quasars, owing to their low average density. 

However, distant stars in
dense regions reveal significant variations on much smaller scales. 
As an example, Fig.~\ref{fig:lmcPlxMap} shows the median parallaxes for about 
2.5~million sources in the area of the LMC.
To remove most foreground stars, we selected sources with
magnitudes between $G=17$ and 19, within 5~deg of the LMC centre 
$(\alpha,\delta)=(78.77^\circ,-69.01^\circ)$, and with proper motions 
$(\mu_{\alpha*}-1.850)^2+(\mu_\delta-0.233)^2<1$~mas$^2$~yr$^{-2}$
(cf.\ \citealt{DPACP-34}). The mean and median values of their parallaxes are 
$-0.014$~mas, roughly consistent with the parallax zero point from quasars 
at the LMC location near the South Ecliptic Pole (Fig.~\ref{fig:qsoPlxVsGetc}, 
right), assuming a true parallax of 0.020~mas for the LMC \citepads{2001ApJ...553...47F}.
The quasi-regular triangular pattern in Fig.~\ref{fig:lmcPlxMap} has a period of about 1~deg and a typical
 amplitude of about $\pm 0.03$~mas. The left part of the circular area seems
 to be offset by 0.02~mas from the rest with a straight and rather sharp boundary.
 These patterns are clearly related to \textit{Gaia}'s scanning law with its 
 precessional motion of about 1~deg per revolution. Similar (unphysical) patterns 
 are seen in parallax maps of high-density areas around the Galactic centre, 
 and also in the proper motions. Thus strong correlated errors (or systematics) 
 also exist on spatial scales much below 1~deg.

\begin{figure}
\centering
  \resizebox{0.85\hsize}{!}{\includegraphics{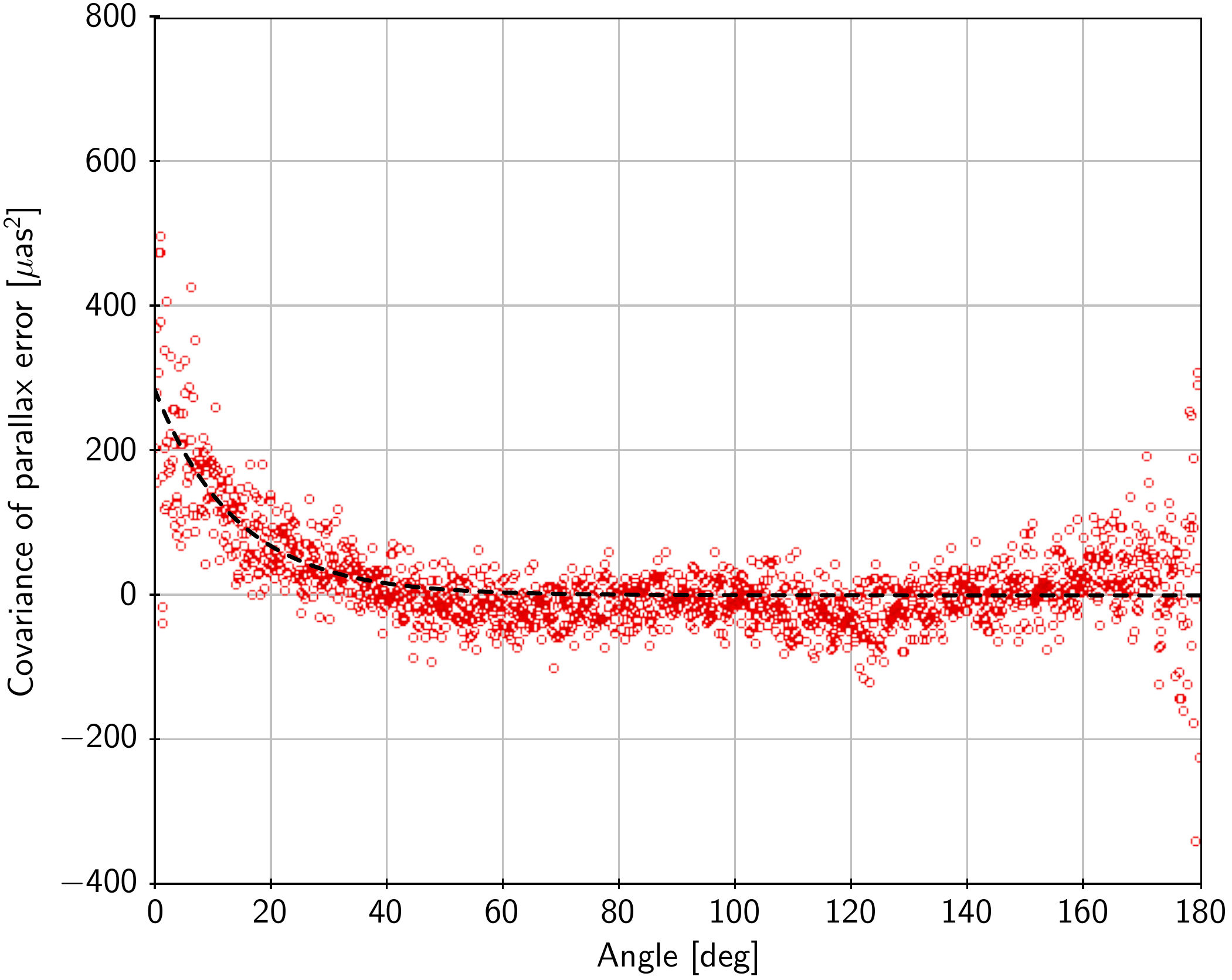}}
  \resizebox{0.85\hsize}{!}{\includegraphics{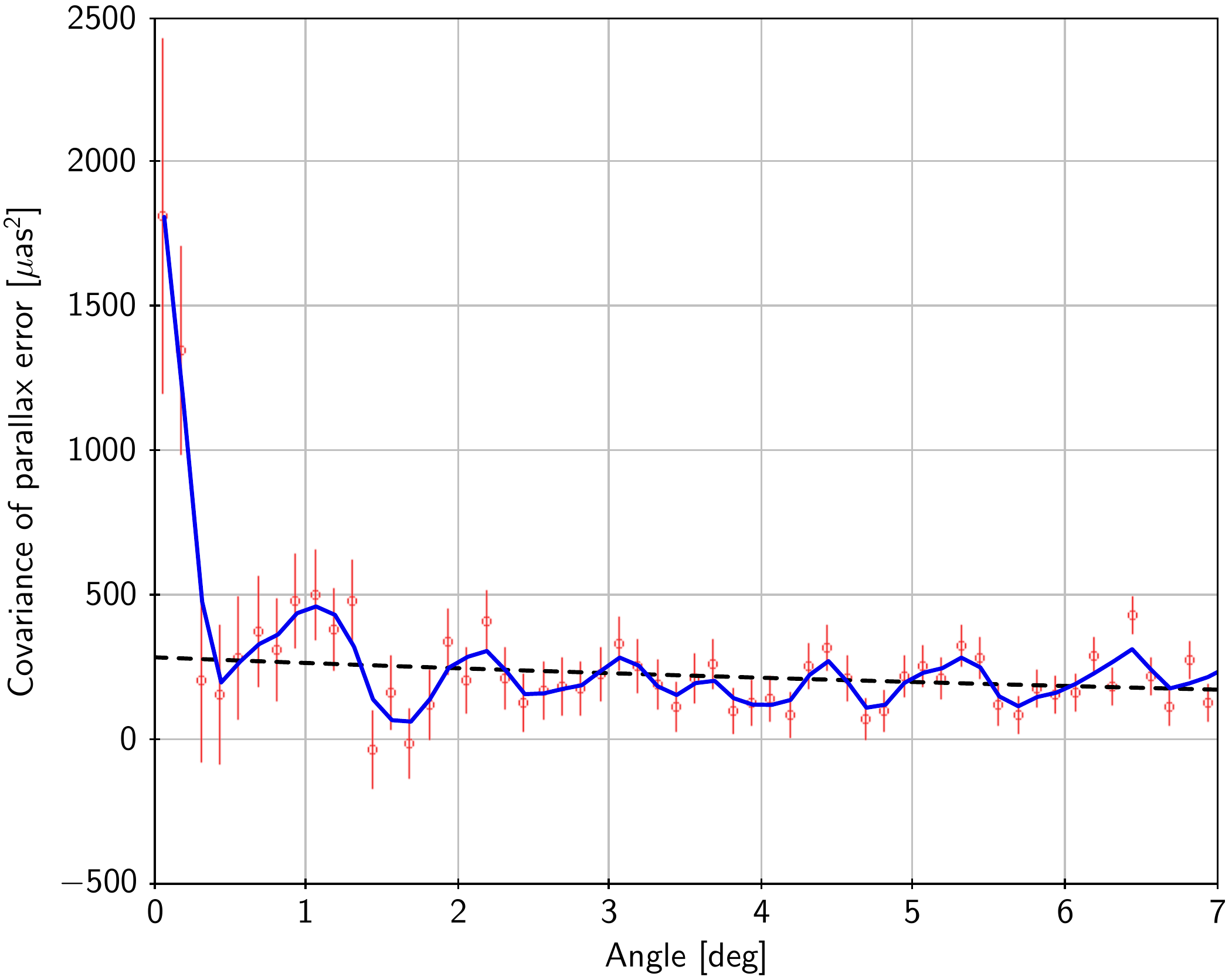}}
    \caption{\textit{Top:} Spatial covariance $V_\varpi(\theta)$ of parallax errors in the 
    high-precision quasar sample. Red circles are the individual estimates, and the dashed black 
    curve shows a fitted exponential. \textit{Bottom:} The same data for separations $<7^\circ$
    with errors bars (68\% confidence intervals) and a running triangular mean (blue curve). 
    The two highest points, for separations $<0.25^\circ$, are outside the plot in the top panel.}
    \label{fig:qsoPlxCov}
\end{figure}

\begin{figure}
\centering
  \resizebox{0.85\hsize}{!}{\includegraphics{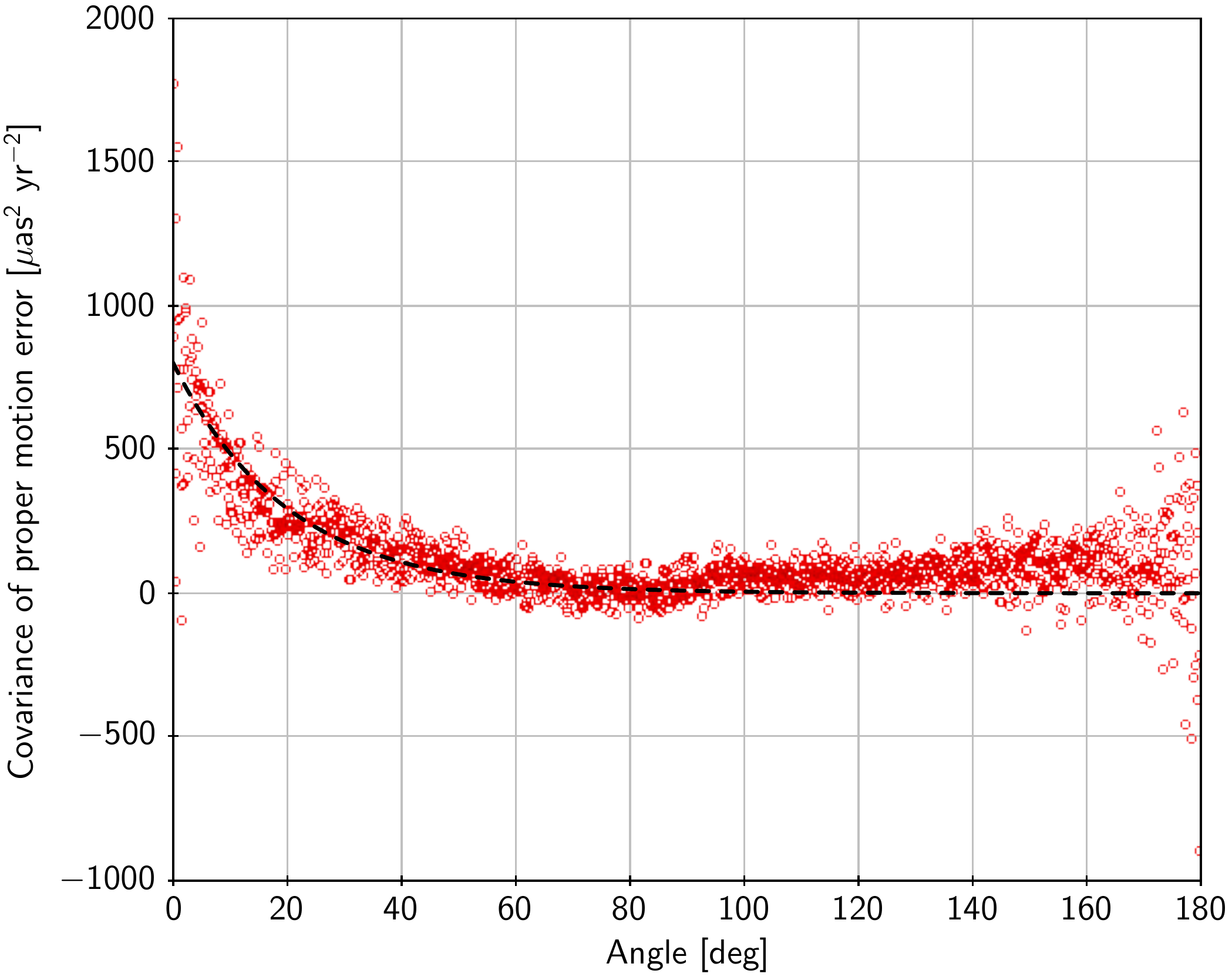}}
  \resizebox{0.85\hsize}{!}{\includegraphics{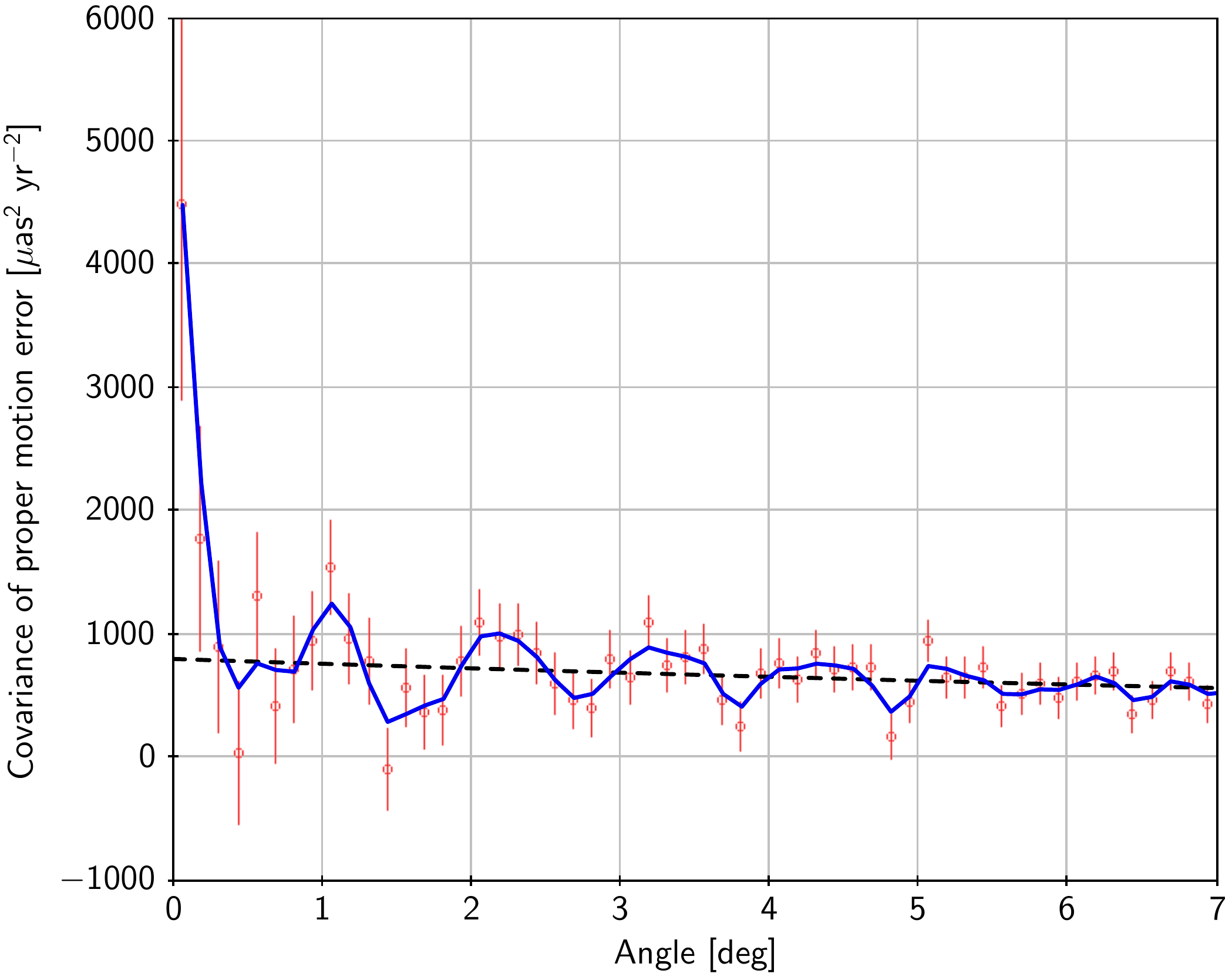}}
    \caption{Same as Fig.~\ref{fig:qsoPlxCov}, but for the proper motions of the
    high-precision quasar sample ($V_\mu(\theta)$). The highest point, for the 
    smallest separation, is outside the plot in the top panel.}
    \label{fig:qsoPmCov}
\end{figure}

A global, quantitative characterisation of these correlations can be obtained by
calculating the covariance of the quasar parallax errors as a function of angular 
separation,
\begin{equation}\label{eq:covPlx}
V_\varpi(\theta) = \langle (\varpi_i-\overline{\varpi})(\varpi_j-\overline{\varpi}) \rangle \, .
\end{equation}
Here $\overline{\varpi}$ is the mean parallax of all the quasars in the sample, and
the average is taken over all non-redundant pairs of quasars $(i>j)$ with angular 
separation $\theta\pm\Delta\theta/2$. Figure~\ref{fig:qsoPlxCov} shows the result 
of this calculation for the high-precision quasar sample, using a bin width of
$\Delta\theta=0.125$~deg. The positive covariance for angles $\lesssim 40$~deg 
is a signature of large-scale systematics
and is reasonably well approximated by the fitted exponential
\begin{equation}\label{eq:covPlxExp}
V_\varpi(\theta) \simeq (285~\mu\text{as}^2)\times\exp(-\theta/14^\circ) \,  ,
\end{equation}
shown by the dashed curve. This function corresponds to errors with an RMS amplitude 
of $285^{1/2}\simeq 17~\mu$as and a characteristic spatial scale of 14~deg, both of which
are consistent with the large-scale patterns seen in Fig.~\ref{fig:qsoPlxMap}.
The dip in $V_\varpi(\theta)$ around $\theta=120$~deg may be related to the basic
angle, although it is centred on a slightly higher value than $\Gamma=106.5$~ deg.

The lower panel of Fig.~\ref{fig:qsoPlxCov} shows $V_\varpi(\theta)$ for $\theta<7$~deg.
The blue curve connects the slightly smoothed values. Although Eq.~(\ref{eq:covPlxExp}),
shown by the dashed curve, well describes the mean covariance averaged over a few degrees,
the detailed curve shows multiple oscillations around the exponential with a period of
about 1~deg, and for the smallest angles ($<0.125$~deg), the covariance becomes
much larger, about 1850~$\mu$as$^2$ (with a large statistical uncertainty), 
corresponding to an RMS amplitude of 43~$\mu$as. These features are clearly 
produced by small-scale patterns similar to what is seen in the LMC area (Fig.~\ref{fig:lmcPlxMap}).

Qualitatively similar correlations on both large and small angular scales are found
by analysing the proper motions of the quasars. We define
\begin{equation}\label{eq:covPm}
V_\mu(\theta) =\frac{1}{2} \langle \vec{\mu}_i'\vec{\mu}_j \rangle \, ,
\end{equation}
where $\vec{\mu}_i=\vec{p}_i\mu_{\alpha*i}+\vec{q}_i\mu_{\delta i}$ is the proper motion 
vector of source $i$, with unit vectors $\vec{p}_i$ and $\vec{q}_i$ towards increasing 
$\alpha$ and $\delta$, respectively (e.g.\ Eq.~3 in \citeads{2016A&A...595A...4L}). 
The prime denotes the scalar product. The vector formulation was chosen in order to combine 
the two components of proper motion in a frame-independent way. For small separations
$\vec{p}_i\simeq\vec{p}_j$ and $\vec{q}_i\simeq\vec{q}_j$, which gives
$V_\mu(\theta) \simeq (\mu_{\alpha*i}\mu_{\alpha*j}+\mu_{\delta i}\mu_{\delta j})/2$;
thus $V_\mu$ is the covariance averaged between the two components of the proper motion.
Figure~\ref{fig:qsoPmCov} shows $V_\mu(\theta)$ for the high-precision quasar sample. 
The dashed curve is the fitted exponential
\begin{equation}\label{eq:covPmExp}
V_\mu(\theta) \simeq (800~\mu\text{as}^2\text{yr}^{-2})\times\exp(-\theta/20^\circ) \, .
\end{equation}
The value at $\theta=0$ corresponds to an RMS amplitude of about $28~\mu$as~yr$^{-1}$ 
for the large-scale systematics. 
At small separations similar features are seen as for $V_\varpi(\theta)$, including the
1~deg oscillations; for $\theta<0.125$~deg the covariance is 
4400~$\mu\text{as}^2\text{yr}^{-2}$, corresponding to an RMS value of 
66~$\mu\text{as}~\text{yr}^{-1}$ per component of the proper motions. Again, this
is consistent with small-scale proper motion patterns seen, for example, in the LMC
\citep{DPACP-34}. 

The RMS values derived above and summarised in Table~\ref{tab:syst} for the
different angular scales can be interpreted as the noise floor when averaging 
the parallaxes or proper motions for a large number of sources in areas of the
corresponding sizes. The numbers should be seen as indicative and not
necessarily as representative for sources that are much brighter than the quasars. 

\begin{table*}
\caption{Summary of estimated systematics for faint sources ($G\gtrsim 16$~mag).\label{tab:syst}}
\small
\begin{tabular}{lccll}
\hline\hline
\noalign{\smallskip}
Angular scale & Parallax & Proper motion & 
Remark & Reference \\
 & [$\mu$as] &  [$\mu$as~yr$^{-1}$] \\
\noalign{\smallskip}
\hline
\noalign{\smallskip}
global & $-29$\phantom{$-$} &  10 & offset (zero point or spin) &
Fig.~\ref{fig:qsoPlxHist} and  \citet{DPACP-30}\\
$\sim\,$90~deg & $5$ & $13$ & RMS value &
Fig.~\ref{fig:qsoPlxVsGetc} and \citet{DPACP-30} \\
$\sim\,$14--20~deg & $17$ & $28$ & RMS value &
Eqs.~(\ref{eq:covPlxExp}) and (\ref{eq:covPmExp}) \\
$<\,$1~deg  & $43$ & $66$ & RMS value &
Figs.~\ref{fig:qsoPlxCov} and \ref{fig:qsoPmCov} \\
\noalign{\smallskip}
\hline
\end{tabular}
\tablefoot{Columns 2 and 3 give the estimated offset or RMS level of systematics
in parallax and proper motion (per component) based on the analysis described in the reference.
The RMS values should be interpreted as the noise floor when averaging many 
sources at the given angular scale. For bright sources (especially $G\lesssim 13$~mag),
the systematics may be significantly larger; see for example Fig.~\ref{fig:spin2}.}
\end{table*}

\begin{figure}
\center
  \resizebox{0.9\hsize}{!}{\includegraphics{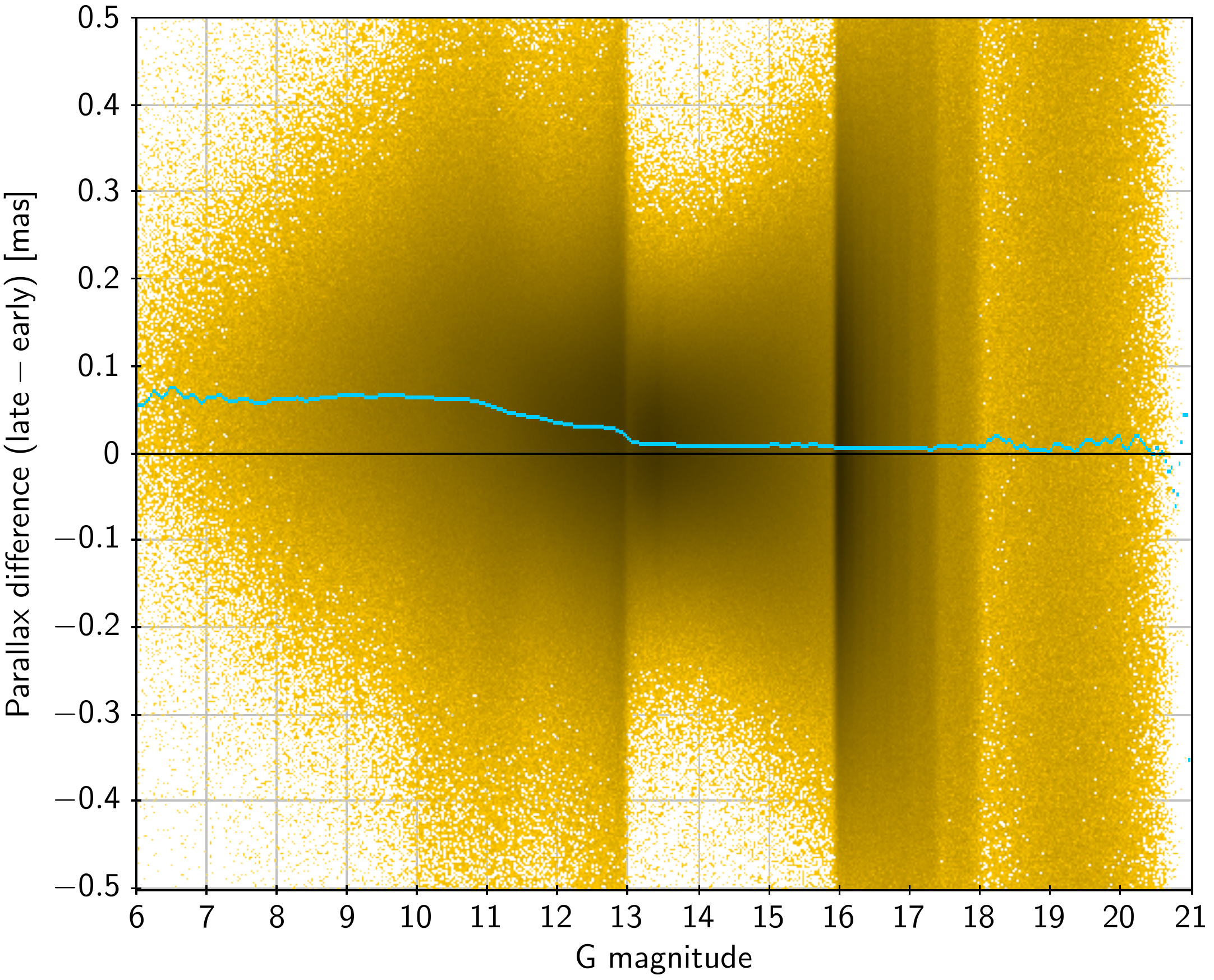}}\\
    \caption{Difference in parallax between the ``late'' and ``early'' solutions as a 
    function of magnitude. The cyan curve is the median. Only results for primary sources are 
    plotted; discontinuities in the density of points at $G=13$, 16, etc.\ are caused by the
    way the primary sources are selected.}
    \label{fig:plxDiffSplitFovVsG}
\end{figure}

\subsection{Split-field solutions}
\label{sec:split}

The internal consistency of the astrometric solution can be examined by comparing 
solutions based on complementary subsets of the observations. The observations
can for example be divided depending on the CCD strip in the astrometric field (AF).
Normally a source is observed in nine consecutive CCD strips, denoted AF1--AF9, 
as its image moves over the focal plane (see e.g. Fig.~3 in the AGIS paper), 
thus generating up to nine AL observations per field-of-view transit. The 
photon noise component is strictly independent between the nine observations,
while systematic errors depending on the calibration model may be partly similar. 

We have made two separate primary solutions using only the AF2--AF5 and
AF6--AF9 strips, respectively; these are called the ``early'' and ``late'' solutions. 
The same set of 
primary sources was used in both solutions as for the final primary solution 
(step~4 of Sect.~\ref{sec:agis22}), and the calibration and attitude models
were also the same. (Naturally, the calibration model only included the relevant 
CCDs, and the normalised AL field angle $\tilde{\eta}$ was similarly re-defined 
for the spin-related distortion model.) Although the early and late solutions
are partly affected by similar systematics from deficiencies in the calibration or
attitude models, the differences in the resulting astrometric parameters may give 
a realistic impression of the magnitude and general character of the systematics, 
and a very good check of the random errors. It should be noted that the differences
can never be interpreted as corrections to the published data: indeed we do not 
know which of the two solutions, if any, is better in terms of systematics.

\begin{figure*}
\centering
  \resizebox{0.33\hsize}{!}{\includegraphics{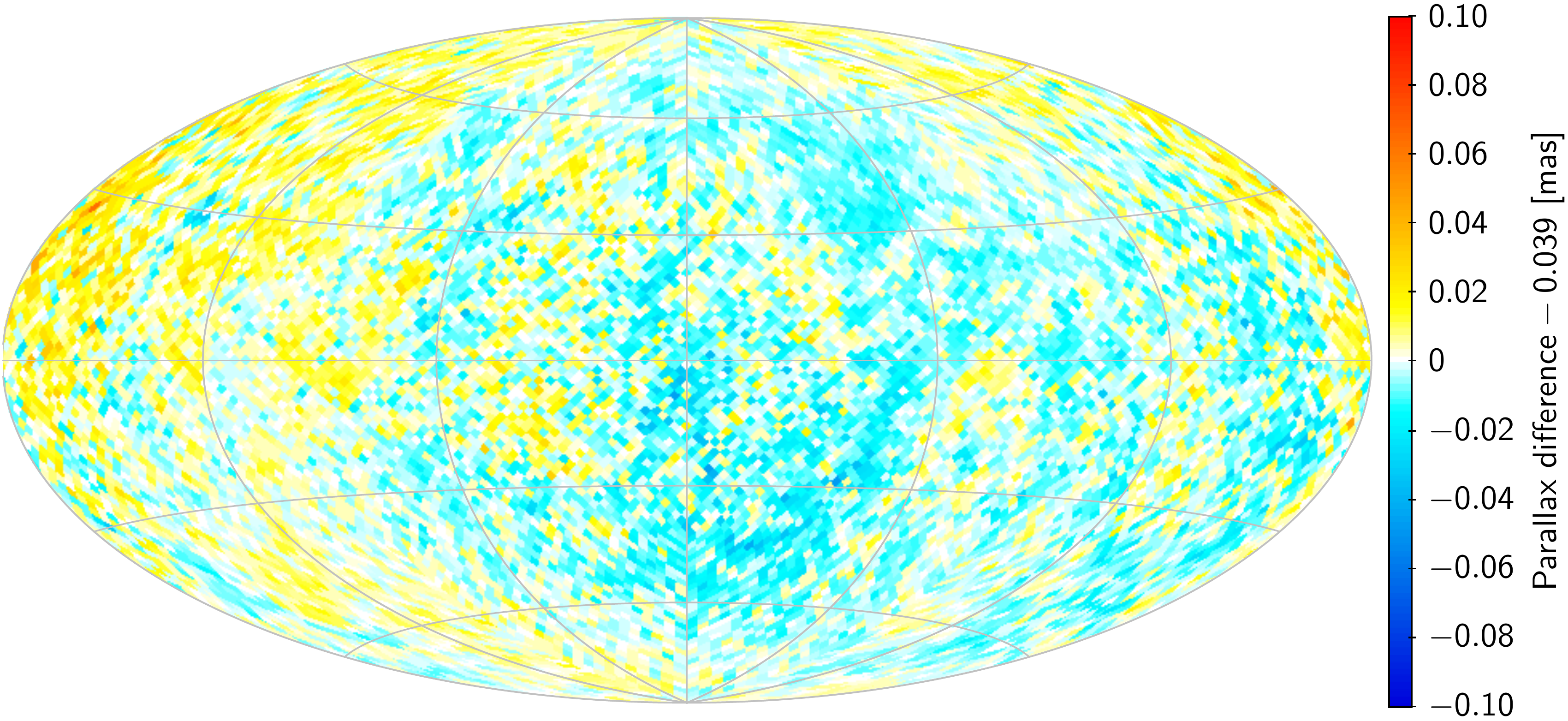}}
  \resizebox{0.33\hsize}{!}{\includegraphics{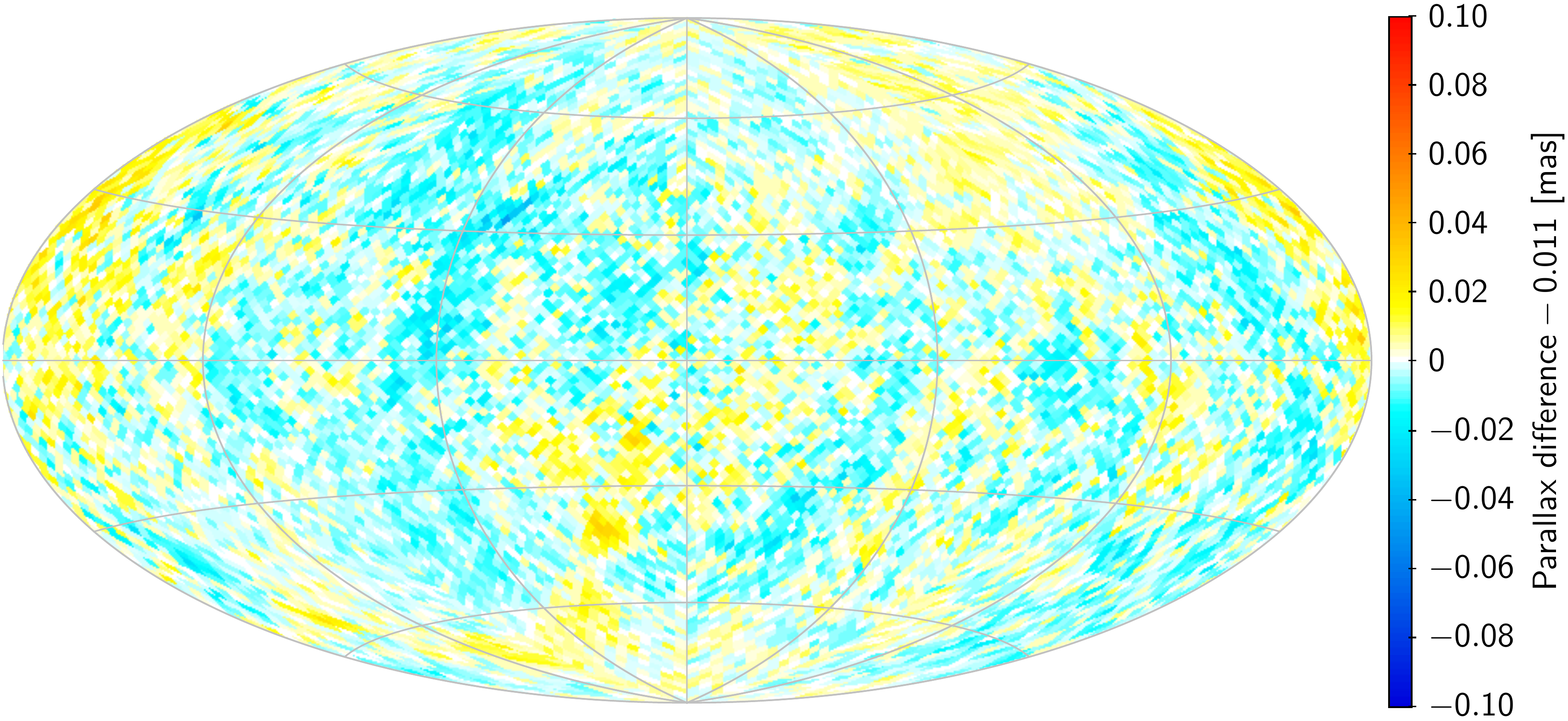}}
  \resizebox{0.33\hsize}{!}{\includegraphics{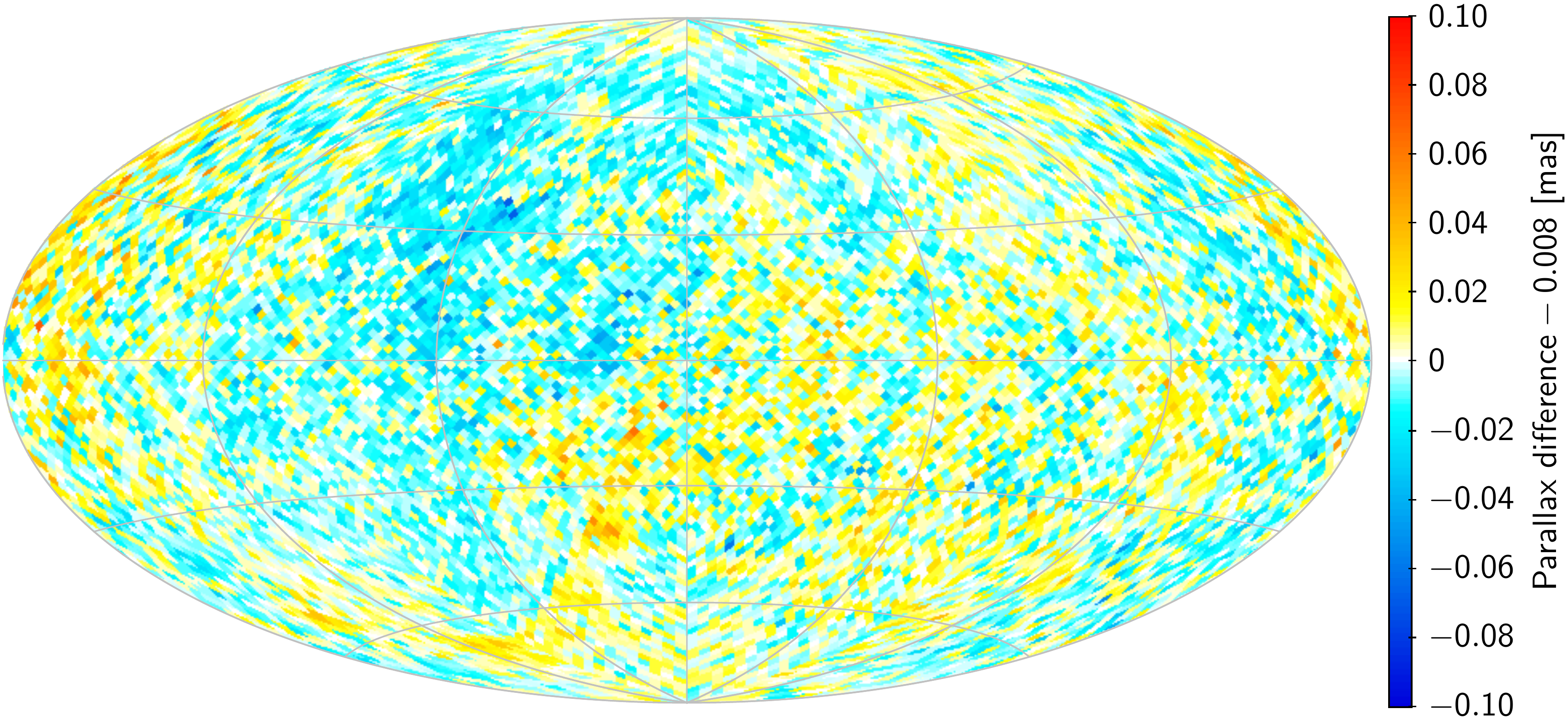}}
    \caption{Maps of the median difference in parallax between the ``late'' and ``early'' solutions,
    subdivided by magnitude. In each map, the global median was subtracted to remove the
    major part of the magnitude dependence seen in Fig.~\ref{fig:plxDiffSplitFovVsG}. 
    \textit{Left:} Magnitude range $G<13$~mag.
    \textit{Middle:} $13 < G < 16$~mag.
    \textit{Right:} $16 < G < 19$~mag.}
    \label{fig:mapPlxDiffSplitFov}
\end{figure*}

Figure~\ref{fig:plxDiffSplitFovVsG} shows the difference in parallax as a function
of magnitude. The scatter is broadly consistent with the combined formal uncertainties
of the two solutions. The median difference exhibits a strong dependence on magnitude, 
which is clearly related to the window class (steps at $G\simeq 13$ and 16) and the
use of gates for $G\lesssim 12$. To explore the spatial variations, maps of the median
parallax differences are shown in Fig.~\ref{fig:mapPlxDiffSplitFov} for three magnitude
ranges that roughly correspond to window classes 0, 1, and 2 (see 
footnote~\ref{footnote1}). In each map, the median parallax difference for sources
in the magnitude range was subtracted in order to eliminate the magnitude effect.
The spatial variations are similar in the middle and right maps, but distinctly different
in the left map (window class 0).
The RMS amplitude of the variations shown in these maps, that
is,\ of the median differences
at a pixel size of 3.36~deg$^2$, is 0.010, 0.008, and 0.013~mas, respectively.

The split-field solutions generally support the findings in Sects.~\ref{sec:frame} and
\ref{sec:corr}, viz.\ the presence of a magnitude-dependent systematic error, probably
mainly affecting the bright ($G\lesssim 13$) sources, and spatial variations of a few
tens of $\mu$as on a scale of several degrees.

\begin{figure}
\centering
  \resizebox{0.8\hsize}{!}{\includegraphics{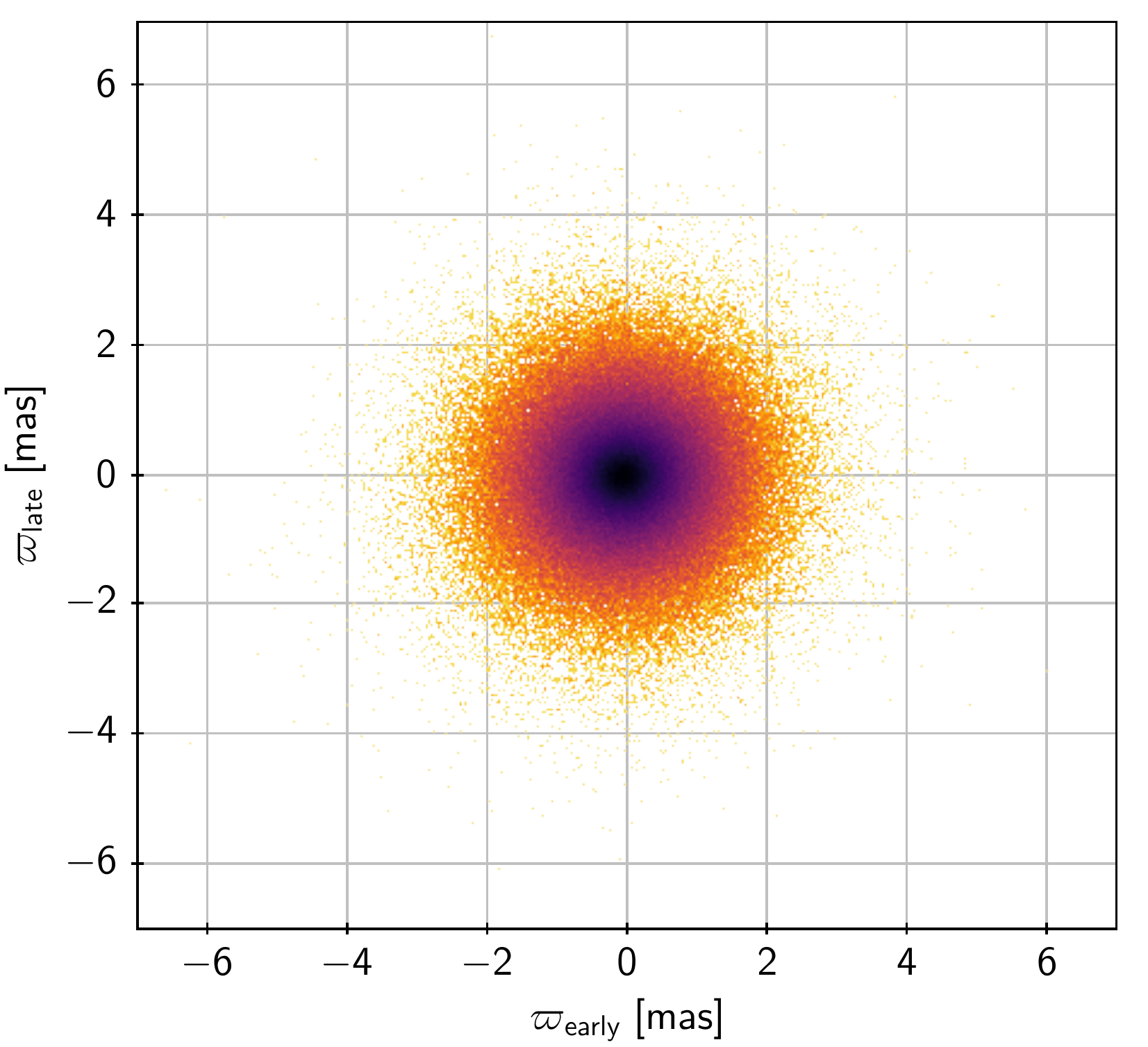}}
    \caption{Split-field parallax solutions for the quasar sample.}
    \label{fig:qsoPlxEarlyLateScatter}
\end{figure}

About 477 000 of the quasars from Sect.~\ref{sec:zeropoint} have accepted 
solutions in both the ``early'' and ``late'' solutions. The median parallax is $-0.034$~mas
for the early solution and $-0.022$~mas for the late solution. A scatter plot of the parallaxes 
(Fig.~\ref{fig:qsoPlxEarlyLateScatter}) shows that the parallax errors are practically
independent in the two solutions; the correlation coefficient is $+0.0245$.

\section{Conclusions} 
\label{sec:concl}

Compared with \textit{Gaia} DR1 \citep{2016A&A...595A...4L}, the second release 
contains a vastly increased number of sources with full astrometric data, including 
parallaxes and proper motions. For the bright ($G \lesssim 12$~mag) sources where 
such data were included already in the first release, the present results are generally 
more accurate and fully independent of the \textsc{Hipparcos} and \textit{Tycho} 
catalogues. The reference frame, \textit{Gaia}-CRF2, is entirely defined by \textit{Gaia} 
observations of quasars, including the optical counterparts of VLBI sources 
in a prototype version of the ICRF3.

In spite of these improvements, we recall that the astrometric
results in \textit{Gaia} DR2 are based on less than two years of observations and
very preliminary calibrations that have not yet benefited from the iterative improvement
of the pre-processing of the CCD measurements. As a consequence, random 
and systematic errors are both considerably higher than can be expected for the final 
mission products. 

In this release, all sources beyond the solar system are still treated as single stars, 
that is,\ as point objects whose motions can be described by the basic five-parameter 
model. For unresolved binaries (separation $\lesssim 100$~mas), the photocentre
is consistently observed and the astrometric parameters thus refer to the position
and motion of the photocentre in the wavelength band of the $G$ magnitude. 
Orbital motion and photometric variablility may bias the astrometric parameters 
for such sources. Resolved or partially resolved binaries cause a different
kind of errors, for example\ when the different observations of a source variously refer to 
one or the other of the components, or to the photocentre, depending on the direction 
of the scan. In some cases, this is known to produce spurious results, for instance, very large 
positive or negative parallaxes (\citeads{DPACP-39}; see also Appendix~\ref{sec:clean}). 
These limitations will be eliminated in future releases.

The random errors, as described by the formal uncertainties in the \textit{Gaia}
Archive, are summarised in Tables~\ref{tab:statP5} and \ref{tab:statP2}. 
An attempt to quantify the systematic errors, mainly based on the analysis of quasar 
data, is given in Table~\ref{tab:syst}. The main weaknesses identified through the 
internal validation process (Sect.~\ref{sec:valid}) are listed below. A more extensive
discussion is found in the paper by \citet{DPACP-39} on the catalogue 
validation.

\textit{Parallax zero point.} Although the measurement principle of \textit{Gaia}
should give absolute parallaxes, the results for quasars very clearly indicate a
global zero point of about $-0.03$~mas (i.e.\ 0.03~mas should be added to the
published values). There are, however, variations of a similar size depending on
magnitude, colour, and position (Figs.~\ref{fig:qsoPlxVsGetc} and \ref{fig:qsoPlxMap}).
On small scales, the zero-point variations may present quasi-periodic patterns as in 
Fig.~\ref{fig:lmcPlxMap}. A different zero point may apply to bright sources (see below).
Therefore, in any scientific usage of samples of \textit{Gaia} DR2 parallaxes for which 
the zero point is important (e.g.\ period-luminosity relations, or other luminosity calibrations), 
the zero point itself might be treated as an adjustable parameter. This will not 
always be possible, for example\ for very small samples, or when the distance is nearly constant
in the sample, as in a star cluster.

\textit{Formal errors.} The DOF bug resulted in significantly under-estimated
uncertainties for the bright ($G\lesssim 13$~mag) sources, which has been 
approximately corrected in the \textit{Gaia} Archive as described in 
Appendix~\ref{sec:correction}. Nevertheless, the quasar sample shows that the
uncertainties of the parallaxes and proper motions of the faint quasars away from the
Galactic plane are under-estimated by 8--12\%. For brighter sources, and closer to 
the Galactic plane, the uncertainties may be more severely under-estimated  
\citep{DPACP-39}. 

\textit{Bright sources.} The instrument calibration is still very provisional and 
particularly problematic
for the bright ($G\lesssim 13$~mag) sources. This manifests itself in larger 
uncertainties compared with the slightly fainter stars, and possibly in a 
systematic rotation of the proper
motion system of the bright sources relative to the quasars (Fig.~\ref{fig:spin2}).
The bright sources also behave distinctly different in the split-field solutions
(Sect.~\ref{sec:split}), suggesting that the parallax zero point could also be
different for $G\lesssim 13$.

\textit{Spurious large positive and negative parallaxes.} The release contains 
a small number of sources with very large positive or negative parallaxes,
for example,\ exceeding $\pm 1$~arcsec. 
These are most likely produced by cross-matching issues, where the different 
observations of the same nominal source were matched to different physical
sources. In such cases, the proper motion will in general also be corrupted.
This, and the related question how the data might be ``cleaned'', is 
discussed in Appendix~\ref{sec:clean}.
No filtering was made in the \textit{Gaia} Archive based on the sizes of the 
parallaxes and proper motions.

A summary of the astrometric properties of \textit{Gaia} DR2 is given in
Appendix~\ref{sec:properties}.

\begin{acknowledgements}
This work presents results from the European Space Agency (ESA) space mission \textit{Gaia}. 
\textit{Gaia} data are being processed by the \textit{Gaia} Data Processing and Analysis Consortium (DPAC). 
Funding for the DPAC is provided by national institutions, in particular the institutions participating in the 
\textit{Gaia} MultiLateral Agreement (MLA). 
The \textit{Gaia} mission website is \url{https://www.cosmos.esa.int/gaia}. 
The \textit{Gaia} Archive website is \url{https://archives.esac.esa.int/gaia}.
This work was financially supported by 
the European Space Agency (ESA) in the framework of the \textit{Gaia} project;
the Centre National d'Etudes Spatiales (CNES);
the National Science Foundation of China (NSFC) through grants 11703065 and 11573054;
German Aerospace Agency (Deutsches Zentrum f\"{u}r Luft- und Raumfahrt e.V., DLR) through grants 50QG0501, 50QG0601, 50QG0602, 50QG0701, 50QG0901, 50QG1001, 50QG1101, 50QG1401, 50QG1402, 50QG1403, and 50QG1404 and the Centre for Information Services and High Performance Computing (ZIH) at the Technische Universit\"{a}t (TU) Dresden through a generous allocation of computer time;
the Agenzia Spaziale Italiana (ASI) through contracts I/037/08/0, I/058/10/0, 2014-025-R.0, and 2014-025-R.1.2015 to the Italian Istituto Nazionale di Astrofisica (INAF), contract 2014-049-R.0/1/2 to INAF dedicated to the Space Science Data Centre (SSDC, formerly known as the ASI Sciece Data Centre, ASDC), and contracts I/008/10/0, 2013/030/I.0, 2013-030-I.0.1-2015, and 2016-17-I.0 to the Aerospace Logistics Technology Engineering Company (ALTEC S.p.A.), and INAF;
the Netherlands Organisation for Scientific Research (NWO) through grant NWO-M-614.061.414 and the Netherlands Research School for Astronomy (NOVA);
the Spanish Ministry of Economy (MINECO/FEDER, UE) through grants ESP2014-55996-C2-1-R, ESP2014-55996-C2-2-R, ESP2016-80079-C2-1-R, and ESP2016-80079-C2-2-R, the Spanish Ministerio de Econom\'{\i}a, Industria y Competitividad through grant AyA2014-55216, the Spanish Ministerio de Educaci\'{o}n, Cultura y Deporte (MECD) through grant FPU16/03827, the Institute of Cosmos Sciences University of Barcelona (ICCUB, Unidad de Excelencia 'Mar\'{\i}a de Maeztu') through grant MDM-2014-0369, the Xunta de Galicia and the Centros Singulares de Investigaci\'{o}n de Galicia for the period 2016-2019 through the Centro de Investigaci\'{o}n en Tecnolog\'{\i}as de la Informaci\'{o}n y las Comunicaciones (CITIC), the Red Espa\~{n}ola de Supercomputaci\'{o}n (RES) computer resources at MareNostrum, and the Barcelona Supercomputing Centre - Centro Nacional de Supercomputaci\'{o}n (BSC-CNS) through activities AECT-2016-1-0006, AECT-2016-2-0013, AECT-2016-3-0011, and AECT-2017-1-0020;
the Swedish National Space Board (SNSB/Rymdstyrelsen);
the Swiss State Secretariat for Education, Research, and Innovation through the ESA PRODEX programme, the Mesures d’Accompagnement, the Swiss Activit\'es Nationales Compl\'ementaires, and the Swiss National Science Foundation;
the United Kingdom Science and Technology Facilities Council (STFC) through grant ST/L006553/1, the United Kingdom Space Agency (UKSA) through grant ST/N000641/1 and ST/N001117/1, as well as a Particle Physics and Astronomy Research Council Grant PP/C503703/1.
The unpublished prototype version of ICRF3 was kindly provided by the IAU Working Group on ICRF3.
Diagrams were produced using the astronomy-oriented data handling and visualisation software TOPCAT
\citepads{2005ASPC..347...29T}.
This research has made use of the SIMBAD database, operated at CDS, Strasbourg, France.
We thank A.G.A.~Brown and C.~Jordi for valuable feedback during the preparation of this paper, and 
the referee, V.V.~Makarov, for constructive comments on the original version of the manuscript.
\end{acknowledgements}

\bibliographystyle{aa} 
\bibliography{refs} 

\begin{thebibliography}{29}
\expandafter\ifx\csname natexlab\endcsname\relax\def\natexlab#1{#1}\fi

\bibitem[{{Arenou et al.}(2018)}]{DPACP-39}
{Arenou et al.} 2018, Gaia Data Release 2: Catalogue validation, \aap, this issue

\bibitem[{{Butkevich} {et~al.}(2017){Butkevich}, {Klioner}, {Lindegren},
  {Hobbs}, \& {van Leeuwen}}]{2017A&A...603A..45B}
{Butkevich}, A.~G., {Klioner}, S.~A., {Lindegren}, L., {Hobbs}, D., \& {van
  Leeuwen}, F. 2017, \aap, 603, A45

\bibitem[{{Casta{\~n}eda et al.}(in prep.)}]{DPACP-45}
{Casta{\~n}eda et al.} Cross match of Gaia observations, in prep.

\bibitem[{{Evans et al.}(2018)}]{DPACP-40}
{Evans et al.} 2018, Gaia Data Release 2: Photometric content and validation, \aap, this issue

\bibitem[{{Fabricius} {et~al.}(2016){Fabricius}, {Bastian}, {Portell},
  {Casta{\~n}eda}, {Davidson}, {Hambly}, {Clotet}, {Biermann}, {Mora},
  {Busonero}, {Riva}, {Brown}, {Smart}, {Lammers}, {Torra}, {Drimmel},
  {Gracia}, {L{\"o}ffler}, {Spagna}, {Lindegren}, {Klioner}, {Andrei}, {Bach},
  {Bramante}, {Br{\"u}semeister}, {Busso}, {Carrasco}, {Gai}, {Garralda},
  {Gonz{\'a}lez-Vidal}, {Guerra}, {Hauser}, {Jordan}, {Jordi}, {Lenhardt},
  {Mignard}, {Messineo}, {Mulone}, {Serraller}, {Stampa}, {Tanga}, {van
  Elteren}, {van Reeven}, {Voss}, {Abbas}, {Allasia}, {Altmann}, {Anton},
  {Barache}, {Becciani}, {Berthier}, {Bianchi}, {Bombrun}, {Bouquillon},
  {Bourda}, {Bucciarelli}, {Butkevich}, {Buzzi}, {Cancelliere}, {Carlucci},
  {Charlot}, {Collins}, {Comoretto}, {Cross}, {Crosta}, {de Felice}, {Fienga},
  {Figueras}, {Fraile}, {Geyer}, {Hernandez}, {Hobbs}, {Hofmann}, {Liao},
  {Licata}, {Martino}, {McMillan}, {Michalik}, {Morbidelli}, {Parsons},
  {Pecoraro}, {Ramos-Lerate}, {Sarasso}, {Siddiqui}, {Steele},
  {Steidelm{\"u}ller}, {Taris}, {Vecchiato}, {Abreu}, {Anglada}, {Boudreault},
  {Cropper}, {Holl}, {Cheek}, {Crowley}, {Fleitas}, {Hutton}, {Osinde},
  {Rowell}, {Salguero}, {Utrilla}, {Blagorodnova}, {Soffel}, {Osorio},
  {Vicente}, {Cambras}, \& {Bernstein}}]{2016A&A...595A...3F}
{Fabricius}, C., {Bastian}, U., {Portell}, J., {et~al.} 2016, \aap, 595, A3

\bibitem[{{Freedman} {et~al.}(2001){Freedman}, {Madore}, {Gibson}, {Ferrarese},
  {Kelson}, {Sakai}, {Mould}, {Kennicutt}, {Ford}, {Graham}, {Huchra},
  {Hughes}, {Illingworth}, {Macri}, \& {Stetson}}]{2001ApJ...553...47F}
{Freedman}, W.~L., {Madore}, B.~F., {Gibson}, B.~K., {et~al.} 2001, \apj, 553,
  47

\bibitem[{{Gaia Collaboration} {et~al.}(2016{\natexlab{a}}){Gaia
  Collaboration}, {Brown}, {Vallenari}, {Prusti}, {de Bruijne}, {Mignard},
  {Drimmel}, {Babusiaux}, {Bailer-Jones}, {Bastian}, \&
  et~al.}]{2016A&A...595A...2G}
{Gaia Collaboration}, {Brown}, A.~G.~A., {Vallenari}, A., {et~al.}
  2016{\natexlab{a}}, \aap, 595, A2

\bibitem[{{Gaia Collaboration} {et~al.}(2016{\natexlab{b}}){Gaia
  Collaboration}, {Prusti}, {de Bruijne}, {Brown}, {Vallenari}, {Babusiaux},
  {Bailer-Jones}, {Bastian}, {Biermann}, {Evans}, \&
  et~al.}]{2016A&A...595A...1G}
{Gaia Collaboration}, {Prusti}, T., {de Bruijne}, J.~H.~J., {et~al.}
  2016{\natexlab{b}}, \aap, 595, A1

\bibitem[{{Gaia Collaboration et al.}(2018{\natexlab{a}})}]{DPACP-36}
{Gaia Collaboration et al.} 2018{\natexlab{a}}, Gaia Data Release 2: Summary of the
  contents and survey properties, \aap, this issue

\bibitem[{{Gaia Collaboration et al.}(2018{\natexlab{b}})}]{DPACP-30}
{Gaia Collaboration et al.} 2018{\natexlab{b}}, Gaia Data Release 2:
  The celestial reference frame (Gaia-CRF2), \aap, this issue

\bibitem[{{Gaia Collaboration et al.}(2018{\natexlab{c}})}]{DPACP-34}
{Gaia Collaboration et al.} 2018{\natexlab{c}}, Gaia Data Release 2: The kinematics 
  of globular clusters and dwarf galaxies around the Milky Way, \aap, this issue

\bibitem[{{Gaia Collaboration et al.}(2018{\natexlab{d}})}]{DPACP-31}
{Gaia Collaboration et al.} 2018{\natexlab{d}}, Gaia Data Release 2: Observational 
  Hertzsprung-Russell diagrams, \aap, this issue

\bibitem[{{Hambly et al.}(2018)}]{DPACP-29}
{Hambly et al.} 2018, Gaia Data Release 2: Calibration and mitigation of electronic 
offset effects in the data, \aap, this issue

\bibitem[{{Klioner}(2003)}]{2003AJ....125.1580K}
{Klioner}, S.~A. 2003, \aj, 125, 1580

\bibitem[{{Klioner}(2004)}]{2004PhRvD..69l4001K}
{Klioner}, S.~A. 2004, \prd, 69, 124001

\bibitem[{{Lindegren} {et~al.}(2016){Lindegren}, {Lammers}, {Bastian},
  {Hern{\'a}ndez}, {Klioner}, {Hobbs}, {Bombrun}, {Michalik}, {Ramos-Lerate},
  {Butkevich}, {Comoretto}, {Joliet}, {Holl}, {Hutton}, {Parsons},
  {Steidelm{\"u}ller}, {Abbas}, {Altmann}, {Andrei}, {Anton}, {Bach},
  {Barache}, {Becciani}, {Berthier}, {Bianchi}, {Biermann}, {Bouquillon},
  {Bourda}, {Br{\"u}semeister}, {Bucciarelli}, {Busonero}, {Carlucci},
  {Casta{\~n}eda}, {Charlot}, {Clotet}, {Crosta}, {Davidson}, {de Felice},
  {Drimmel}, {Fabricius}, {Fienga}, {Figueras}, {Fraile}, {Gai}, {Garralda},
  {Geyer}, {Gonz{\'a}lez-Vidal}, {Guerra}, {Hambly}, {Hauser}, {Jordan},
  {Lattanzi}, {Lenhardt}, {Liao}, {L{\"o}ffler}, {McMillan}, {Mignard}, {Mora},
  {Morbidelli}, {Portell}, {Riva}, {Sarasso}, {Serraller}, {Siddiqui}, {Smart},
  {Spagna}, {Stampa}, {Steele}, {Taris}, {Torra}, {van Reeven}, {Vecchiato},
  {Zschocke}, {de Bruijne}, {Gracia}, {Raison}, {Lister}, {Marchant},
  {Messineo}, {Soffel}, {Osorio}, {de Torres}, \&
  {O'Mullane}}]{2016A&A...595A...4L}
{Lindegren}, L., {Lammers}, U., {Bastian}, U., {et~al.} 2016, \aap, 595, A4

\bibitem[{{Lindegren} {et~al.}(2012){Lindegren}, {Lammers}, {Hobbs},
  {O'Mullane}, {Bastian}, \& {Hern{\'a}ndez}}]{2012A&A...538A..78L}
{Lindegren}, L., {Lammers}, U., {Hobbs}, D., {et~al.} 2012, \aap, 538, A78 (the
  AGIS paper)

\bibitem[{{Michalik} {et~al.}(2015){Michalik}, {Lindegren}, {Hobbs}, \&
  {Butkevich}}]{2015A&A...583A..68M}
{Michalik}, D., {Lindegren}, L., {Hobbs}, D., \& {Butkevich}, A.~G. 2015, \aap,
  583, A68

\bibitem[{{Mora} {et~al.}(2016){Mora}, {Biermann}, {Bombrun}, {Boyadjian},
  {Chassat}, {Corberand}, {Davidson}, {Doyle}, {Escolar}, {Gielesen},
  {Guilpain}, {Hernandez}, {Kirschner}, {Klioner}, {Koeck}, {Laine},
  {Lindegren}, {Serpell}, {Tatry}, \& {Thoral}}]{2016SPIE.9904E..2DM}
{Mora}, A., {Biermann}, M., {Bombrun}, A., {et~al.} 2016, in \procspie, Vol.
  9904, Space Telescopes and Instrumentation 2016: Optical, Infrared, and
  Millimeter Wave, 99042D

\bibitem[{{Riello et al.}(2018)}]{DPACP-44}
{Riello et al.} 2018, Gaia Data Release 2: Processing of the photometric data, \aap, this issue

\bibitem[{{Rodgers} \& {Eggen}(1974)}]{1974PASP...86..742R}
{Rodgers}, A.~W. \& {Eggen}, O.~J. 1974, \pasp, 86, 742

\bibitem[{{Sartoretti et al.}(2018)}]{DPACP-47}
{Sartoretti et al.} 2018, Gaia Data Release 2: Processing the spectroscopic data, \aap, this issue

\bibitem[{{Secrest} {et~al.}(2015){Secrest}, {Dudik}, {Dorland}, {Zacharias},
  {Makarov}, {Fey}, {Frouard}, \& {Finch}}]{2015ApJS..221...12S}
{Secrest}, N.~J., {Dudik}, R.~P., {Dorland}, B.~N., {et~al.} 2015, \apjs, 221,
  12

\bibitem[{{Secrest} {et~al.}(2016){Secrest}, {Dudik}, {Dorland}, {Zacharias},
  {Makarov}, {Fey}, {Frouard}, \& {Finch}}]{2016yCat..22210012S}
{Secrest}, N.~J., {Dudik}, R.~P., {Dorland}, B.~N., {et~al.} 2016, VizieR
  Online Data Catalog, 222

\bibitem[{{Soffel} {et~al.}(2003){Soffel}, {Klioner}, {Petit}, {Wolf},
  {Kopeikin}, {Bretagnon}, {Brumberg}, {Capitaine}, {Damour}, {Fukushima},
  {Guinot}, {Huang}, {Lindegren}, {Ma}, {Nordtvedt}, {Ries}, {Seidelmann},
  {Vokrouhlick{\'y}}, {Will}, \& {Xu}}]{2003AJ....126.2687S}
{Soffel}, M., {Klioner}, S.~A., {Petit}, G., {et~al.} 2003, \aj, 126, 2687

\bibitem[{{Taylor}(2005)}]{2005ASPC..347...29T}
{Taylor}, M.~B. 2005, in Astronomical Society of the Pacific Conference Series,
  Vol. 347, Astronomical Data Analysis Software and Systems XIV, ed.
  P.~{Shopbell}, M.~{Britton}, \& R.~{Ebert}, 29

\bibitem[{{van Leeuwen}(2007)}]{2007ASSL..350.....V}
{van Leeuwen}, F., ed. 2007, Astrophysics and Space Science Library, Vol. 350,
  {Hipparcos, the New Reduction of the Raw Data}

\bibitem[{{van Leeuwen} {et~al.}(2017){van Leeuwen}, {Evans}, {De Angeli},
  {Jordi}, {Busso}, {Cacciari}, {Riello}, {Pancino}, {Altavilla}, {Brown},
  {Burgess}, {Carrasco}, {Cocozza}, {Cowell}, {Davidson}, {De Luise},
  {Fabricius}, {Galleti}, {Gilmore}, {Giuffrida}, {Hambly}, {Harrison},
  {Hodgkin}, {Holland}, {MacDonald}, {Marinoni}, {Montegriffo}, {Osborne},
  {Ragaini}, {Richards}, {Rowell}, {Voss}, {Walton}, {Weiler}, {Castellani},
  {Delgado}, {H{\o}g}, {van Leeuwen}, {Millar}, {Pagani}, {Piersimoni},
  {Pulone}, {Rixon}, {Suess}, {Wyrzykowski}, {Yoldas}, {Alecu}, {Allan},
  {Balaguer-N{\'u}{\~n}ez}, {Barstow}, {Bellazzini}, {Belokurov},
  {Blagorodnova}, {Bonfigli}, {Bragaglia}, {Brown}, {Bunclark}, {Buonanno},
  {Burgon}, {Campbell}, {Collins}, {Cross}, {Ducourant}, {van Elteren},
  {Evans}, {Federici}, {Fern{\'a}ndez-Hern{\'a}ndez}, {Figueras}, {Fraser},
  {Fyfe}, {Gebran}, {Heyrovsky}, {Holl}, {Holland}, {Iannicola}, {Irwin},
  {Koposov}, {Krone-Martins}, {Mann}, {Marrese}, {Masana}, {Munari}, {Ortiz},
  {Ouzounis}, {Peltzer}, {Portell}, {Read}, {Terrett}, {Torra}, {Trager},
  {Troisi}, {Valentini}, {Vallenari}, \& {Wevers}}]{2017A&A...599A..32V}
{van Leeuwen}, F., {Evans}, D.~W., {De Angeli}, F., {et~al.} 2017, \aap, 599,
  A32

\bibitem[{{Wenger} {et~al.}(2000){Wenger}, {Ochsenbein}, {Egret}, {Dubois},
  {Bonnarel}, {Borde}, {Genova}, {Jasniewicz}, {Lalo{\"e}}, {Lesteven}, \&
  {Monier}}]{2000A&AS..143....9W}
{Wenger}, M., {Ochsenbein}, F., {Egret}, D., {et~al.} 2000, \aaps, 143, 9

\end{thebibliography}

\appendix

\section{DOF bug and how it was corrected}
\label{sec:correction}

\subsection{Background}

A necessary but not sufficient condition for correctly estimated formal uncertainties
in a least-squares solution is that the residuals have expected sizes in relation to
the assumed uncertainties of the observations. Given the considerable gap, illustrated in
Fig.~\ref{fig:sigmaAlVsG}, between the formal uncertainties of the observations derived 
from the image parameter determination and the actual scatter of residuals, it is clear 
that some re-weighting of the observations is necessary in order to achieve the required 
consistency. As explained in Sect.~3.6 of the AGIS paper, the re-weighting is done by 
quadratically adding the excess
noise $\epsilon$ to the formal uncertainty of the observation $\sigma_\eta$.
The excess noise has two components: the excess source noise $\epsilon_i$, which
is the same for all observations of a given source $i$, and the excess attitude noise
$\epsilon_a(t)$, which is a function of time but the same for all sources at a given 
time. Briefly, $\epsilon_i$ and $\epsilon_a(t)$ are globally adjusted to make the
weighted sum of squared residuals 
\begin{equation}\label{eq:Q}
Q=\sum_l \frac{R_l^2}{\sigma_{\eta,l}^2+\epsilon_i^2+\epsilon_a(t_l)^2}\le\nu
\end{equation}
for all the sources. Here $R_l$ is the AL residual of observation $l$, the sum
is taken over all the accepted observations of the source, and $\nu$ is the number of
degrees of freedom, that is,\ the number of accepted AL observations minus 5.
(In this and the following equation, we disregard the outlier treatment for simplicity.)
The non-negative quantity $\epsilon_i$, given in the \textit{Gaia} Archive as
\gacs{astrometric\_excess\_noise}, is a useful characteristic of the source, since
it should only be zero if all the observations fit the single-star model well enough,
given the level of excess attitude noise set by the majority of other sources.  

An alternative measure of how well the single-star model fits a given source is the quantity 
\gacs{astrometric\_chi2\_al}, also given in the \textit{Gaia} Archive. It is calculated as  
\begin{equation}\label{eq:chi2}
\chi^2=\sum_l \frac{R_l^2}{\sigma_{\eta,l}^2+\epsilon_a(t_l)^2} \, .
\end{equation}
We note that the excess source noise is not included in the denominator, otherwise
we would always have $\chi^2\le\nu$. The a~posteriori mean error of unit weight 
$u=(\chi^2/\nu)^{1/2}$  , also known as the unit weight error, is a more useful
goodness-of-fit statistic, since it is expected to be around unity in well-behaved 
cases.

\begin{figure}
\centering
  \resizebox{0.9\hsize}{!}{\includegraphics{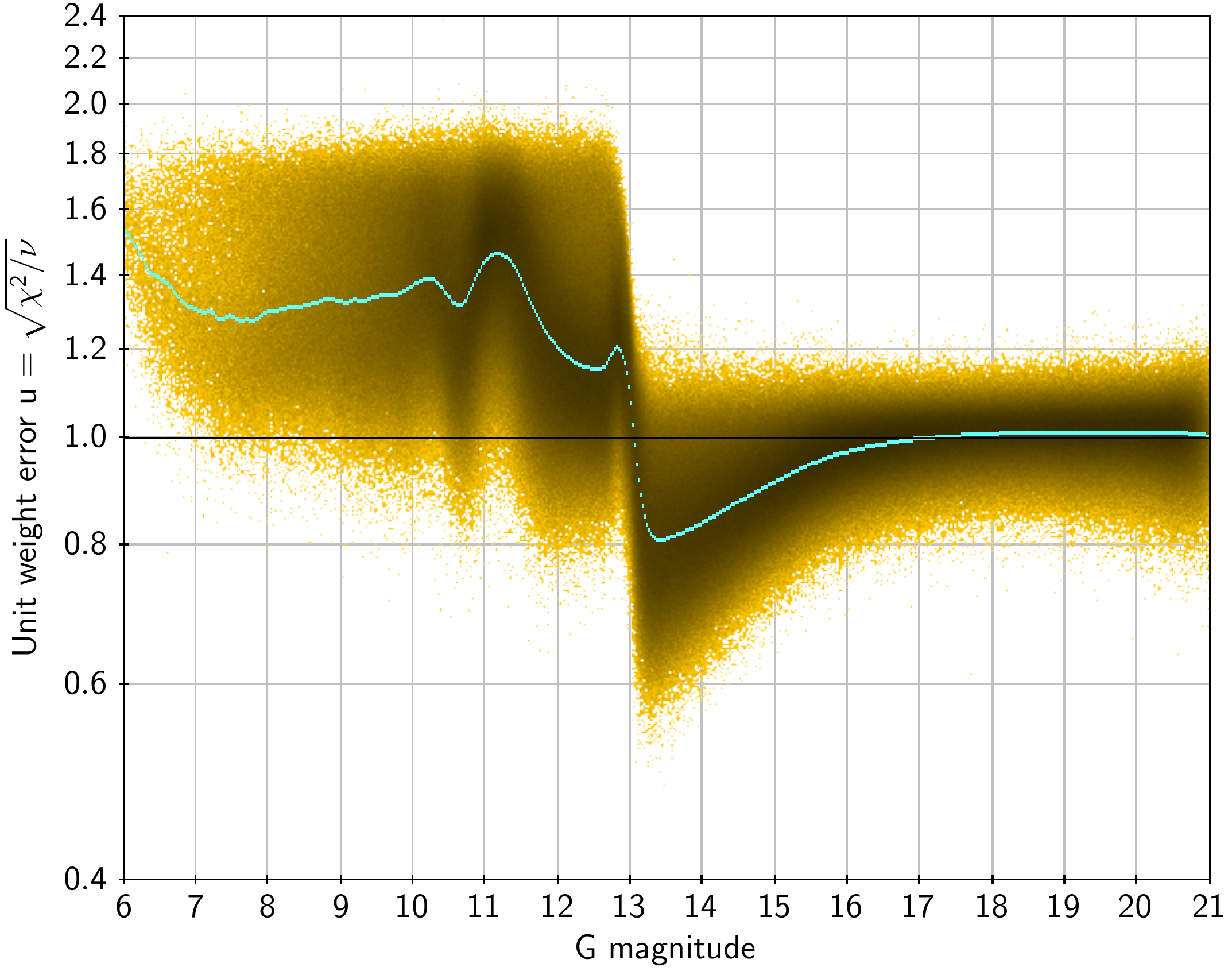}}
    \caption{Residual statistics for sources with five-parameters solutions
    and excess source noise equal to zero. 
    The yellow dots show individual values of the unit weight error $u=(\chi^2/\nu)^{1/2}$ for
    the sources brighter than $G=11$, and for a gradually decreasing random
    fraction of the fainter sources. The cyan curve is the running median.}
    \label{fig:uweVsG}
\end{figure}

So far, we have described how the weighting scheme was intended to work. Now we consider some actual
statistics in \textit{Gaia} DR2. In Fig.~\ref{fig:uweVsG} we have plotted the unit weight error
$u=\sqrt{\gacs{astrometric\_chi2\_al}/(\gacs{astrometric\_n\_good\_obs\_al}-5)}$
as a function of magnitude for a random subset of sources with zero excess source noise. 
According to what was said above, we expect this $u$ to be on average around 1.0. 
As shown by the cyan curve, which is a running median, this is actually the case
only for sources fainter than $G\simeq 17$. For the brighter sources, there are strong
deviations: at $G\lesssim 13$, corresponding to window class 0, the median $u$ is
in the range 1.2--1.4, while at intermediate magnitudes it is $<\,$1, with a minimum of 
0.8 at $G\simeq 13.4$.

This unexpected behaviour of $u$ for sources brighter than $G\simeq 17$ was traced 
to a bug in the source update algorithm. This bug, which we refer to as the ``DOF bug'',
directly affected sources with observations in window class 0, and indirectly other 
sources as well, as explained below.
Observations in window class 0 are special in that they provide measurements
in both the along-scan (AL) and across-scan (AC) directions, while window class 
1 and 2 only give AL measurements. However, the estimation of the excess source noise $\epsilon_i$ 
by means of Eq.~(\ref{eq:Q}) should only use the more precise AL observations. 
The source update algorithm correctly neglected the AC observations
when computing the sum $Q$, but erroneously included them in the degrees of freedom (DOF) $\nu$. 
As a result, the excess source noise was seriously under-estimated for sources with
$G\lesssim 13$, and in fact set to zero for about 80\% of them. Since these
sources usually have an equal number of AC and AL observations, this explains
why $u$ is roughly a factor 1.4 too large for these sources. It also means 
that the formal uncertainties are under-estimated in this magnitude range.

The too low values of $u$ for the somewhat fainter sources ($G\simeq 13$--17)
are an indirect effect of the same bug. In the primary solution, the excess attitude noise 
was estimated as a function of time by analysis of the residuals after taking into 
account the excess source noise. As is evident from Eq.~(\ref{eq:Q}), the 
under-estimated excess source noise for many sources then had to be compensated for
by a higher excess attitude noise. As a result, the excess attitude noise was
generally over-estimated, leading to under-estimated $\chi^2$ in Eq.~(\ref{eq:chi2})
at intermediate magnitudes ($G\simeq 13$--17). Even fainter sources are not significantly 
affected by the over-estimated excess attitude noise, as their error budget is in any 
case dominated by the photon noise.  

\subsection{How the bug was corrected}

In order to evaluate the impact of the DOF bug on the astrometric results, a new primary 
solution was computed after having corrected the software for the DOF bug. The
astrometric parameters, their formal uncertainties, and other statistics were
compared with the corresponding data from the original (uncorrected) solution. It was found that
the astrometric parameters themselves are only very marginally affected by the bug.
This was expected, as the observations had not changed, only their relative weights.
The most serious impact is on the formal uncertainties, which are under-estimated for
bright sources and slightly over-estimated at intermediate magnitudes. Very nearly
the same correction factor applies to the uncertainties of all five astrometric parameters 
of a given source.

The DOF bug was discovered very late in the data processing cycle and at a
stage when it was judged too risky to re-compute, re-validate, and replace the complete 
astrometric solution in \textit{Gaia} DR2. Instead it was decided not to touch the
astrometric parameters themselves, but apply a statistical correction to their
formal uncertainties. For each source with a five-parameter solution, a correction factor 
$F$ was computed as described below and applied to the formal uncertainties 
$\varsigma_{\alpha*}$, etc.\ of the original solution. The \textit{Gaia} Archive then 
contains the corrected uncertainties
\begin{equation}\label{eq:FF}
\sigma_{\alpha*} = F\varsigma_{\alpha*} \, , \quad
\sigma_{\delta} = F\varsigma_{\delta} \, , \quad \text{(etc.)}
\end{equation}
The correction factor was computed as
\begin{equation}\label{eq:X1}
F = \frac{1+0.8R}{\sqrt{1+(\text{0.025~mas}/\varsigma_\varpi)^2}} \, ,
\end{equation}
where
\begin{equation}\label{eq:X2}
R = \frac{\gacs{astrometric\_n\_obs\_ac}}{\gacs{astrometric\_n\_obs\_al}}
\end{equation}
is the ratio of the number of AC to AL observations.
Since $0\le R \le 1$ and $\varsigma_\varpi\ge 0.015$~mas in \textit{Gaia} DR2, 
the factor $F$ is constrained to the range 0.5 to 1.8. 

\begin{figure}
\center
  \resizebox{0.9\hsize}{!}{\includegraphics{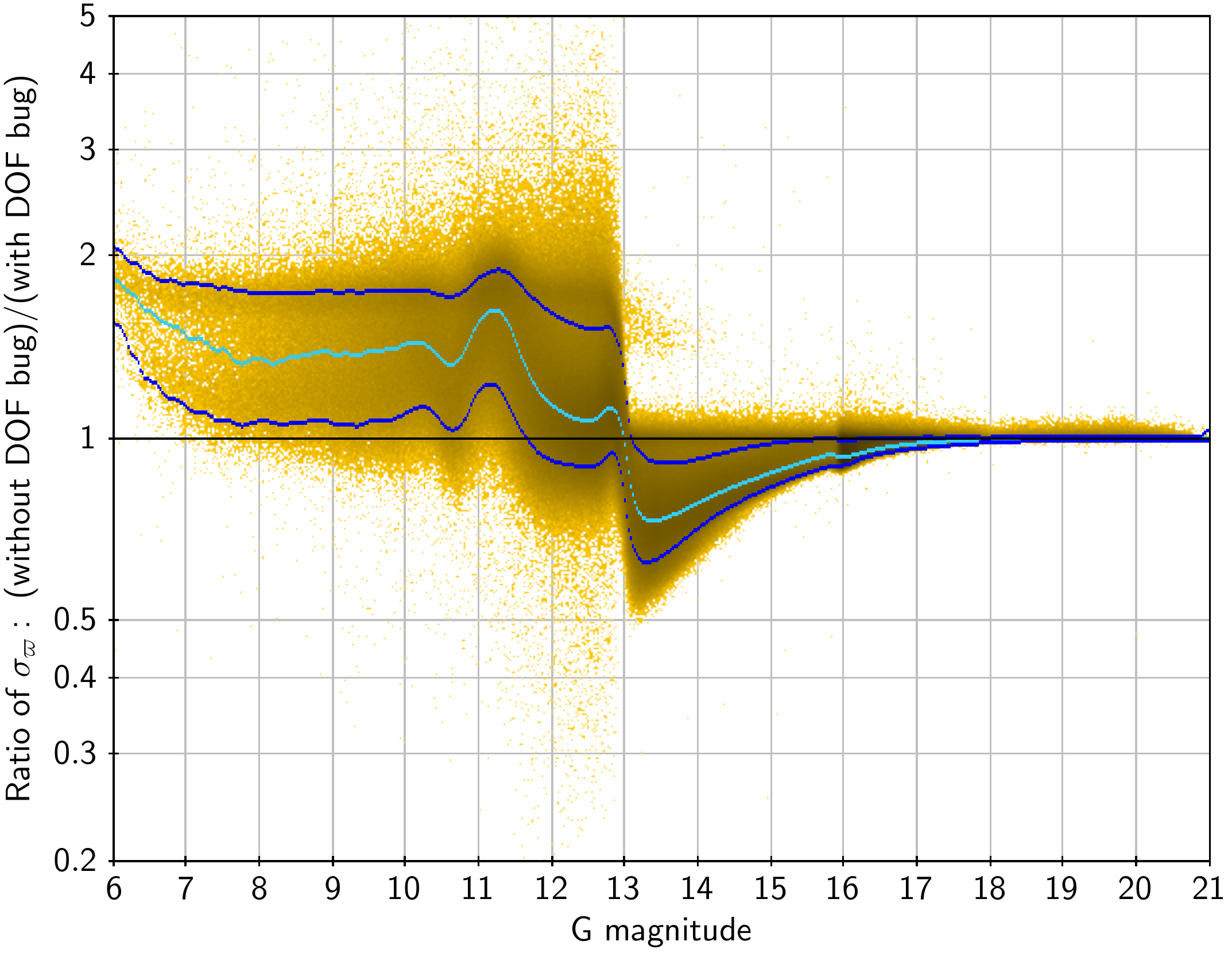}}\\
  \resizebox{0.9\hsize}{!}{\includegraphics[width=59mm]{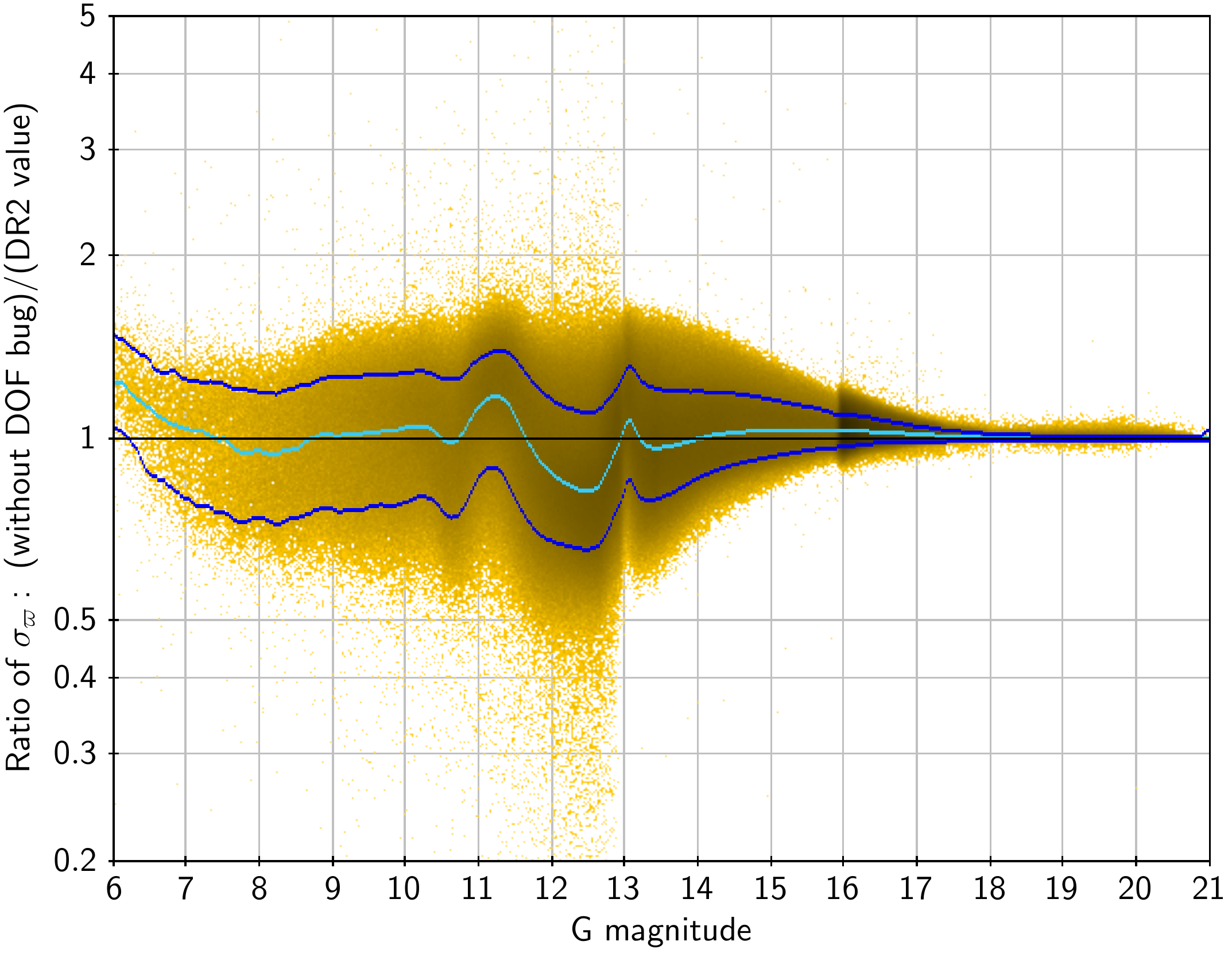}}
    \caption{Ratio of parallax uncertainties before (top) and after (bottom)
    applying the statistical correction factor $F$ from Eq.~(\ref{eq:X1}). The yellow
    dots are for the individual primary sources, the cyan curve is the median, and
    the blue curves are the 10th and 90th percentiles.}
    \label{fig:plxErrorRatio}
\end{figure}

Equation~(\ref{eq:X1}) was derived by comparing the parallax uncertainties in the
primary solution with the DOF bug fixed with the corresponding values in the
original solution. The top diagram in Fig.~\ref{fig:plxErrorRatio} shows the ratio 
$(\sigma_\varpi)_\text{no~bug}/\varsigma_\varpi$
as a function of magnitude before the correction; the bottom diagram shows the
same ratio after the correction by $F$, that is,\ 
$(\sigma_\varpi)_\text{no~bug}/\sigma_\varpi$, where $\sigma_\varpi=F\varsigma_\varpi$. 
The constants 0.8 and 0.025~mas in Eq.~(\ref{eq:X1}) were adjusted to make the median
curve in the bottom diagram as close to unity as possible. The diagrams illustrate the statistical 
nature of the correction: although the median ratio is roughly correct after
correction, the uncertainties could still be significantly wrong for some sources.

Should there ever be a need to undo the correction, it is possible to compute $F$ in terms of
the published parallax uncertainty $\sigma_\varpi$ as
\begin{equation}\label{eq:X3}
F = (1+0.8R)\sqrt{\frac{2}{1+\sqrt{1+4(1+0.8R)^2\left(\frac{0.025~\text{mas}}{\sigma_\varpi}\right)^2}}} \, ,
\end{equation}
from which $\varsigma_{\alpha*} = \sigma_{\alpha*}/F$, etc.

\subsection{Secondary effects on other statistics}

The DOF bug also affected the excess source noise and its significance, 
but there was no simple way to correct this, and the uncorrected values are
therefore left in the \textit{Gaia} Archive. Typically the excess source noise 
may be under-estimated by 0.15--0.3~mas for $G\lesssim 13$, 
and not at all or by less than 0.15~mas for fainter sources. The astrometric 
$\chi^2$ is affected by the over-estimated excess attitude noise, and is 
therefore generally under-estimated at all magnitudes; again no correction was
made for this quantity. To single out ``bad'' solutions using any of these 
statistics can in any case only be done in an ad~hoc fashion by considering 
the overall distributions of the quantities at the relevant magnitudes. An
example is given in Appendix~\ref{sec:clean}.

\section{Astrometric properties of \textit{Gaia} DR2} 
\label{sec:properties}

\begin{figure}
\center
  \resizebox{0.9\hsize}{!}{\includegraphics{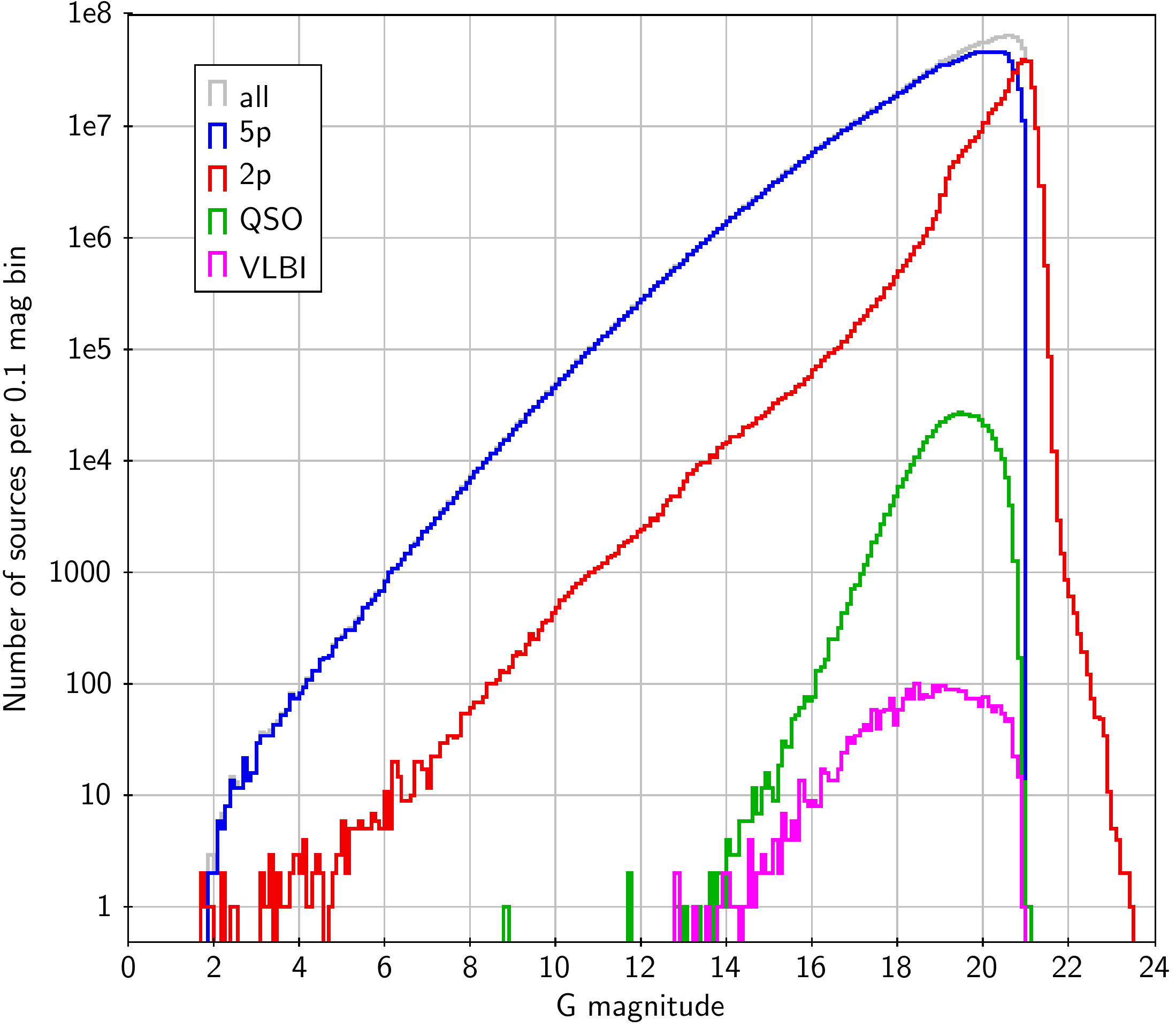}}
    \caption{Magnitude distribution of sources in \textit{Gaia} DR2. 
    Grey: All sources.
    Blue: Sources with a full astrometric solution (five parameters).
    Red: Sources with a fall-back solution (position only).
    Green: Quasar candidates from the AllWISE AGN catalogue.
    Magenta: VLBI sources from the ICRF3 prototype.}
    \label{fig:histNsrc}
\end{figure}

This appendix gives statistics for the most important astrometric characteristics of
\textit{Gaia} DR2. Figure~\ref{fig:histNsrc} shows the distribution of sources according 
to $G$ magnitude (\gacs{photometric\_g\_mean\_mag}). In all statistics, it is
necessary to separate the two kinds of solutions: full (five-parameter) solutions
with positions, parallaxes, and proper motions; and fall-back (two-parameter) 
solutions with only positions. The subsets of the sources used to define the reference 
frame (Sect.~\ref{sec:frame}) are shown by the green and magenta histograms
in Fig.~\ref{fig:histNsrc}.

Subsequent tables and figures illustrate the variation of various quality indicators
with magnitude and position.
The quantities considered are listed below with a brief explanation.
\begin{itemize}
\item \gacs{ra\_error} = standard uncertainty in right ascension at epoch J2015.5, 
$\sigma_{\alpha*}=\sigma_\alpha\cos\delta$
\item  \gacs{dec\_error} = standard uncertainty in declination at epoch J2015.5, 
$\sigma_{\delta}$
\item  \gacs{parallax\_error} = standard uncertainty in parallax, 
$\sigma_{\varpi}$
\item \gacs{pmra\_error} = standard uncertainty of proper motion in right ascension, 
$\sigma_{\mu\alpha*}=\sigma_{\mu\alpha}\cos\delta$
\item \gacs{pmdec\_error} = standard uncertainty of proper motion in declination, 
$\sigma_{\mu\delta}$
\item semi-major axis of error ellipse in position at epoch J2015.5, $\sigma_\text{pos,max}$,
see Eq.~(\ref{eq:sigmaPos})
\item semi-major axis of error ellipse in proper motion, $\sigma_\text{pm,max}$,
see Eq.~(\ref{eq:sigmaPm})
\item \gacs{astrometric\_excess\_noise} = excess source noise, $\epsilon_i$: this
is the extra noise per observation that must be postulated to explain the scatter
of residuals in the astrometric solution for the source 
\item  \gacs{visibility\_periods\_used} = number of visibility periods of the source,
i.e.\ groups of observations separated by at least four days (Sect.~\ref{sec:fallback})
\item  \gacs{astrometric\_matched\_observations} = number of field-of-view
transits of the source used in the astrometric solution
\item \gacs{astrometric\_n\_good\_obs\_al} = number of good
CCD observations AL of the source used in the astrometric solution 
\item fraction of bad CCD observations AL of the source = 
$\gacs{astrometric\_n\_bad\_obs\_al}/\gacs{astrometric\_n\_obs\_al}$
\item \gacs{parallax\_pmra\_corr} = correlation coefficient between $\varpi$ and
$\mu_{\alpha*}$, $\rho(\varpi,\mu_{\alpha *})$
\item \gacs{parallax\_pmdec\_corr} = correlation coefficient between $\varpi$ and
$\mu_\delta$, $\rho(\varpi,\mu_{\delta})$
\item \gacs{pmra\_pmdec\_corr} = correlation coefficient between $\mu_{\alpha*}$ 
and $\mu_\delta$, $\rho(\mu_{\alpha *},\mu_{\delta})$.
\end{itemize}
The meaning of ``good'' and ``bad'' CCD observations requires an explanation. In AGIS an
ill-fitting observation is never downright rejected, but its statistical weight is reduced
by a factor $0 < w \le 1$ depending on the size of the post-fit residual in relation to
the expected uncertainty -- see Eq.~(66) in the AGIS paper. Somewhat arbitrarily we 
count an observation as ``good'' if $w\ge 0.2$ and ``bad'' if $w<0.2$. This corresponds
to a limit of 4.83 standard deviations for a ``good'' residual.

The semi-major axes of the error ellipses in position and proper motion are not given
in the \textit{Gaia} Archive but can be calculated as 
\begin{equation}\label{eq:sigmaPos}
\sigma_\text{pos,\,max} = \sqrt{\frac{1}{2}(C_{00}+C_{11}) +
\frac{1}{2}\sqrt{(C_{11}-C_{00})^2+4C_{01}^2}} \, 
\end{equation}
and
\begin{equation}\label{eq:sigmaPm}
\sigma_\text{pm,\,max} = \sqrt{\frac{1}{2}(C_{33}+C_{44}) +
\frac{1}{2}\sqrt{(C_{44}-C_{33})^2+4C_{34}^2}} \, 
\end{equation}
(cf.\ Eq.~9 in \citeads{2016A&A...595A...4L}), where $C_{ij}$ are
elements of the $5\times 5$ covariance matrix; specifically
\begin{equation}\label{eq:Cij}
\left.\begin{aligned}
C_{00}&=\gacs{ra\_error}\times\gacs{ra\_error}\\
C_{01}&=\gacs{ra\_error}\times\gacs{dec\_error}\times\gacs{ra\_dec\_corr}\\
C_{11}&=\gacs{dec\_error}\times\gacs{dec\_error}\\
C_{33}&=\gacs{pmra\_error}\times\gacs{pmra\_error}\\
C_{34}&=\gacs{pmra\_error}\times\gacs{pmdec\_error}\times\gacs{pmra\_pmdec\_corr}\\
C_{44}&=\gacs{pmdec\_error}\times\gacs{pmdec\_error}
\end{aligned}~\right\}
\end{equation} 

\begin{figure*}
\centering
  \resizebox{0.33\hsize}{!}{\includegraphics{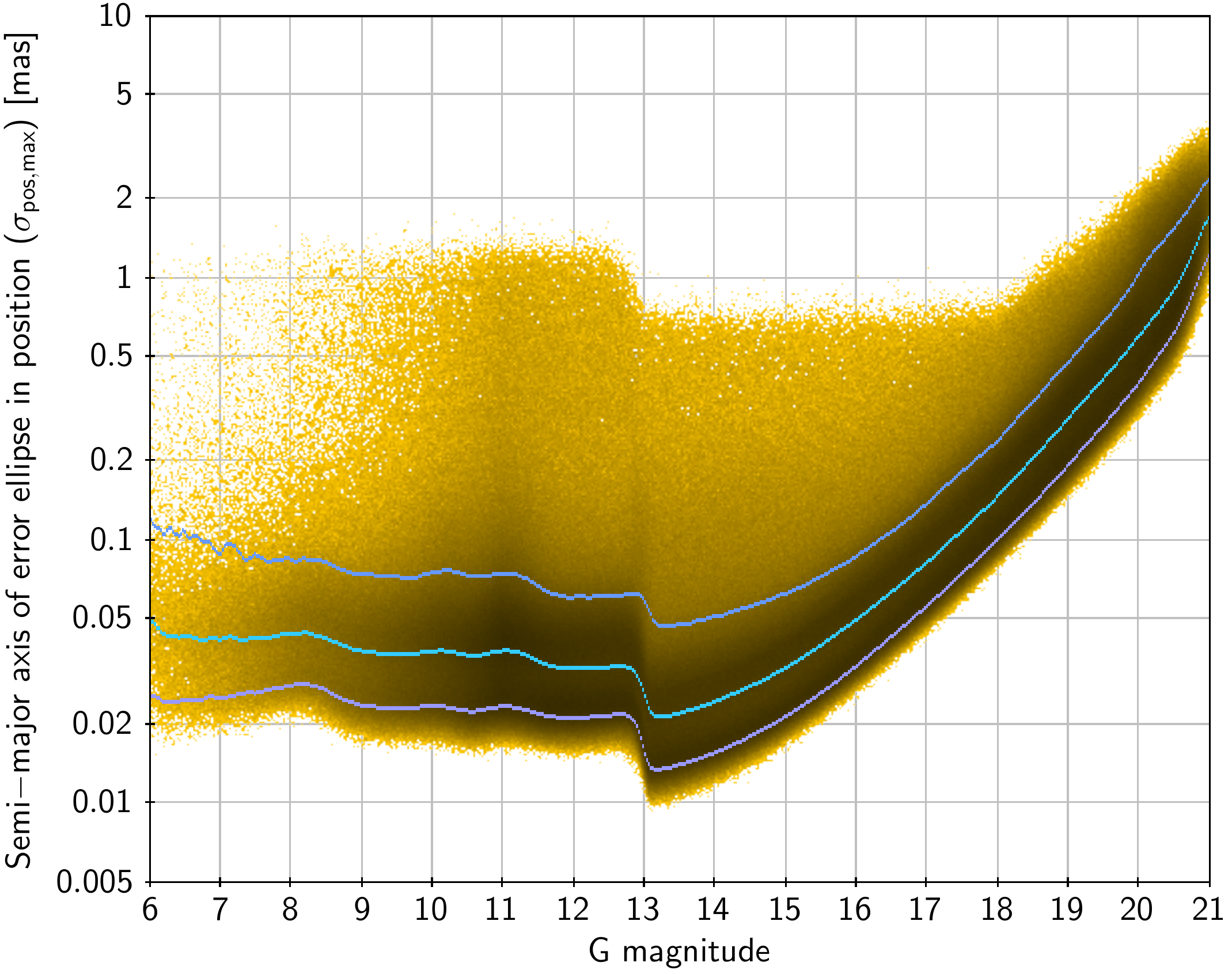}}
  \resizebox{0.33\hsize}{!}{\includegraphics{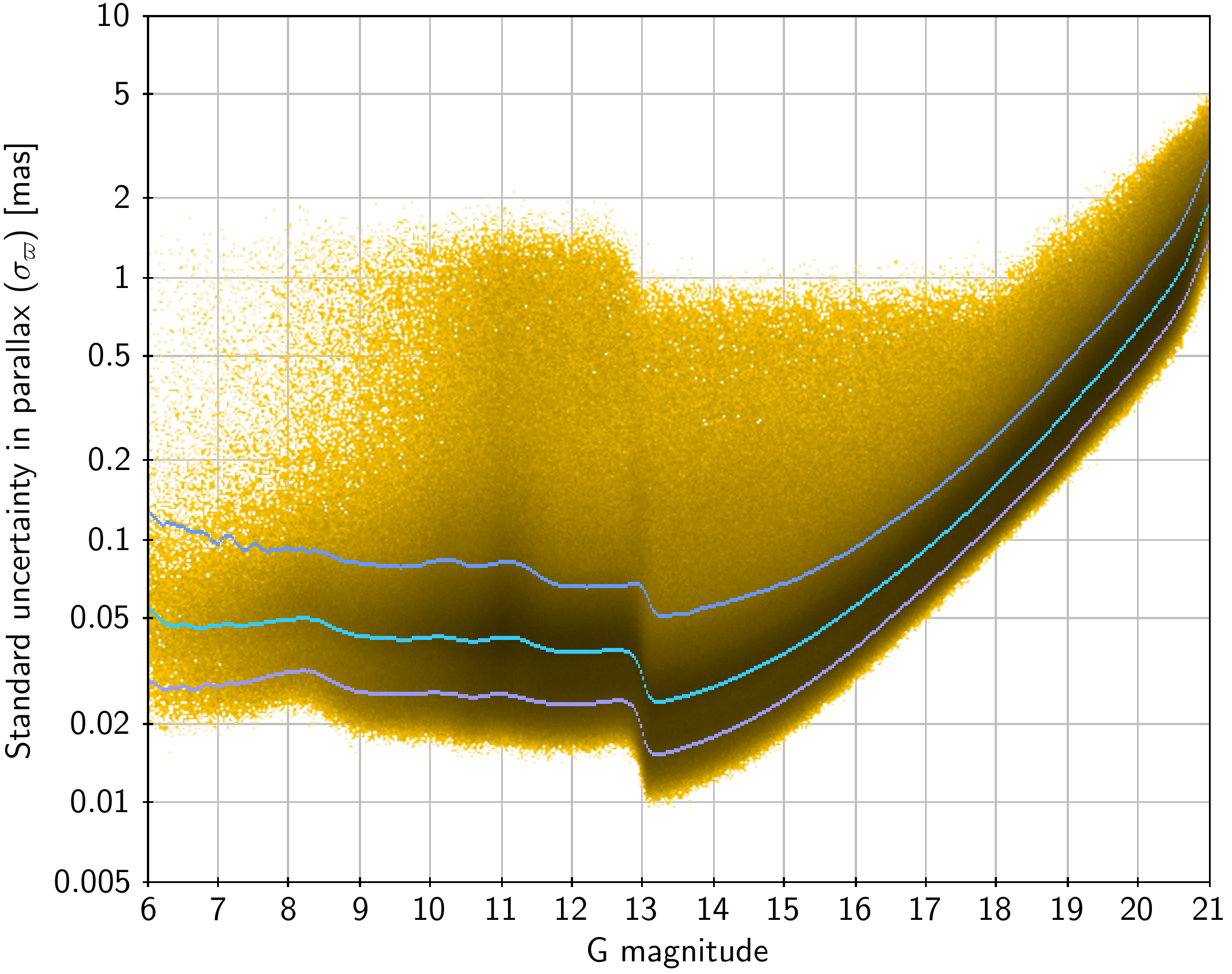}}
  \resizebox{0.33\hsize}{!}{\includegraphics{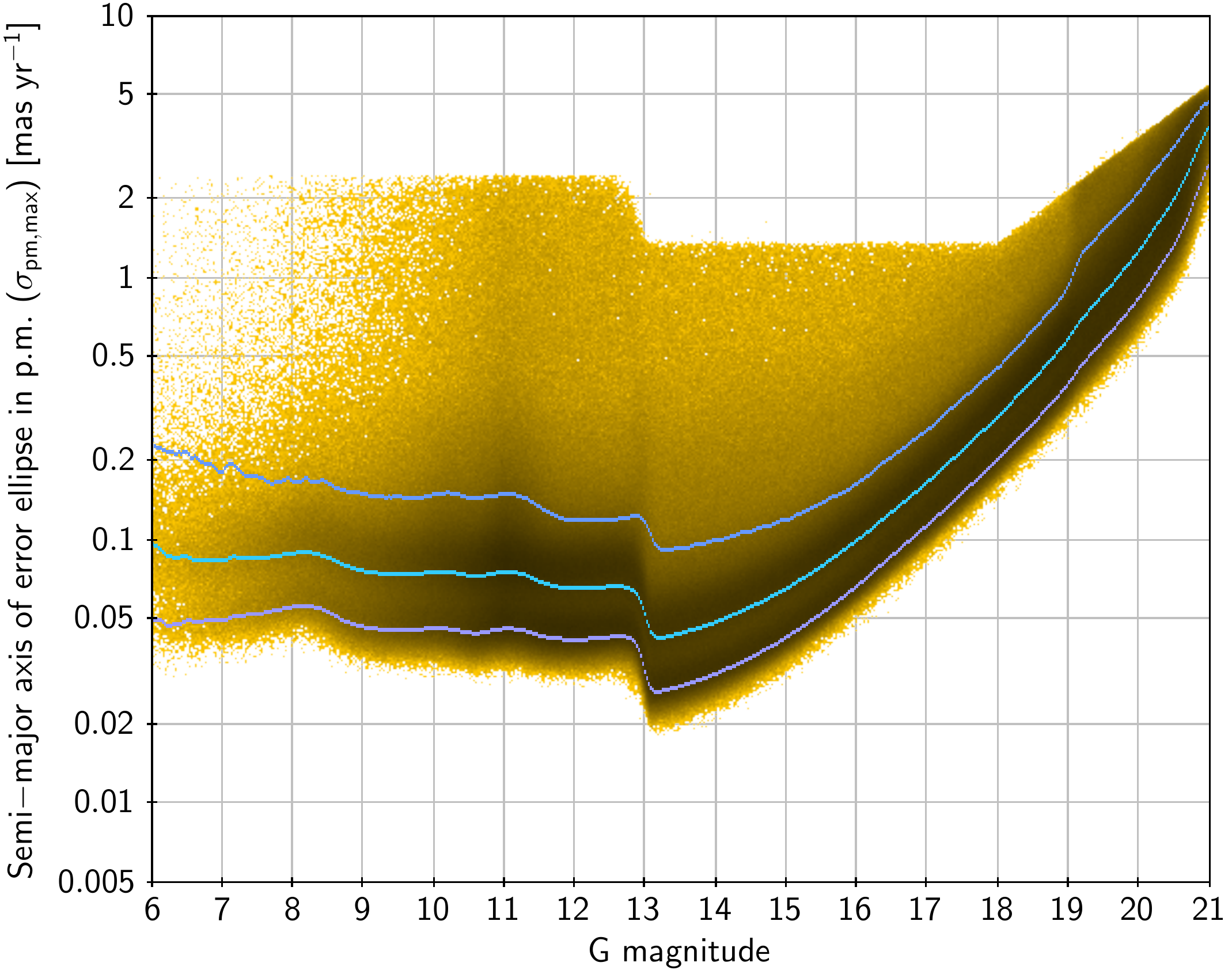}}
    \caption{Formal uncertainties versus the $G$ magnitude for sources with
    a five-parameter astrometric solution. 
    \textit{Left:} Semi-major axis of the error ellipse in position at epoch J2015.5.
    \textit{Middle:} Standard deviation in parallax.
    \textit{Right:} Semi-major axis of the error ellipse in proper motion.
    The yellow dots show individual values for a representative selection of the 
    sources; the cyan curve is the median uncertainty and the blue curves are the
    10th and 90th percentiles. The plotted sample 
    contains all sources for $G<11$, and a geometrically decreasing random 
    fraction of the fainter sources with roughly uniform distribution in $G$.}
    \label{fig:sigmasVsG}
\end{figure*}

\begin{figure*}
\centering
  \resizebox{0.33\hsize}{!}{\includegraphics{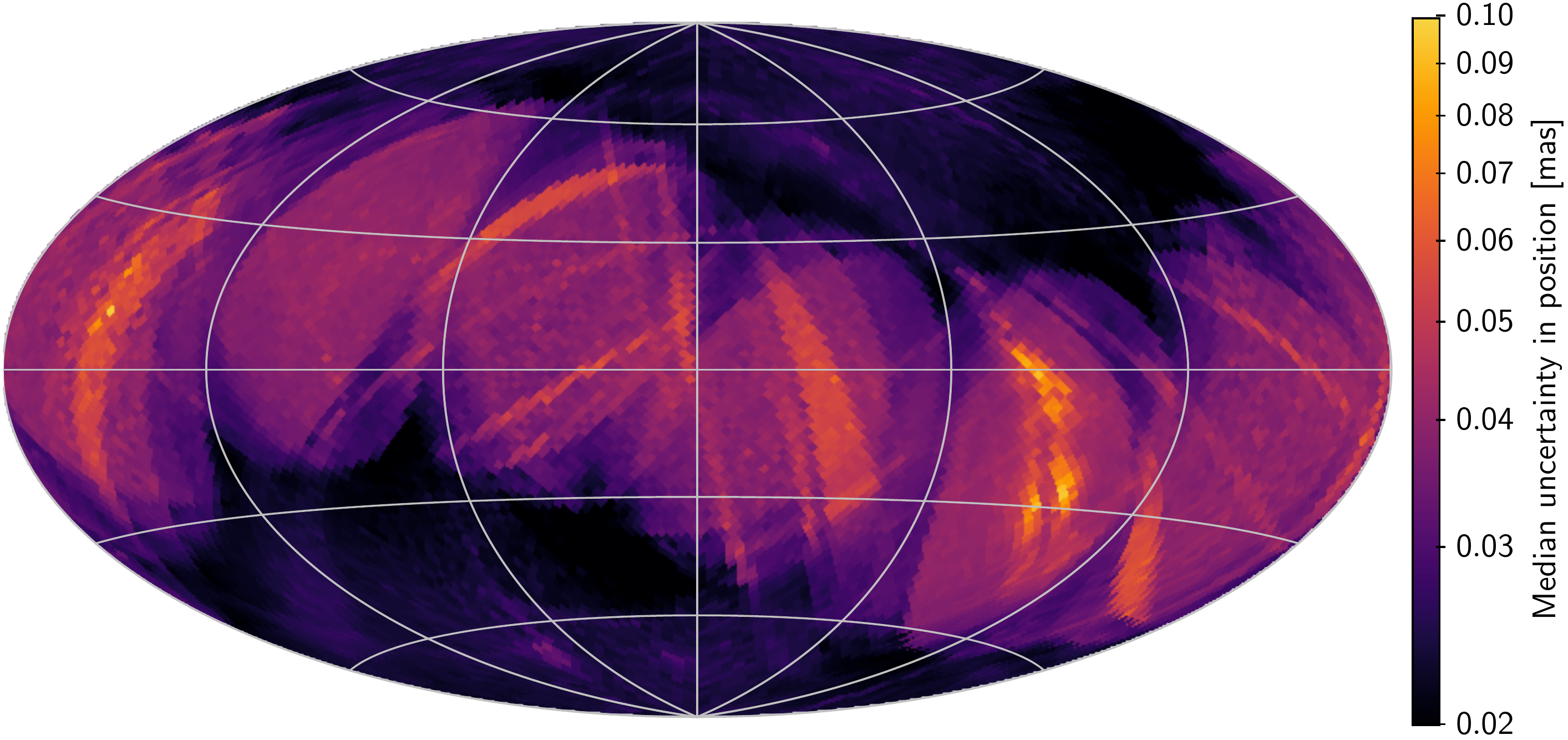}}
  \resizebox{0.33\hsize}{!}{\includegraphics{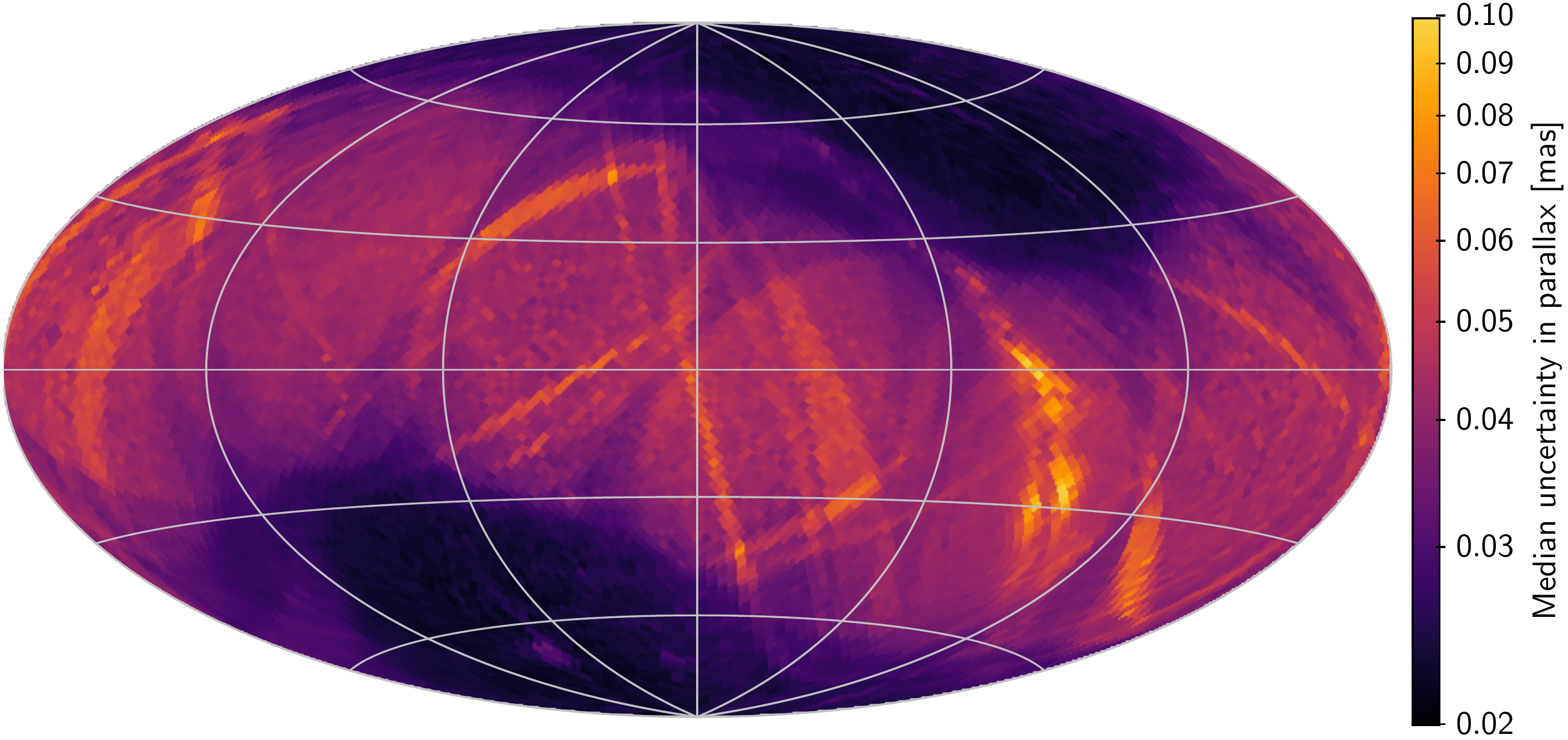}}
  \resizebox{0.33\hsize}{!}{\includegraphics{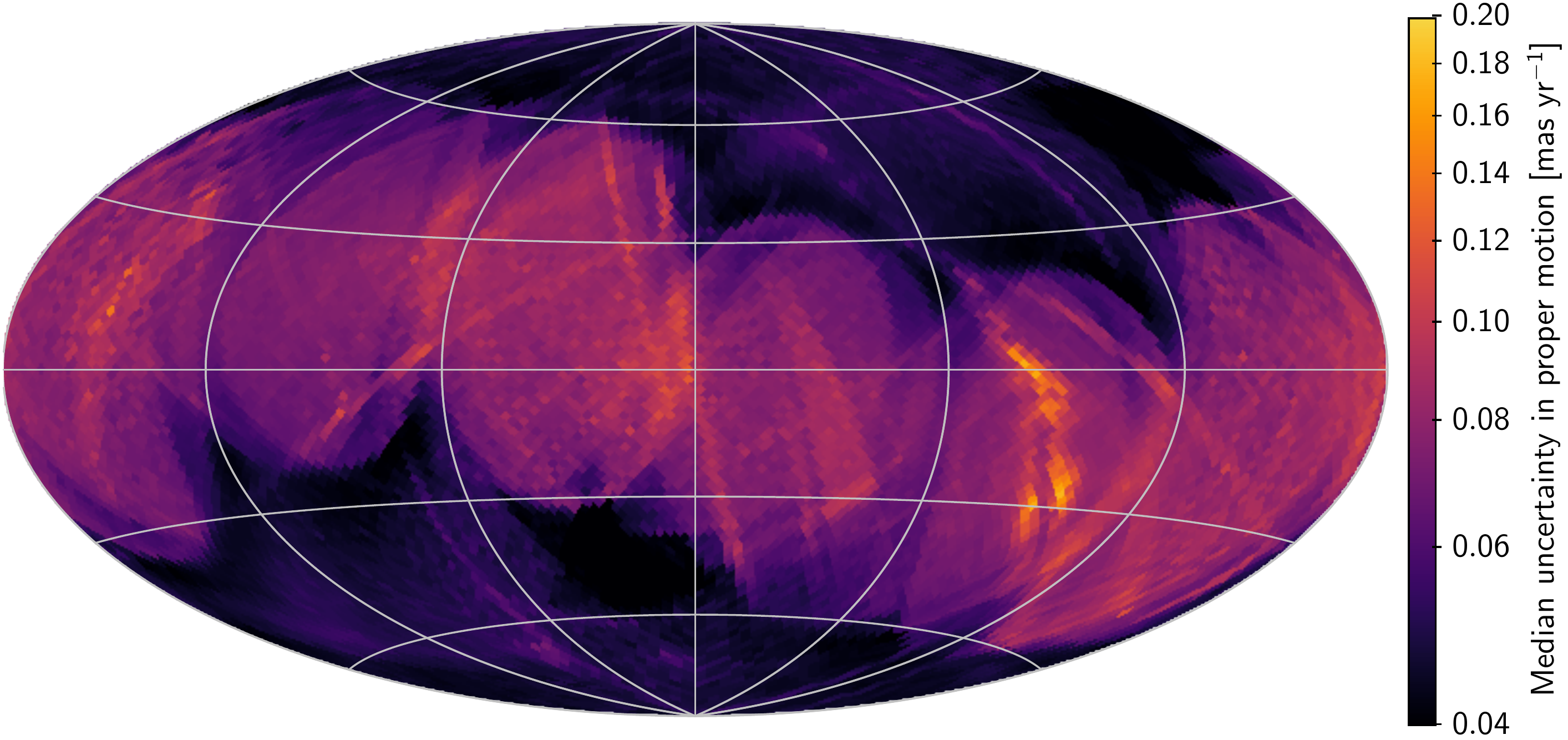}}
    \caption{Formal uncertainties at $G\simeq 15$ for sources with
    a five-parameter astrometric solution.
    \textit{Left:} Semi-major axis of the error ellipse in position at epoch J2015.5.
    \textit{Middle:} Standard deviation in parallax.
    \textit{Right:} Semi-major axis of the error ellipse in proper motion.
    This and all other full-sky maps in this paper use a Hammer--Aitoff projection in 
    equatorial (ICRS) coordinates with $\alpha=\delta=0$ at the centre, north up, 
    and $\alpha$ increasing from right to left.}
    \label{fig:sigmasVsPos}
\end{figure*}

\begin{figure*}
\centering
  \resizebox{0.33\hsize}{!}{\includegraphics{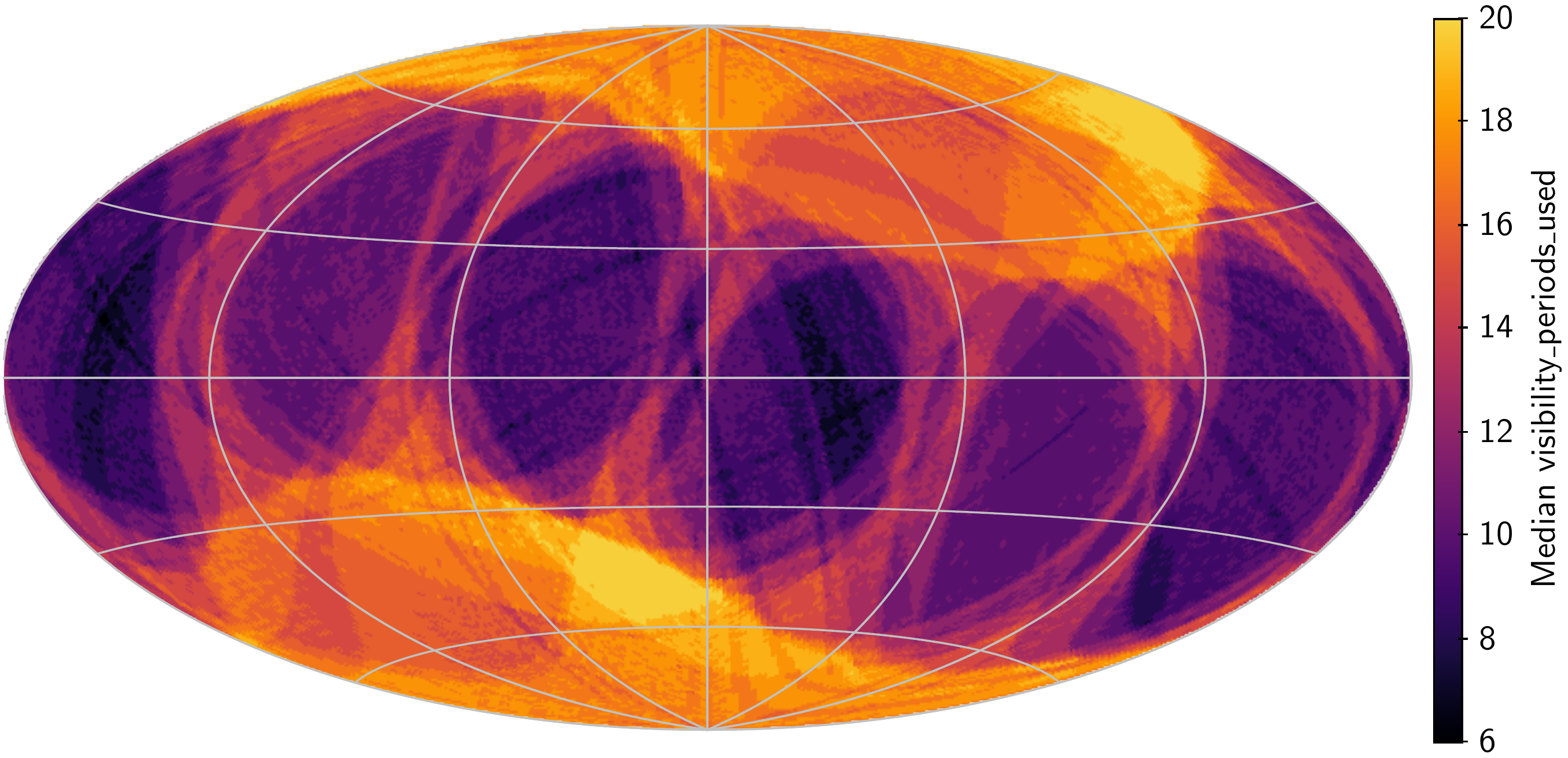}}
  \resizebox{0.33\hsize}{!}{\includegraphics{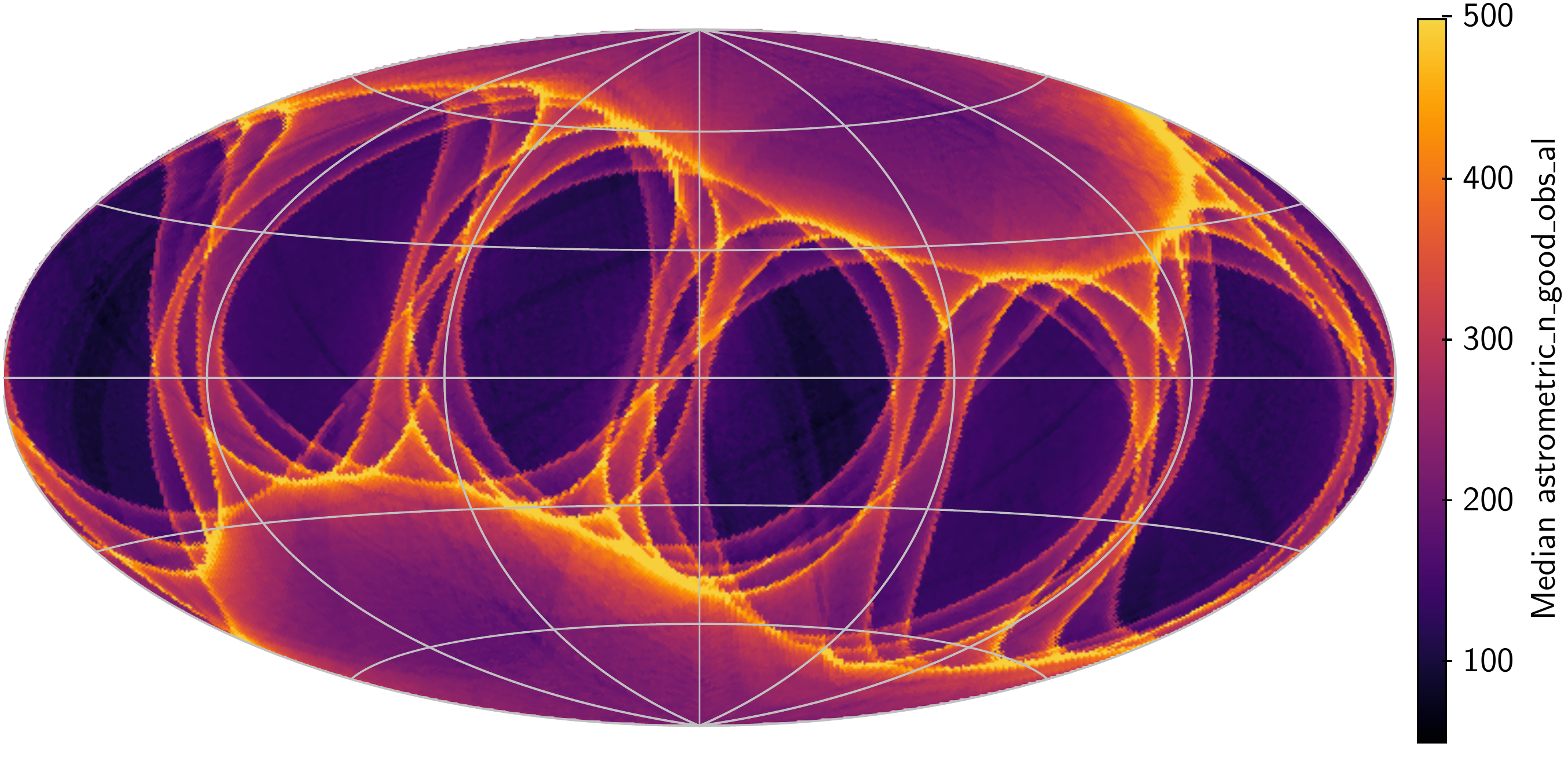}}
  \resizebox{0.33\hsize}{!}{\includegraphics{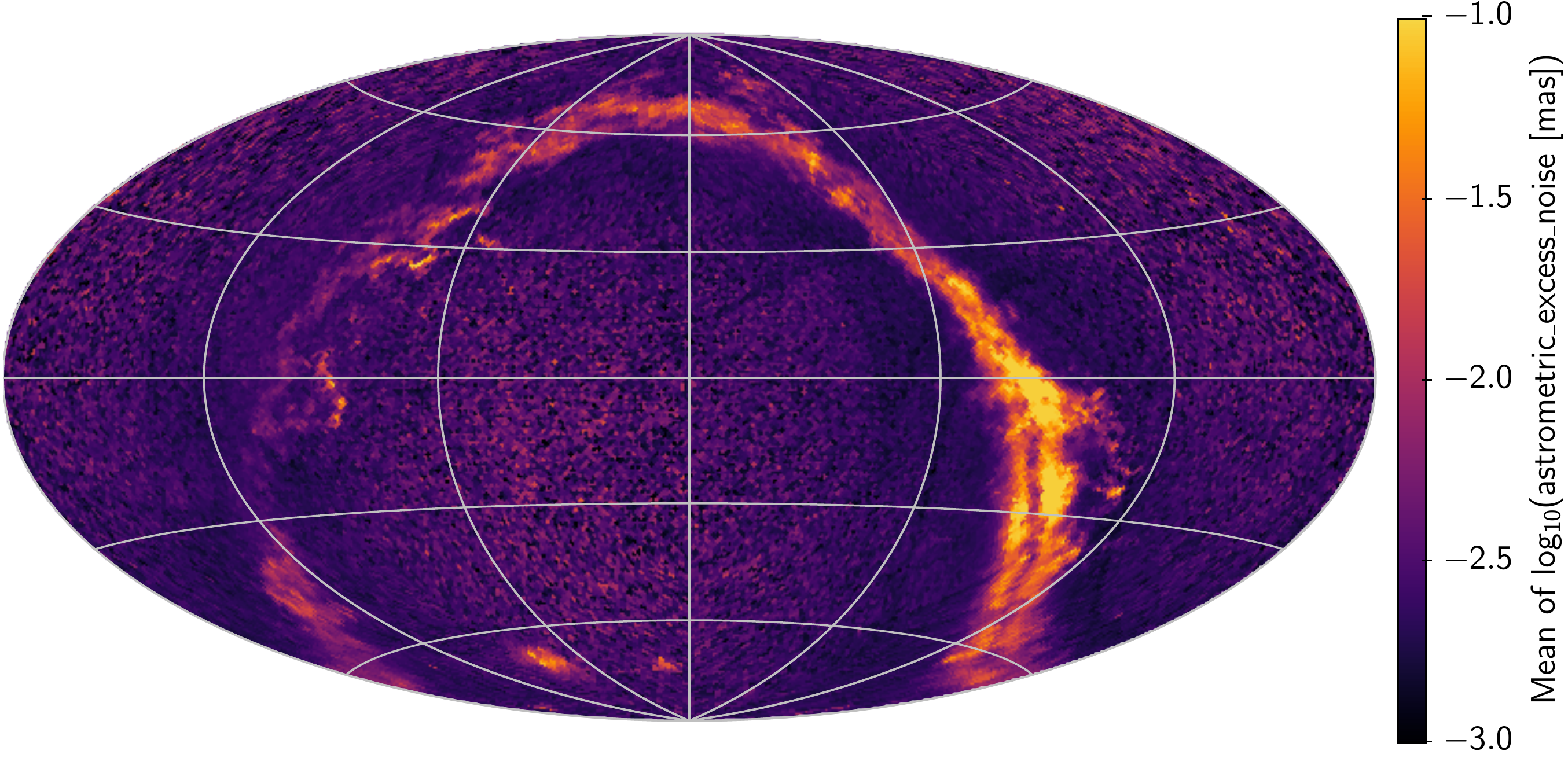}}
    \caption{Observation statistics at $G\simeq 15$ for sources with
    a five-parameter astrometric solution. These statistics are main factors
    governing the formal uncertainties of the astrometric data.
    \textit{Left:} Number of visibility periods used.
    \textit{Middle:} Number of good CCD observations AL. A map of the number of 
    used field-of-view transits is very similar, with a factor nine smaller numbers.
    \textit{Right:} Mean excess source noise.}
    \label{fig:obsVsPos}
\end{figure*}

\begin{figure*}
\centering
  \resizebox{0.33\hsize}{!}{\includegraphics{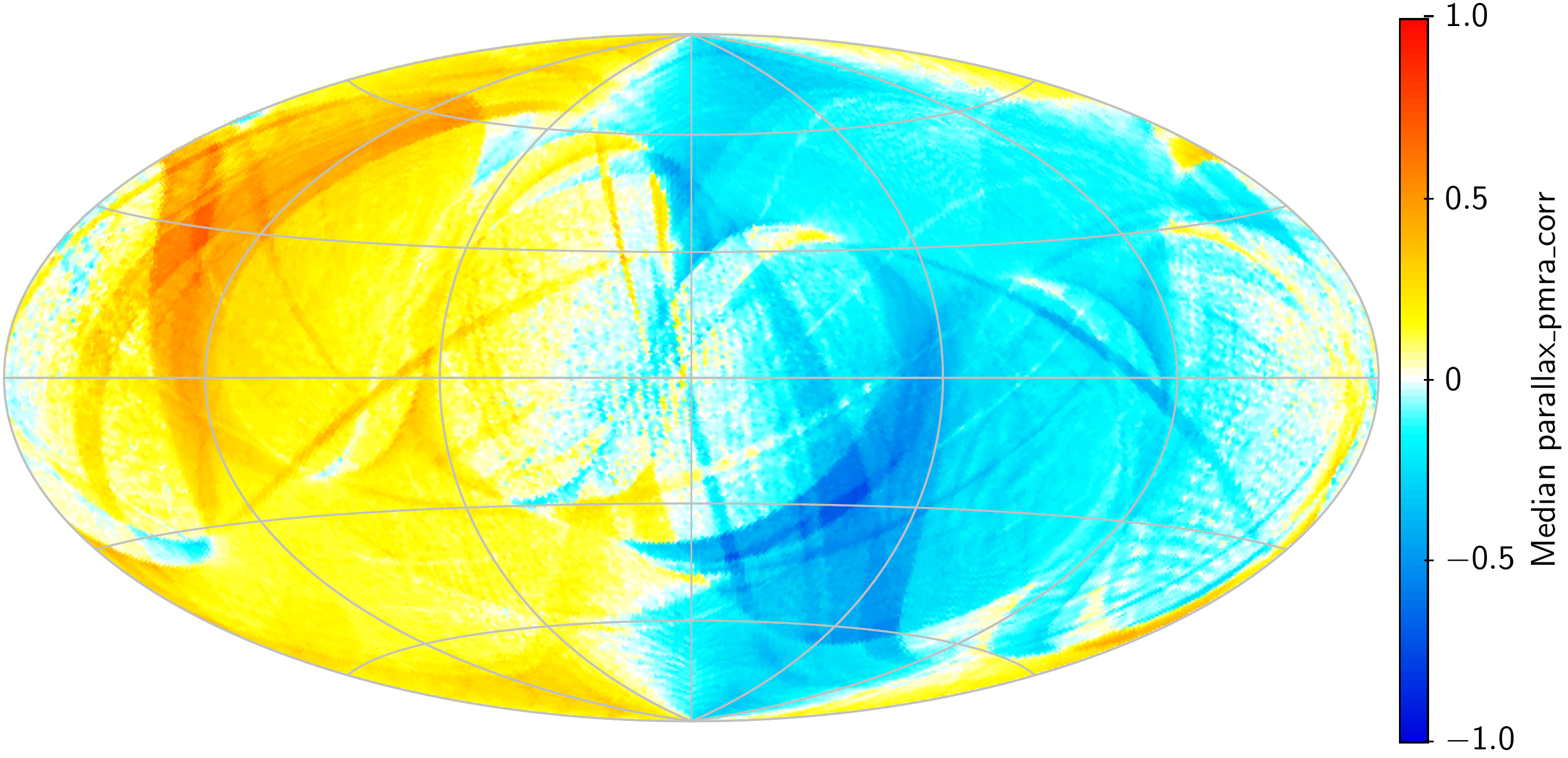}}
  \resizebox{0.33\hsize}{!}{\includegraphics{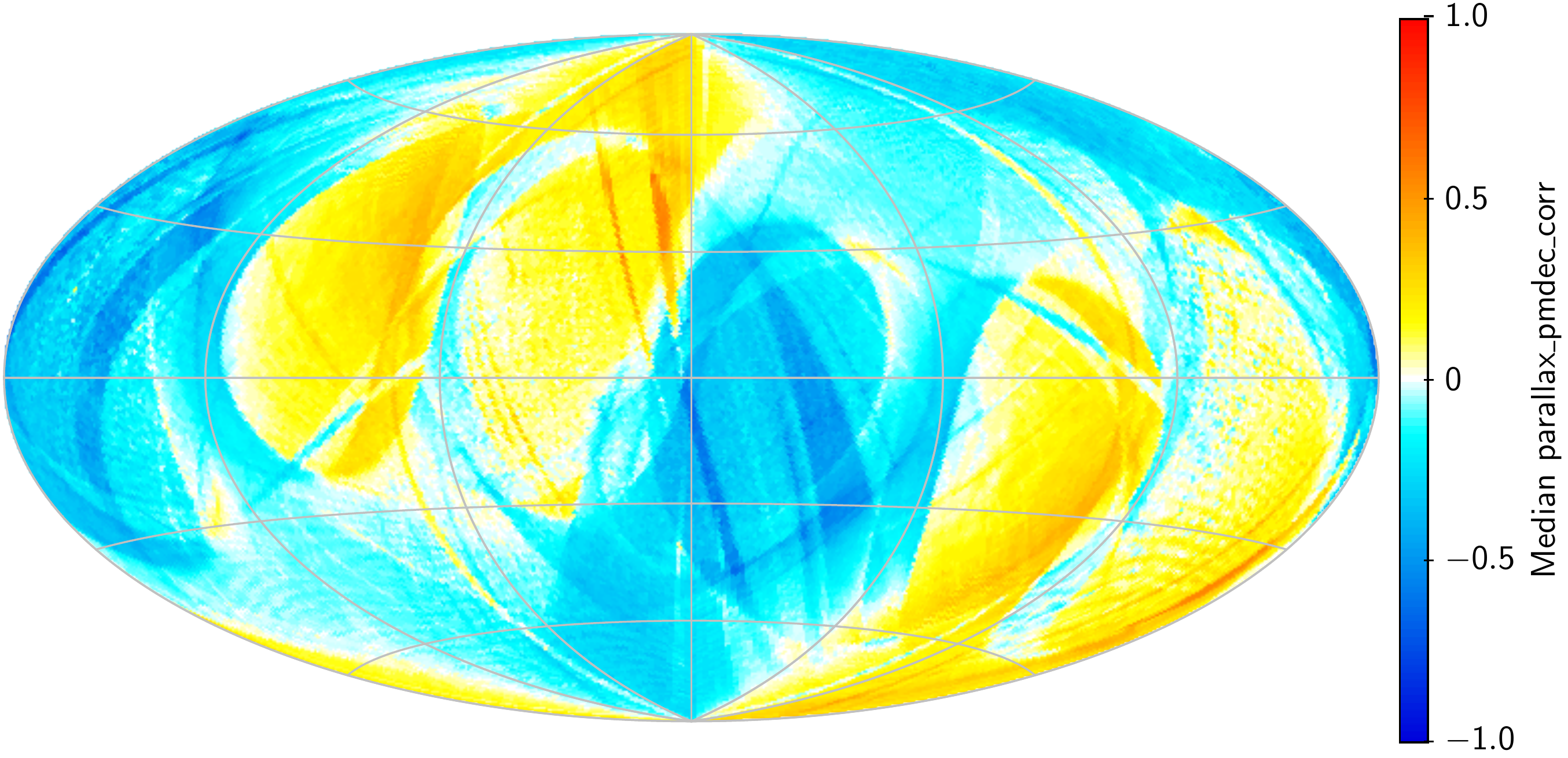}}
  \resizebox{0.33\hsize}{!}{\includegraphics{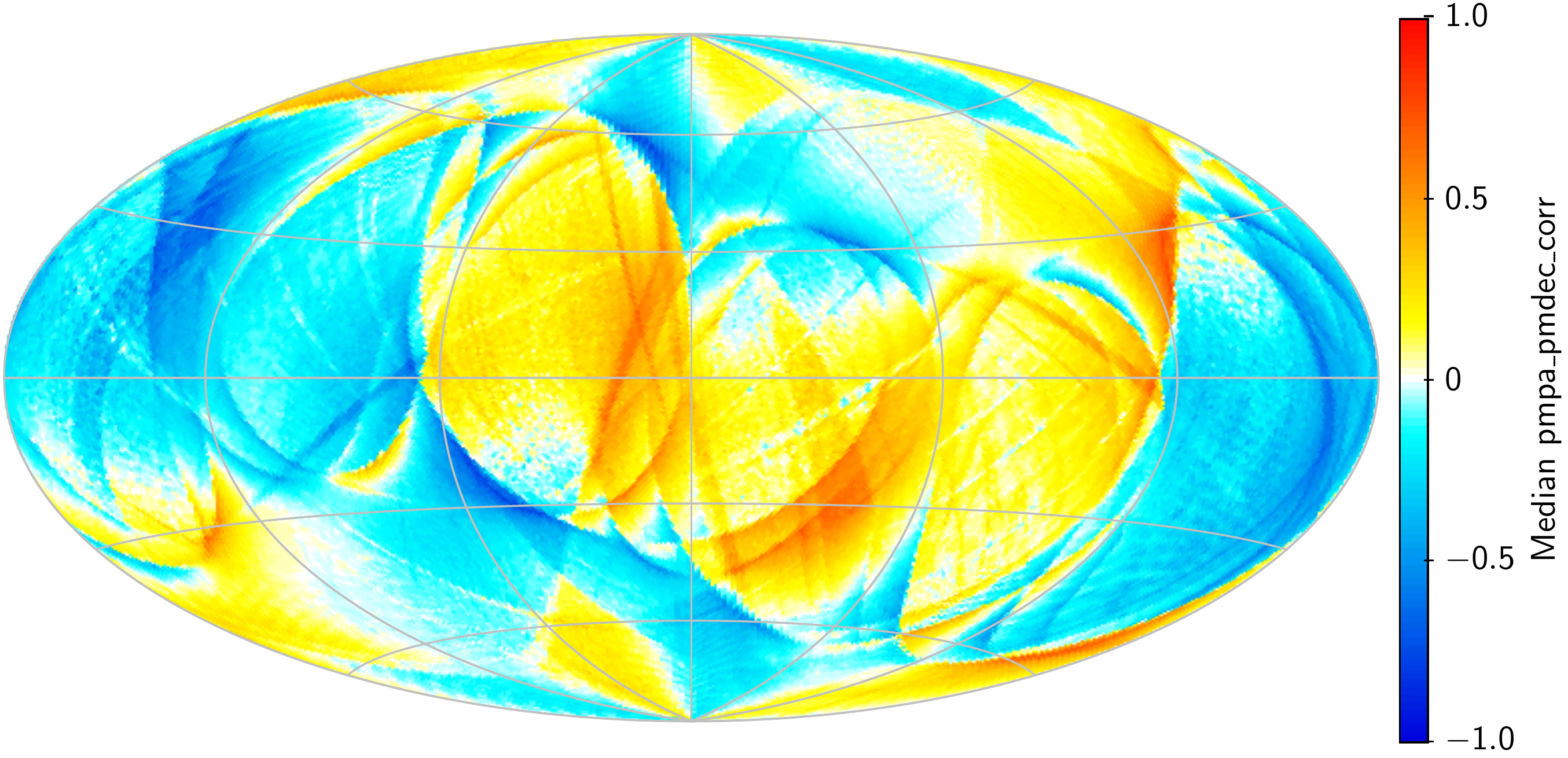}}
    \caption{Correlation coefficients at $G\simeq 15$ for sources with
    a five-parameter astrometric solution. Maps of the correlations at other 
    magnitudes are very similar to these.
    \textit{Left:} Correlation between $\varpi$ and $\mu_{\alpha*}$.
    \textit{Middle:} Correlation between $\varpi$ and $\mu_\delta$.
    \textit{Right:} Correlation between $\mu_{\alpha*}$ and $\mu_\delta$.}
    \label{fig:corrVsPos}
\end{figure*}

Table~\ref{tab:statP5} gives the median uncertainties of the
astrometric parameters, and some other statistics, at selected magnitudes.
At any magnitude there is a considerable scatter among the individual sources,
as illustrated in Fig.~\ref{fig:sigmasVsG}, and a systematic variation with position,
as illustrated in Fig.~\ref{fig:sigmasVsPos} for $G\simeq 15$. The latter figure is 
fairly representative for all magnitudes after appropriate scaling. Additional statistics
at $G\simeq 15$ are shown in Figs.~\ref{fig:obsVsPos} and \ref{fig:corrVsPos}. 
Table~\ref{tab:statP2} gives statistics for the fall-back (two-parameter) solutions.

\begin{table*}[t]
\caption{Summary statistics for the 1332~million sources in \textit{Gaia} DR2 
with a full astrometric solution (five astrometric parameters).
 \label{tab:statP5}}
\small
\begin{tabular}{lrrrrrrrrrrl}
\hline\hline
\noalign{\smallskip}
& \multicolumn{10}{c}{Median value at magnitude $G$} \\
Quantity & $\le 12$ & $13$ & $14$ & $15$ & $16$ & $17$ & $18$ & $19$ & $20$ & $21$ & Unit \\
\noalign{\smallskip}\hline\noalign{\smallskip}
Fraction of sources with 5-param.\ solution
  &  99.1  &  99.0  &  99.0  &  99.0  &  98.9  &  98.6  &  97.6  &  94.5  &  82.9  &  15.9  & \%\\
Standard uncertainty in $\alpha$ ($\sigma_{\alpha*}$) at J2015.5
  & 0.033  & 0.023  & 0.023  & 0.031  & 0.047  & 0.077  & 0.137  & 0.268  & 0.548  & 1.457  & mas\\
Standard uncertainty in $\delta$ ($\sigma_{\delta}$) at J2015.5
  & 0.030  & 0.022  & 0.020  & 0.027  & 0.041  & 0.069  & 0.123  & 0.242  & 0.490  & 1.559  & mas\\
Standard uncertainty in $\varpi$ ($\sigma_{\varpi}$)
  & 0.041  & 0.029  & 0.028  & 0.038  & 0.057  & 0.094  & 0.165  & 0.317  & 0.651  & 2.104  & mas\\
Standard uncertainty in $\mu_{\alpha*}$ ($\sigma_{\mu\alpha*}$)
  & 0.068  & 0.047  & 0.047  & 0.063  & 0.096  & 0.158  & 0.280  & 0.550  & 1.164  & 3.114  & mas~yr$^{-1}$\\
Standard uncertainty in $\mu_{\delta}$ ($\sigma_{\mu\delta}$)
  & 0.059  & 0.042  & 0.040  & 0.054  & 0.082  & 0.137  & 0.243  & 0.479  & 1.011  & 3.374  & mas~yr$^{-1}$\\
Fraction with significant excess noise
  &  20.6  &  21.0  &  17.8  &  17.9  &  18.4  &  19.2  &  20.6  &  21.2  &  18.2  &  10.6  & \%\\
Excess source noise (when significant)
  & 0.183  & 0.249  & 0.311  & 0.331  & 0.367  & 0.474  & 0.701  & 1.226  & 2.235  & 4.563  & mas\\
Number of visibility periods used
  &  13  &  13  &  13  &  13  &  13  &  12  &  12  &  12  &  12  &   9  & \\
Number of field-of-view transits used
  &  26  &  26  &  26  &  26  &  25  &  25  &  24  &  23  &  22  &  12  & \\
Number of good CCD observations AL
  &  220  &  227  &  226  &  223  &  218  &  215  &  212  &  199  &  194  &  102  & \\
Fraction of bad CCD observations AL
  &   2.7  &   0.8  &   0.0  &   0.4  &   0.4  &   0.4  &   0.4  &   0.5  &   0.4  &   0.5  & \%\\
\noalign{\smallskip}\hline
\end{tabular}
\tablefoot{The table gives the median quantities at the magnitudes indicated in the
header. For $G\le 12$ the median values were computed using all the sources in that range;
at other magnitudes about 2~million sources were used around the indicated $G$.}
\end{table*}

\begin{table*}[t]
\caption{Summary statistics for the 361~million sources in \textit{Gaia} DR2 with a fall-back solution (position only).\label{tab:statP2}}
\small
\begin{tabular}{lrrrrrrrrrrl}
\hline\hline
\noalign{\smallskip}
& \multicolumn{10}{c}{Median value at magnitude $G$} \\
Quantity & $\le 12$ & $13$ & $14$ & $15$ & $16$ & $17$ & $18$ & $19$ & $20$ & $21$ & Unit \\
\noalign{\smallskip}\hline\noalign{\smallskip}
Fraction of sources with fall-back solution
  &   0.9  &   1.0  &   1.0  &   1.0  &   1.1  &   1.4  &   2.4  &   5.5  &  17.1  &  84.1  & \%\\
Standard uncertainty in $\alpha$ ($\sigma_{\alpha*}$) at J2015.5
  & 0.905  & 0.968  & 0.972  & 0.970  & 0.989  & 1.018  & 1.036  & 1.518  & 2.204  & 3.623  & mas\\
Standard uncertainty in $\delta$ ($\sigma_{\delta}$) at J2015.5
  & 0.876  & 0.924  & 0.931  & 0.921  & 0.936  & 0.956  & 0.971  & 1.399  & 1.897  & 3.387  & mas\\
Number of visibility periods used
  &  10  &  11  &  11  &  10  &   9  &   8  &   7  &   6  &   6  &   6  & \\
Number of field-of-view transits used
  &  16  &  19  &  19  &  18  &  15  &  13  &  11  &   9  &   9  &   8  & \\
Number of good CCD observations AL
  &  133  &  161  &  158  &  149  &  130  &  107  &   96  &   79  &   78  &   62  & \\
Fraction of bad CCD observations AL
  &   2.7  &   0.6  &   0.0  &   0.0  &   0.0  &   0.0  &   0.0  &   0.0  &   0.0  &   0.0  & \%\\
\noalign{\smallskip}\hline
\end{tabular}
\tablefoot{The table gives the median quantities at the magnitudes indicated in the header.}
\end{table*}

\section{Selecting astrometrically ``clean'' subsets} 
\label{sec:clean}

The criterion for an accepted five-parameter solution, Eq.~(\ref{eq:critP5}), was
designed to include as many sources as possible with reasonably reliable astrometry.
Using a stricter criterion would have resulted in a smaller, but possibly more reliable
catalogue. The choice of a relatively lenient criterion presumes that users can and 
should implement additional filters as required by their particular applications, with
due consideration of possible selection biases introduced by the filters. The
\textit{Gaia} Archive contains several statistics that may be useful in this process,
but their interpretation is far from simple. In this appendix we illustrate both the 
benefits and limitations of certain filters in a specific case, namely the construction
of a ``clean'' HR diagram of nearby ($<100$~pc) stars. This should not be seen as
a fixed recipe for selecting sources with the most reliable astrometry, but it may
provide some useful hints for further exploration. Complementing the internal
validation in Sect.~\ref{sec:valid} it also contains a brief discussion of the extremely 
large positive and negative parallaxes in \textit{Gaia} DR2.

\begin{figure*}
\centering
  \resizebox{0.33\hsize}{!}{\includegraphics{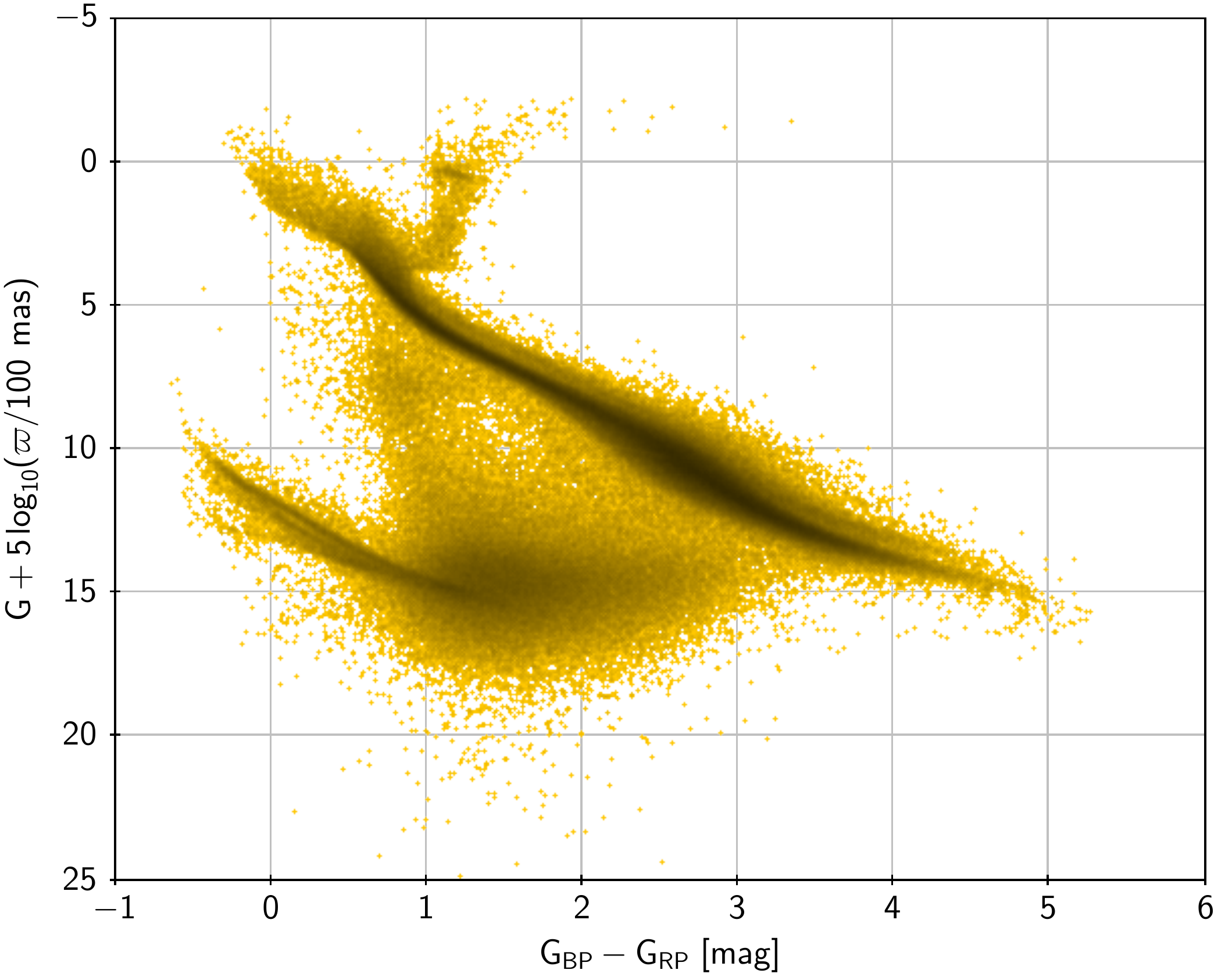}}
  \resizebox{0.33\hsize}{!}{\includegraphics{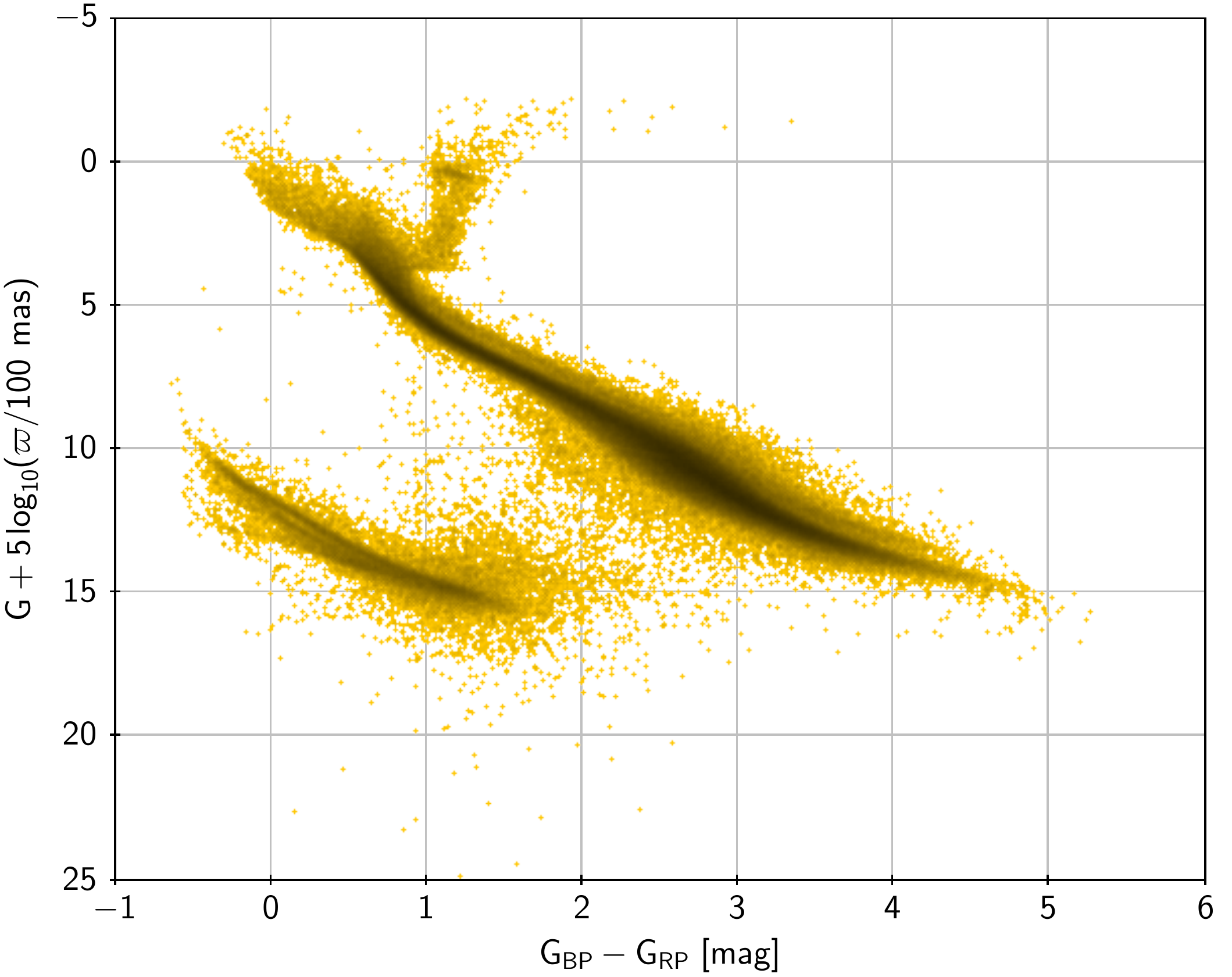}}
  \resizebox{0.33\hsize}{!}{\includegraphics{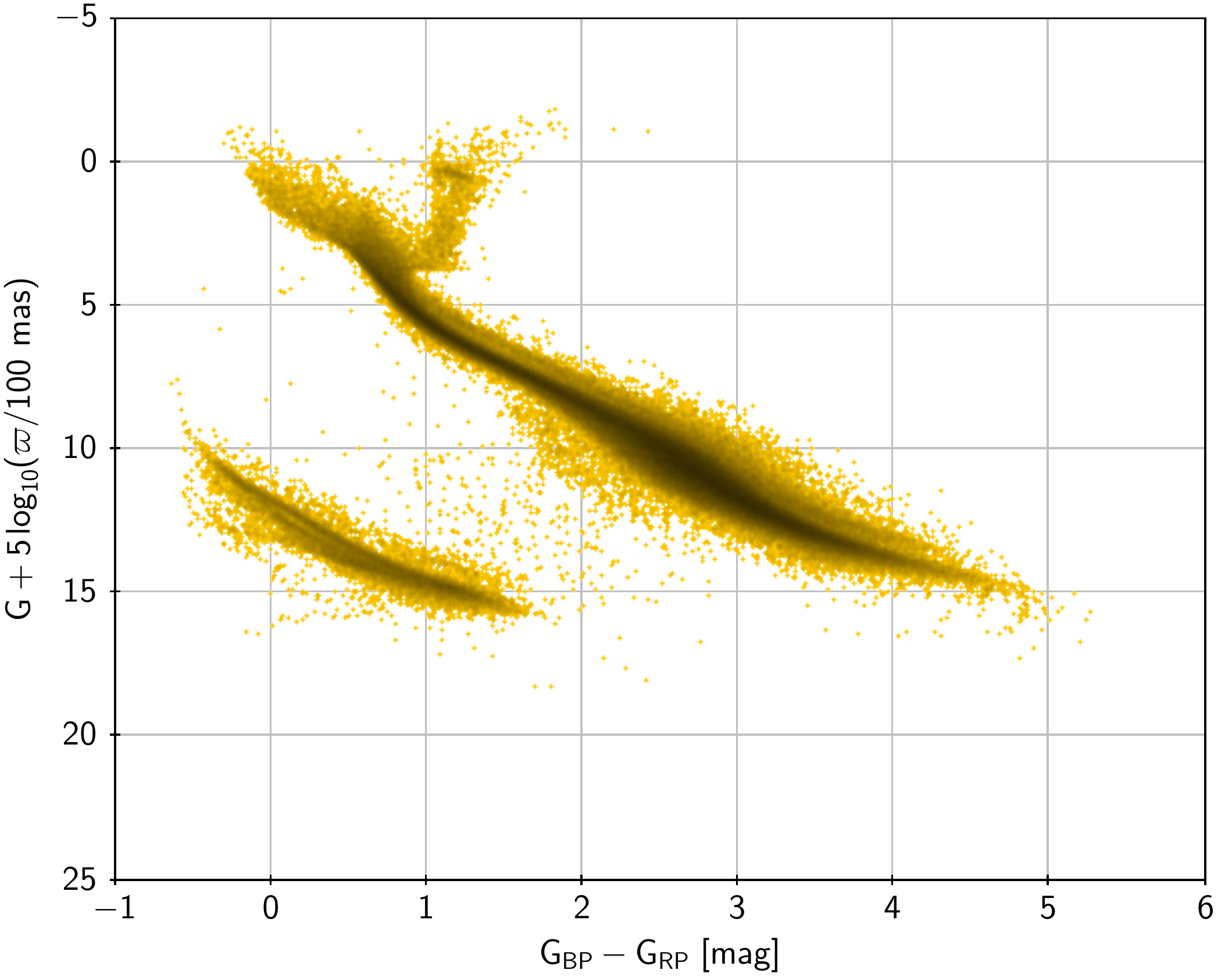}}
    \caption{HR diagram of sources nominally within 100~pc and with relative
    distance error less than 10\%.
    \textit{Left:} Raw diagram (Selection A).
    \textit{Middle:} Sources filtered by unit weight error (Selection B).
    \textit{Right:} Sources filtered by unit weight error and flux excess ratio (Selection C).}
    \label{fig:HR}
\end{figure*}

\begin{figure*}
\centering
\sidecaption
  \resizebox{12cm}{!}{\includegraphics{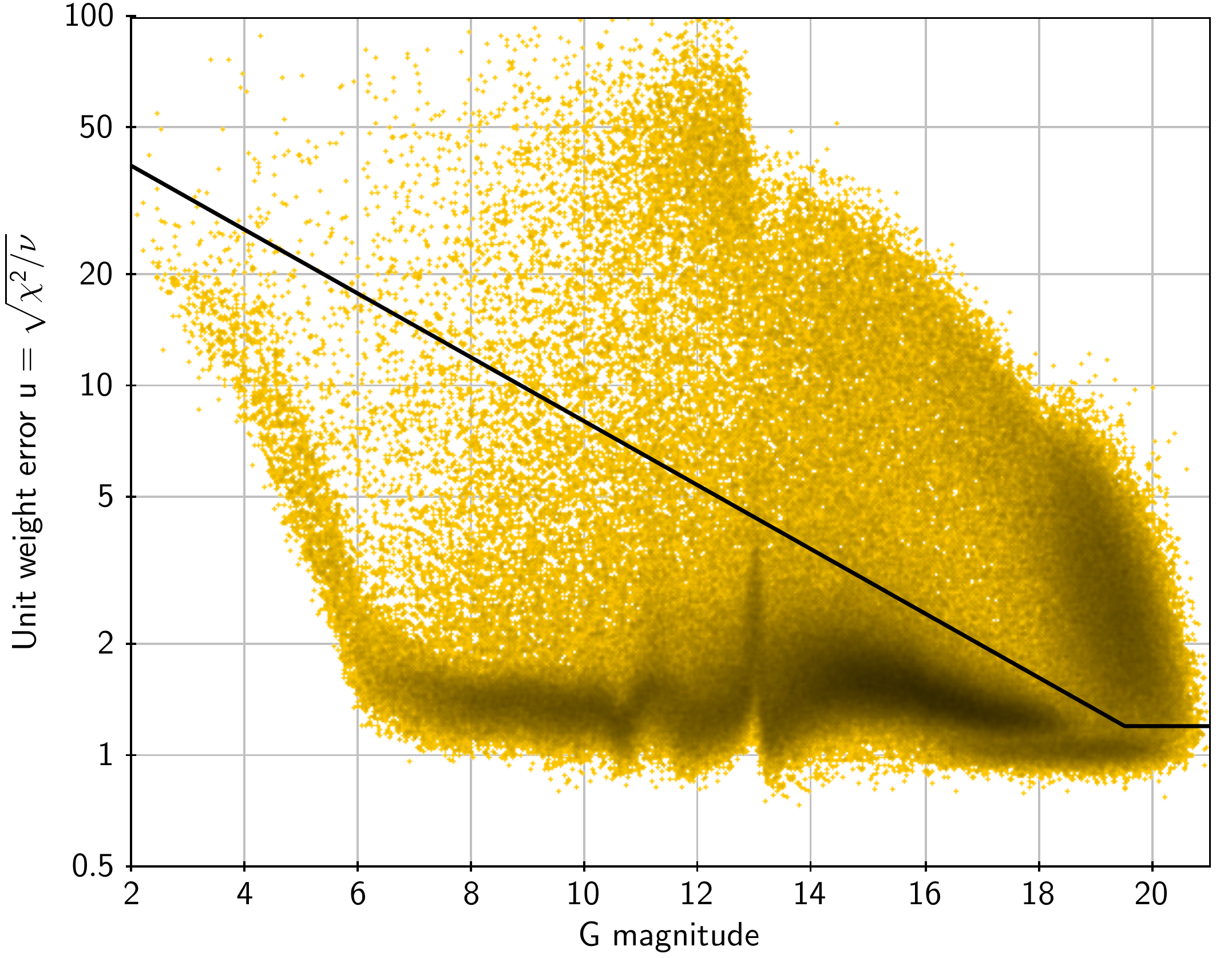}
  \includegraphics{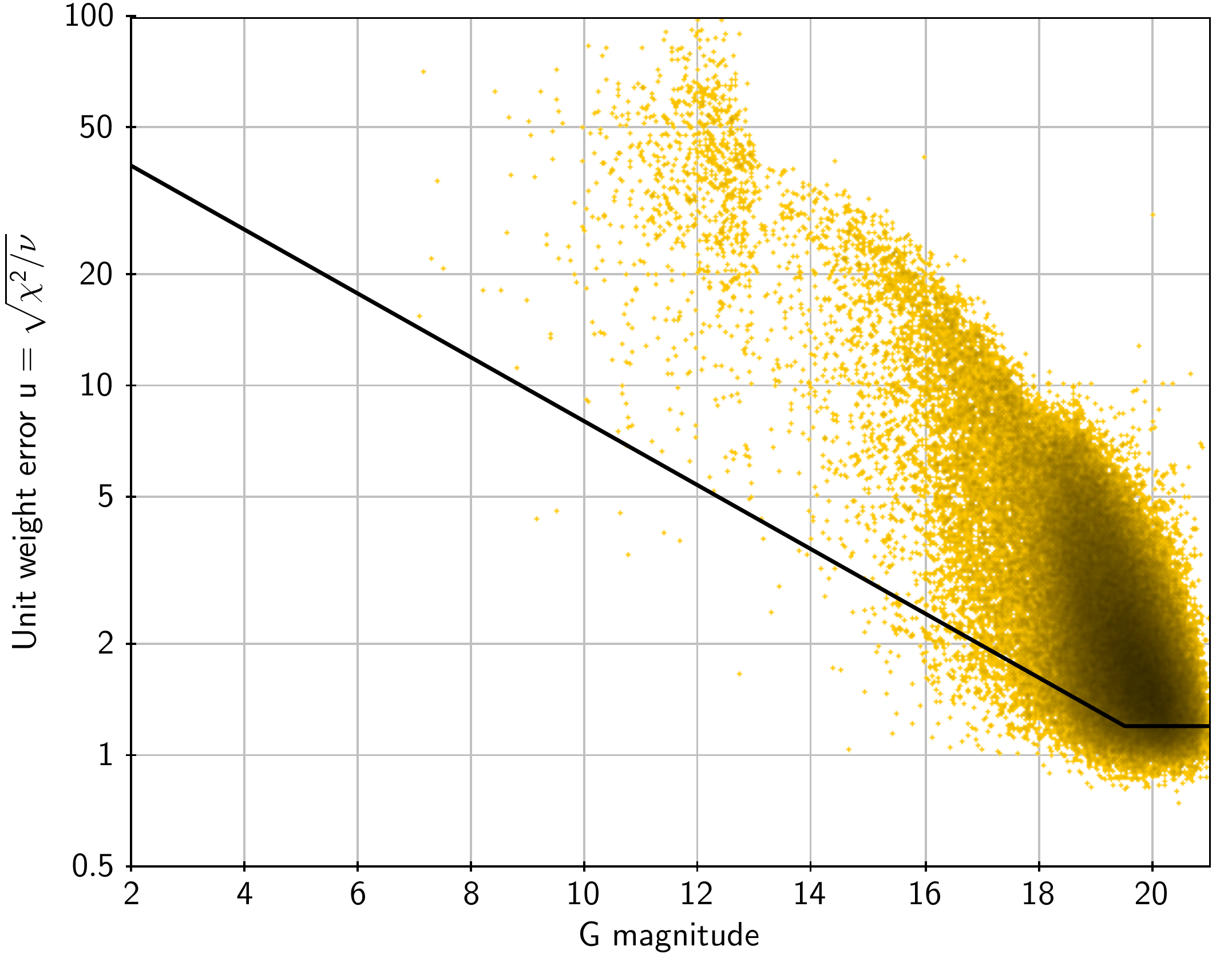}}
    \caption{Unit weight error $u=(\chi^2/\nu)^{1/2}$ versus magnitude for two
    samples. \textit{Left:} Selection A (positive parallaxes). \textit{Right:}
    Selection N (negative parallaxes). The black line is the threshold 
    defined in Eq.~(\ref{eq:fG}).}
    \label{fig:uweVsGposNeg}
\end{figure*}

The left panel of Fig.~\ref{fig:HR} is an HR diagram obtained by plotting
$G+5\log_{10}(\varpi/100~\text{mas})$ versus colour index $G_\text{BP}-G_\text{RP}$. 
(This ignores extinction and takes the distance to be inverse parallax, both
reasonable approximations in the solar neighbourhood.) The criteria used were:
\begin{equation*}\label{eq:selA}
\left.\begin{aligned}
\text{(i)}\quad&\varpi > 10~\text{mas} \\ 
\text{(ii)}\quad&\varpi/\sigma_\varpi > 10 \\ 
\text{(iii)}\quad&\gacs{phot\_bp\_mean\_flux\_over\_error} > 10 \\
\text{(iv)}\quad&\gacs{phot\_rp\_mean\_flux\_over\_error} > 10 
\end{aligned} \quad \right\} \quad \text{(Selection~A)}
\end{equation*}
yielding 338\,833~sources. Nominally, (i) selects sources within 100~pc, (ii)
those with at most 10\% uncertainty in distance (corresponding to $\simeq\,$0.2~mag
in absolute magnitude), and (iii)--(iv) those with at most 10\% uncertainty in the
BP and RP fluxes (corresponding to $\simeq\,$0.1~mag in $G_\text{BP}$ and
$G_\text{RP}$). Taken at face value, this selection should produce a very clean
HR diagram. Indeed, the astrophysically expected features are very prominent in the 
left panel of Fig.~\ref{fig:HR} but many points fall in unexpected places, e.g.\ between 
the main and 
white-dwarf sequences. Selection~A includes three sources with $\varpi>800$~mas, 
i.e.\ nominally closer to the Sun than Proxima~Centauri (which has the fourth largest
parallax in the sample). All three sources are faint ($G>19.7$) and lie in a very crowded 
region within 10~deg of the Galactic centre. This suggests that their large parallaxes 
are spurious, resulting from inconsistent matching of the observations to different 
physical sources. If that is the case, then most likely the proper motions of these sources
are also corrupted.

With a maximum star density of the order of one million per square degree, there is a
non-negligible probability to have a chance configuration of two stars, separated 
by an arcsec or less, which could be mistaken for a single object with a large parallax.
This is more likely to happen in areas that combine a high star density with a
relatively small number of visibility periods, as is the case in the region of the
Galactic centre (Fig.~\ref{fig:obsVsPos}). However, it is reasonable to expect that 
in most such cases of spurious parallaxes, observations do not fit the single-star parallax 
model very well. This should lead to an increased chi-square, or unit weight error
$u=(\chi^2/\nu)^{1/2}$. In the left panel of Fig.~\ref{fig:uweVsGposNeg} this quantity
is plotted versus $G$ for Selection~A. Compared with a similar plot for well-behaved
sources (Fig.~\ref{fig:uweVsG}), there are several noteworthy differences: the strong
rise for $G<6$ caused by uncalibrated CCD saturation; a blob of moderately large 
values of $u$ for $G>18$, possibly extending to much larger values for brighter sources;
and a general scatter of large $u$ at all magnitudes, which could be caused by 
partially resolved or astrometric binaries. If we want to keep the sources with $G<6$
(which include most of the giants) but remove the blob at $G>18$, a possible cut
is given by the black lines, i.e.\ the function
\begin{equation}\label{eq:fG}
u < 1.2\times\max(1,\exp(-0.2(G-19.5))) \, .
\end{equation}
Adding this criterion to Selection~A gives Selection~B with 249\,793 
sources and the much cleaner HR diagram in the middle panel of Fig.~\ref{fig:HR}.
(A similar filtering could be obtained by using the excess source noise instead of $u$,
for example by selecting $\gacs{astrometric\_excess\_noise}<1$~mas, but the 
behaviour of the excess noise for $G\lesssim 15$ is less discriminating due to the 
DOF bug.) 
Selection~B still contains two sources with $\varpi>800$~mas.

Additional scatter in the HR diagram is produced by photometric errors mainly in
the BP and RP bands, affecting in particular faint sources in crowded areas. An 
indicator of possible issues with the BP and RP photometry is the flux excess factor
$E=(I_\text{BP}+I_\text{RP})/I_\text{G}$ (\gacs{phot\_bp\_rp\_excess\_factor}),
where $I_X$ is the photometric flux in band $X$ \citepads{DPACP-40}. Adding
the criterion \citepads{DPACP-31} 
\begin{equation}\label{eq:flx}
1.0+0.015(G_\text{BP}-G_\text{RP})^2 < E < 1.3+0.06(G_\text{BP}-G_\text{RP})^2
\end{equation}
to Selection B gives Selection C with 242\,582 sources and the 
HR diagram in the right panel of Fig.~\ref{fig:HR}. The remaining scatter of points 
between the main and white-dwarf sequences may be partly real, consisting of 
binaries with white-dwarf and main-sequence companions of roughly equal magnitude.
In Selection~C the source with the largest parallax is Proxima~Centauri.

The chance matching mechanism discussed above, where different observations 
of the same \textit{Gaia} source are matched to two (or more) physically distinct 
objects, should produce a roughly equal number of positive and negative spurious 
parallaxes. Further insight into the mechanism can therefore be gained by inspecting 
a sample of sources with significantly negative parallaxes. The selection
\begin{equation*}\label{eq:selN}
\left.\begin{aligned}
\text{(i)}\quad&\varpi < -10~\text{mas} \\ 
\text{(ii)}\quad&\varpi/\sigma_\varpi < -10 
\end{aligned} \quad \right\} \quad \text{(Selection~N)}
\end{equation*}
gives 113\,393 sources with manifestly unphysical parallaxes. A plot of $u$ versus 
$G$ for this sample is shown in the right panel of Fig.~\ref{fig:uweVsGposNeg}.
The similarity to the ``blob'' in the left plot is striking, and supports the idea that 
most of the spurious large (positive or negative) parallaxes can be removed by 
a judicious cut in the $(G,u)$ plane. In fact 90\% of the the sources in Selection~N 
are removed by the cut in Eq.~(\ref{eq:fG}).

\begin{figure}
\centering
  \resizebox{0.85\hsize}{!}{\includegraphics{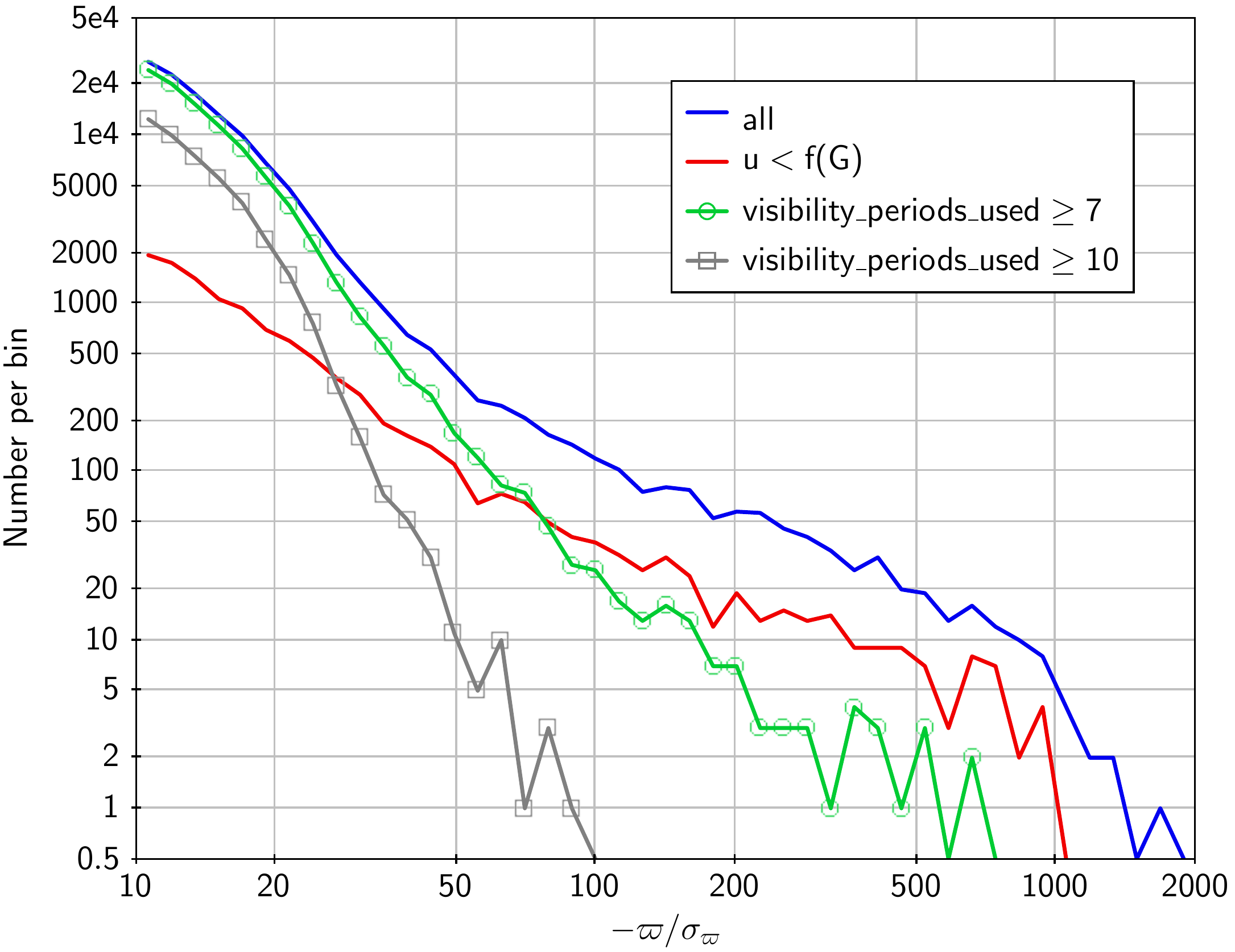}}
    \caption{Distribution of the negative tail of normalised parallaxes.}
    \label{fig:histNegPlx}
\end{figure}

Selection~N includes 61 sources with $\varpi<-800$~mas, the smallest being 
$-1857$~mas. For comparison, if the photometric criteria (iii) and (iv) are removed 
from Selection~A, the number of sources with $\varpi>800$~mas is 46. Conversely, 
if (iii) and (iv) are imposed on Selection~N, the number of sources 
with $\varpi<-800$~mas is reduced to 6. The similar number of very large negative 
and positive parallaxes, when similar criteria are applied, broadly supports the 
hypothesis that most of the spurious large parallaxes result from the previously 
described chance matching of the observations to distinct objects. 
(The same thing can of course happen with the resolved components of a physical 
double star, if the separation is $\lesssim 1$~arcsec.) The probability that it happens 
should decrease steeply with an increased number of available observations, or rather with 
the number of visibility periods (Sect.~\ref{sec:fallback}). That this is indeed the 
case is illustrated in Fig.~\ref{fig:histNegPlx}, where the tail of normalised negative 
parallaxes is plotted for Selection~N and for some subsets of it. Nominally, if the parallax 
errors were truly unbiased and Gaussian, we would expect to have no source at all with 
$-\varpi/\sigma_\varpi>6$. The blue curve shows the distribution for the sources
in Selection~N, which by Eq.~(\ref{eq:critP5}) all have at least six visibility periods. 
Requiring at least 7 or 10 visibility periods (green line/rings, and grey line/squares, 
respectively) drastically reduces the negative tail while retaining 85\% and 41\% of 
the sources. Requiring even more visibility periods only shrinks the sample without
changing the shape of the tail. If these criteria are applied to Selection~A, the
HR diagram gets cleaner at the faint end, but most of the points between the 
main sequence and white dwarfs around colour index 1 are still present. Increasing
the minimum number of visibility periods is therefore efficient for eliminating the 
most extreme spurious parallaxes, but not for cleaning the middle and upper part 
of the HR diagram. The red curve in Fig.~\ref{fig:histNegPlx} shows the distribution 
of negative parallaxes after the cut in Eq.~(\ref{eq:fG}), which is clearly more effective 
in removing the many parallaxes that are only moderately wrong.

\begin{figure*}
\centering
  \resizebox{0.33\hsize}{!}{\includegraphics{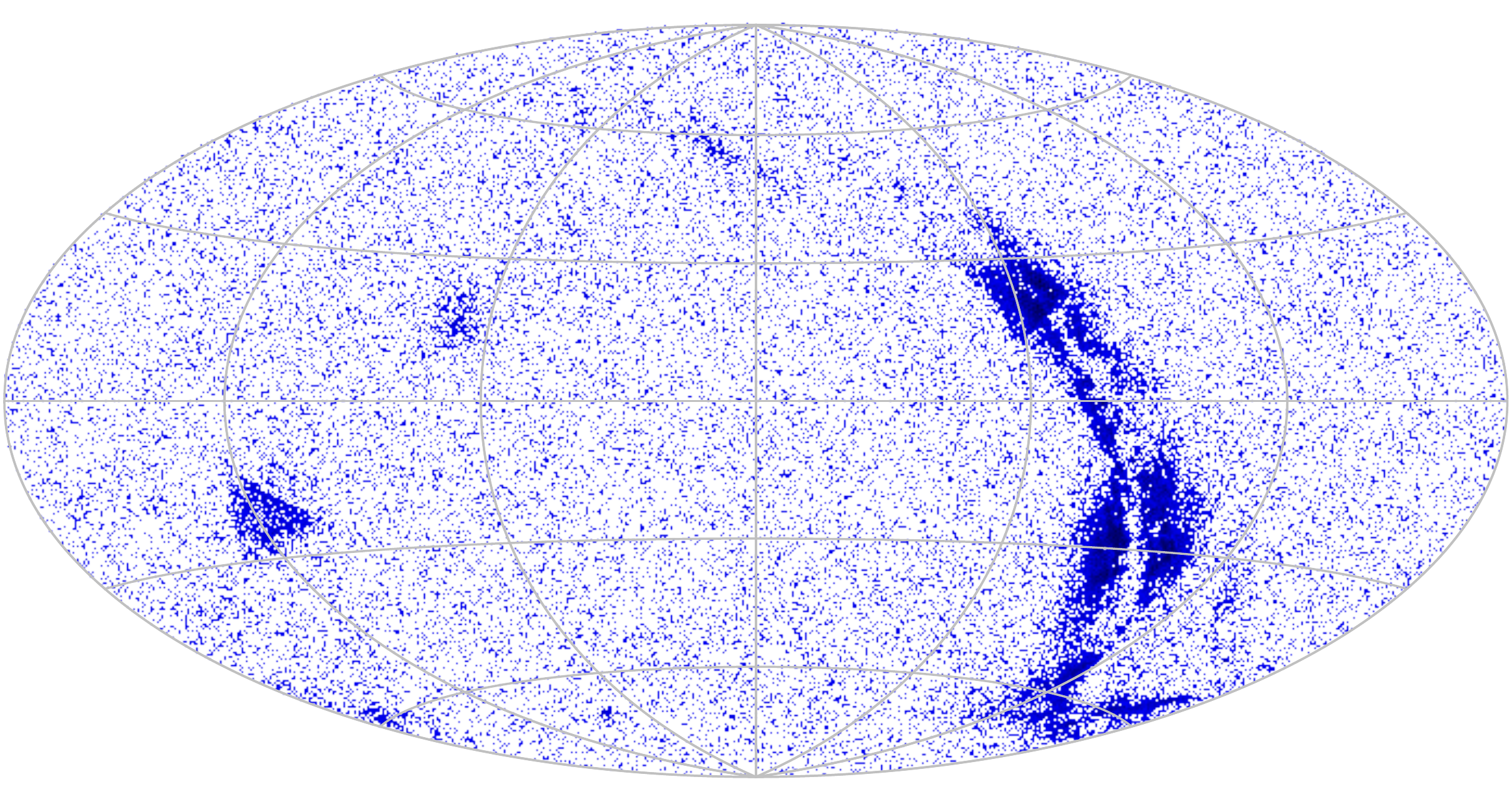}}
  \resizebox{0.33\hsize}{!}{\includegraphics{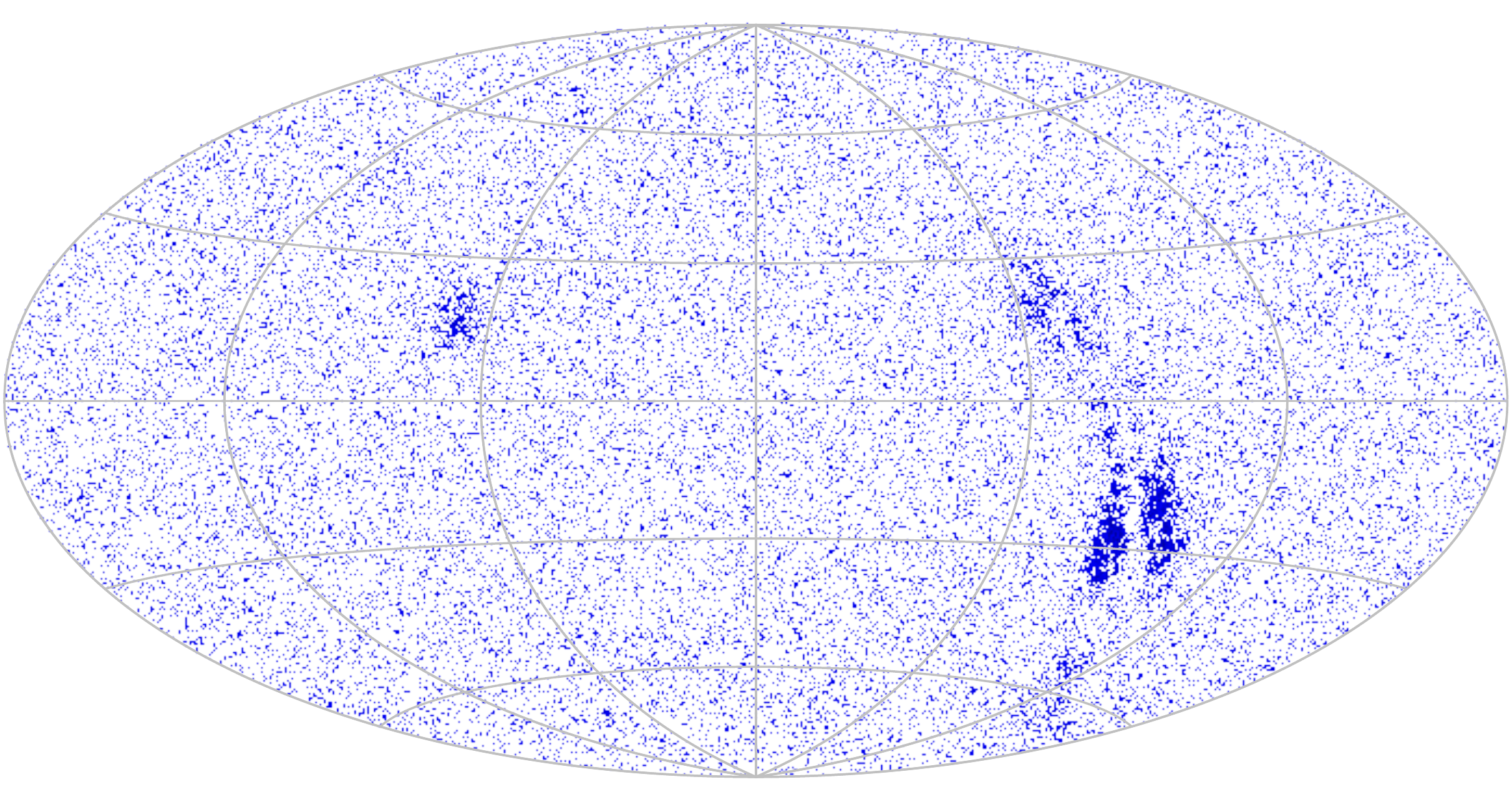}}
  \resizebox{0.33\hsize}{!}{\includegraphics{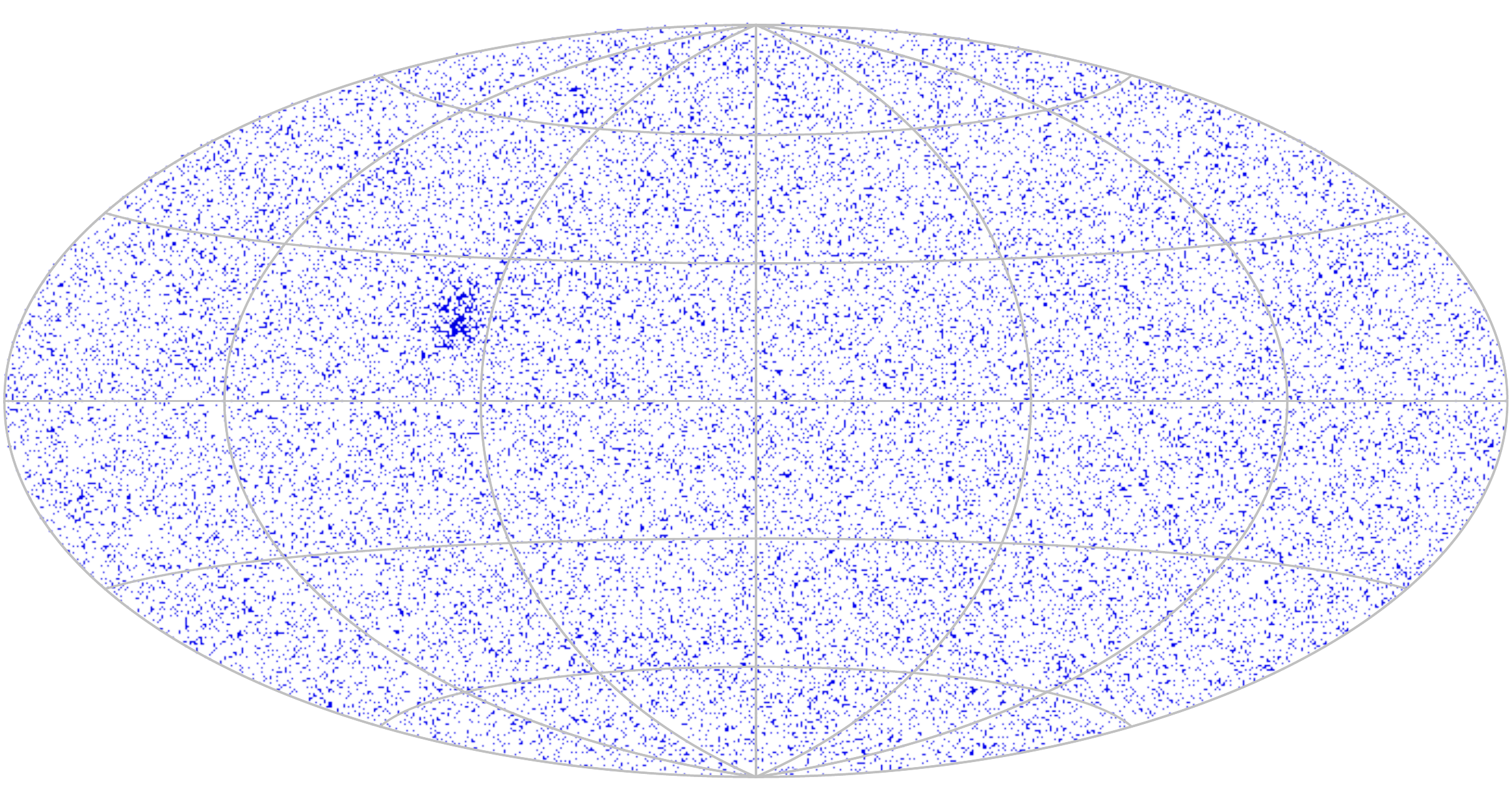}}
    \caption{Distribution in equatorial (ICRS) coordinates of sources formally within 50~pc.
    \textit{Left:} All 73\,246 sources with $\varpi>20$~mas.
    \textit{Middle:} The subset of 39\,478 sources satisfying Eq.~(\ref{eq:fG}).
    \textit{Right:} The subset of 34\,001 sources satisfying both Eqs.~(\ref{eq:fG}) and (\ref{eq:flx}).}
    \label{fig:plxGt20}
\end{figure*}

The effectiveness of the filters described above is also illustrated in Fig.~\ref{fig:plxGt20}. 
The left map shows the celestial distribution of the 73\,246 sources in \textit{Gaia} DR2 that are 
nominally within 50~pc from the Sun, i.e.\ with $\varpi>20$~mas. Stars in this volume should 
have a rather uniform distribution on the sky; yet the map shows strong features 
correlated with the density of faint stars (e.g.\ along the Galactic equator) or related to the 
scanning law (e.g.\ the triangular patch in the left part of the map). Much of these features 
disappear after applying the cut in Eq.~(\ref{eq:fG}), as shown in the middle map. Applying 
in addition the cut in Eq.~(\ref{eq:flx}) leaves 34\,001 sources with a nearly uniform distribution 
(right map). The remaining concentration of points at $(\alpha,\delta)\simeq (67^\circ,+16^\circ)$ 
is the Hyades cluster. It can be inferred that most of the remaining sources are real. Inevitably, 
however, the filtering eliminates also some real sources with valid solutions. In this example the
39\,245 sources removed by Eqs.~(\ref{eq:fG})--(\ref{eq:flx}) include at least some 700 actual 
nearby stars, among them Sirius~B, Kruger~60, Ross~614, $\eta$~Cas, $\pi^3$~Ori, and 
$\delta$~Eri. 

\end{document}